%% file: fmnew20.tex
\magnification=1200
\baselineskip=16truept
\hsize 15.5 truecm \vsize 21.7truecm  
\parindent=18truept
\parskip=3truept plus 2truept minus 3truept
\tolerance=10000
\hoffset= 0.25truecm  
\font\text=cmr10 at 12 truept
 at 11 truept
\font\note=cmr10 at 10 truept
\font\bf=cmbx10
\font\it=cmti10

\def\pni{\par \noindent}
\def\vsh{\smallskip}
\def\vs{\medskip}
\def\vvs{\bigskip}
\def\vsp{\par}
\def\vsn{\vsh\pni}
\def\cen{\centerline}
    
\def\eg{{\it e.g.}\ } \def\ie{{\it i.e.}\ }

   \def\rl{\item{-- \ }}
    
    \def\rf#1{\item{{#1}.\ }}  
\def\e{{\rm e}}
\def\exp{{\rm exp}}
\def\ds{\displaystyle}
\def\dis{\displaystyle}
\def\q{\quad}	 \def\qq{\qquad}

\def\l{\left} \def\r{\right}

\def\d{\partial}
   \def\dt{\partial t}
\def\dx{\partial x}     
\def\rec#1{{1\over{#1}}}

\def\barr{\widetilde}
\def\epsilons{{\tilde \epsilon(s)}}
\def\sigmas{{\tilde \sigma (s)}}

\def\Js{{\tilde J(s)}}
\def\Gs{{\tilde G(s)}}
\def\Fs{{\tilde F(s)}}
 
\def\NN{{\rm I\hskip-2pt N}}
\def\RR{\vbox {\hbox to 8.9pt {I\hskip-2.1pt R\hfil}}}
\def\CC{{\rm C\hskip-4.8pt \vrule height 6pt width 12000sp\hskip 5pt}}

\pageno=-1
\cen{CISM  \ LECTURE \ NOTES}
\cen{International Centre for Mechanical Sciences}
\cen{Palazzo del Torso, Piazza Garibaldi, Udine, Italy}
\vs\hrule\vvs
\vskip 0.5 truecm
\cen{{\bf FRACTIONAL \ CALCULUS :}}
\vsh
\cen{{\bf Some Basic Problems in Continuum and Statistical Mechanics}}
\vskip 0.5 truecm
\cen{Francesco MAINARDI}
\vsh
\cen{Department of Physics, University of Bologna}
\cen{Via Irnerio 46, I-40126 Bologna, Italy}
\cen{{\it e-mail: mainardi@bo.infn.it}}
\vskip 1.0 truecm
\cen{{\bf {\TeX} \ PRE-PRINT \ 58 pages : pp. 291-348}}
\vskip 1truecm
\def\indice{\leaders\hbox to 1 em {\hss.\hss}\hfill}
\def\hb#1{\hbox to 1.25 truecm{p. \hss#1}}
\line{\ \ \ \ ABSTRACT\indice  \hb{291}}
\line{1. LINEAR VISCOELASTICITY AND FRACTIONAL CALCULUS
   \indice \hb{292}}
\line{2. THE BASSET PROBLEM VIA  FRACTIONAL CALCULUS
    \indice \hb{303}}
\line{3.  BROWNIAN MOTION AND FRACTIONAL CALCULUS
  \indice   \hb{311}}
\line{4. THE FRACTIONAL DIFFUSION-WAVE EQUATION
  \indice \hb{321}}
\line{\ \ \ \  APPENDIX: THE WRIGHT FUNCTION
  \indice \hb{333}}
\line{\ \ \ \ REFERENCES \indice \hb{340}}
\vskip 1.0truecm
\hrule\vs
The paper is based on the lectures  delivered by the author
at the CISM Course
{\it Scaling Laws and Fractality in Continuum Mechanics: A Survey of the
Methods based on Renormalization Group and Fractional Calculus},
held at the seat of CISM, Udine, from 23 to 27 September 1996,
under the direction of Professors A. Carpinteri  and F. Mainardi.
\vsn
This {\TeX} pre-print is a revised version (November 2001)
of the chapter published in
 \vsn
 {\bf A. Carpinteri and F. Mainardi (Editors):
 Fractals and Fractional Calculus in Continuum Mechanics,
 Springer Verlag, Wien and New York 1997, pp. 291-348.}
  \vsn
Such book is the volume No. 378 of the series CISM COURSES AND LECTURES
[ISBN 3-211-82913-X]
\vfill\eject
$\null$
\vskip 5.0truecm
\cen{PREFACE to the REVISED VERSION}
\vsp
In this revised version a number of misprints have been corrected and
several improvements have been introduced.  All the plots have been
re-drawn by using the MATLAB system; for this the Author is grateful to
his students: D. Moretti, G. Pagnini, P. Paradisi, D. Piazza and
D. Turrini.
Up to some extent the references have been
up-dated to the year 2000. For further information about the
applications of fractional calculus  we recommend the recent
treatises
\pni
\rl
  R. Hilfer (Editor):
  {\it Applications of Fractional Calculus in Physics},
World Scientific, Singapore, 2000.
\par
\rl I. Podlubny:
 {\it Fractional Differential Equations},
Academic Press, San Diego, 1999.
\pni
To be informed on the  developing subject of the applications
of  fractional calculus in modelling  various phenomena,
we  suggest the interested readers to visit the WEB site
{\tt http://www.fracalmo.org}
devoted to the {\tt fra}ctional {\tt cal}culus {\tt mo}delling.
\pni
This 2012 E-print version for arXiv reproduces the 2001 Version. 
Since that time several papers of the author were published on related topics, see the home page
{\tt http://www.fracalmo.org/mainardi}. In particular we point out the book 
  F. Mainardi: {\it Fractional Calculus and Waves in Linear Viscoelasticity}, Imperial
College Press, London (2010), pp. 340, ISBN 978-1-84816-329-4, see:
{\tt http://www.icpress.co.uk/mathematics/p614.html}
\vskip 2.5 truecm
\cen{\copyright\ 1997, 2001, 2012 ~Prof. ~Francesco ~Mainardi - Bologna - Italy}
\vs
\cen{fmcism20.tex (old version), fmnew20.tex  (revised version) in plain
\TeX, 58 pages.}
\vfill\eject
\input epsf
\input fmmacro2.tex
\input fmhead
\HeadLinea{Fractional Calculus: Some Basic Problems} {F. Mainardi}
\input fmnew21.tex
\input fmnew22.tex
\input fmnew23.tex
\input fmnew24.tex
\input fmnew2A.tex
\input fmnew2R.tex
\bye

%% file: fmmacro2.tex
\baselineskip=16truept
\hsize 15.5 truecm \vsize 21.7truecm  
\parindent=18truept
\parskip=3truept plus 2truept minus 3truept
\tolerance=10000
\hoffset= 0.25truecm  
\font\text=cmr10 at 12 truept
 at 11 truept
\font\note=cmr10 at 10 truept
\font\bf=cmbx10
\font\it=cmti10

\def\pni{\par \noindent}
\def\vsh{\smallskip}
\def\vs{\medskip}
\def\vvs{\bigskip}
\def\vsp{\par}
\def\vsn{\vsh\pni}
\def\cen{\centerline}
    
\def\eg{{\it e.g.}\ } \def\ie{{\it i.e.}\ }

   \def\rl{\item{-- \ }}
    
    \def\rf#1{\item{{#1}.\ }}  
\def\e{{\rm e}}
\def\exp{{\rm exp}}
\def\ds{\displaystyle}
\def\dis{\displaystyle}
\def\q{\quad}	 \def\qq{\qquad}

\def\l{\left} \def\r{\right}

\def\d{\partial}
   \def\dt{\partial t}
\def\dx{\partial x}     
\def\rec#1{{1\over{#1}}}

\def\barr{\widetilde}
\def\epsilons{{\tilde \epsilon(s)}}
\def\sigmas{{\tilde \sigma (s)}}

\def\Js{{\tilde J(s)}}
\def\Gs{{\tilde G(s)}}
\def\Fs{{\tilde F(s)}}
 
\def\NN{{\rm I\hskip-2pt N}}
\def\RR{\vbox {\hbox to 8.9pt {I\hskip-2.1pt R\hfil}}}
\def\CC{{\rm C\hskip-4.8pt \vrule height 6pt width 12000sp\hskip 5pt}}

 \def\g{\widetilde{g}}
\def\u{\widetilde{u}}	


%% file: fmhead.tex
\def\riga{\vskip .1  truecm   \hrule \vskip .2	    truecm \noindent }
\def\HeadLinea#1#2{	\headline={\vbox to 0pt{\vss\noindent
{\ifnum \pageno=1  \hfill {\bf \folio}		 
\else {\ifodd \pageno			   
{\noindent     \hfill  {\it     #2} \quad {\bf \folio}
 }\riga
\else				 
{\noindent {\bf\folio} \quad  {\it  #1} \hfill 	}
\riga
\fi } \fi }			}}
}
\nopagenumbers
\voffset= 2 truecm

%% file: fmnew21.tex
\magnification=1200
\baselineskip=16truept
\hsize 15.5 truecm \vsize 21.7truecm  
\parindent=18truept
\parskip=3truept plus 2truept minus 3truept
\tolerance=10000
\font\text=cmr10 at 12 truept
 at 11 truept
\font\note=cmr10 at 10 truept
\font\bf=cmbx10
\font\it=cmti10

\def\pni{\par \noindent}
\def\vsh{\smallskip}
\def\vs{\medskip}
\def\vvs{\bigskip}
\def\vsp{\par}
\def\vsn{\vsh\pni}
\def\cen{\centerline}
    
\def\eg{{\it e.g.}\ } \def\ie{{\it i.e.}\ }

   \def\rl{\item{-- \ }}
    
    \def\rf#1{\item{{#1}.\ }}  
\def\e{{\rm e}}
\def\exp{{\rm exp}}
\def\ds{\displaystyle}
\def\dis{\displaystyle}
\def\q{\quad}	 \def\qq{\qquad}

\def\l{\left} \def\r{\right}

\def\d{\partial}
  \def\dt{\partial t}
\def\dx{\partial x}     
\def\rec#1{{1\over{#1}}}

\def\barr{\widetilde}
\def\epsilons{{\widetilde \epsilon(s)}}
\def\sigmas{{\widetilde \sigma (s)}}

\def\Js{{\widetilde J(s)}}
\def\Gs{{\widetilde G(s)}}
\def\Fs{{\wiidetilde F(s)}}
 
\def\NN{{\rm I\hskip-2pt N}}
\def\RR{\vbox {\hbox to 8.9pt {I\hskip-2.1pt R\hfil}}}
\def\CC{{\rm C\hskip-4.8pt \vrule height 6pt width 12000sp
  \hskip 5pt}}

\pageno=291
$\null$
\vskip 1.6truecm
\cen{{\bf FRACTIONAL \ CALCULUS :}}
\vsh
\cen{{\bf Some Basic Problems in Continuum and 
  Statistical Mechanics}}
\vskip 1.4 truecm
\cen{Francesco MAINARDI}
\footnote {} {This research was
partially supported by the Ministry of University
and by the National Research Council (CNR-GNFM).
The author is grateful to Professor Rudolf Gorenflo
for fruitful discussions and comments.}
\vsh
\cen{Department of Physics, University of Bologna}
\cen{Via Irnerio 46, I-40126 Bologna, Italy}
\cen{E-mail: {\tt mainardi@bo.infn.it} URL: {\tt www.fracalmo.org}}
\vskip 1.6 truecm
\line{ABSTRACT \hfill}
\vsn
We review some applications   of fractional
calculus 
developed by
the author (partly in  collaboration with others)
to treat some  basic problems  in continuum and statistical mechanics.
The problems in continuum mechanics concern
mathematical modelling of viscoelastic bodies ($\S 1$),
and unsteady motion of a particle in a viscous fluid, \ie
the {Basset problem}  ($\S 2$).
In the former analysis fractional calculus leads us to introduce
intermediate models of viscoelasticity which
 generalize
the classical spring-dashpot models.
The latter analysis induces us to introduce a hydrodynamic model
suitable to revisit in $\S 3$ the classical theory of
the Brownian motion,
which is a relevant topic  in statistical mechanics.
By the tools of fractional calculus we explain the long tails in the
velocity correlation  and   in the displacement variance.
In $\S 4$ we consider the {fractional diffusion-wave equation},
which is obtained from the classical diffusion 
equation  by replacing the first-order time derivative by
a fractional derivative of order  $\beta $ with $ 0 <\beta  <2\,.$
Led by our analysis we express the fundamental solutions (the {Green
functions}) in terms of two interrelated {auxiliary} functions
 in the similarity  variable,
which turn out to be of Wright type (see Appendix),
and
to distinguish slow-diffusion processes ($0<\beta  <1$) from
intermediate processes ($1<\beta <2$).
\vsh\noindent
\vfill\eject
\line{1. LINEAR VISCOELASTICITY AND FRACTIONAL CALCULUS \hfill} \vsn
\line{1.1 {\it Fundamentals  of Linear Viscoelasticity}\hfill}
\vsp
Viscoelasticity is a property possessed by bodies which, when deformed,
exhibit both viscous and elastic behaviour through
simultaneous dissipation and storage of mechanical energy.
Here, for simplicity, we are restricting the discussion only to the
scalar case, \ie to  one-dimensional problems.
 We denote the
stress by $\sigma= \sigma (x,t)$
and the strain by $\epsilon = \epsilon (x,t)$ where $x$ and $t$
are the space and time variables, respectively.
\vsp
According to the {\it linear} theory  of viscoelasticity,
at a fixed position, the body may be considered  a
linear
system with the stress (or strain) as the excitation function (input)
and the strain (or stress) as the response function (output).
Consequently, the response functions to
an excitation expressed by the Heaviside step function $\Theta(t)$
are known  to	play a fundamental role both from a
mathematical  and physical point of view, see \eg Gross [1],
Bland [2], Caputo \& Mainardi [3], Christensen [4] and Pipkin [5].
\vsp
We denote by $J(t)$ the strain response  to the unit step of
 stress ({\it creep test}),
 and by $G(t)$ the stress response to a unit step of strain
({\it relaxation test}).
These functions $J(t)\,,\, G(t)$  are usually referred to as the
{\it creep compliance} and {\it relaxation modulus}
respectively, or, simply, the {\it material functions}
of the viscoelastic body. In view of the causality requirement,
both the functions are
  causal (\ie  vanishing for $t<0$).
The limiting values
of the material functions
for $t \to 0^+$ and $t \to +\infty$  are related to the
instantaneous (or glass) and equilibrium behaviours of the viscoelastic
body, respectively. As a consequence, it is usual to denote
   $ J_g := J(0^+)$  the {\it glass  compliance},
  $J_e := J(+\infty)$  the {\it equilibrium compliance},
and
 $ G_g := G(0^+) $  the {\it glass  modulus},
 $ G_e := G(+\infty)$ the {\it equilibrium  modulus}.
As a matter of fact, both the material functions are
non-negative. Furthermore, for	$0< t < +\infty\,,$
$J(t)$ is a differentiable {\it increasing} function of time,
      \ie
$$  t\in \RR^+ \,,\q  {dJ \over dt} > 0  \; \Longrightarrow\;
	   0 \le J(0^+) < J(t) < J(+\infty) \le +\infty\,,
$$
while $G(t)$ is a differentiable {\it decreasing} function of time,
 \ie
$$ t\in \RR^+\,, \q   {dG \over dt} < 0   \; \Longrightarrow\;
	   +\infty \ge G(0^+) > G(t) >	G(+\infty) \ge 0\,.
$$
The above characteristics of monotonicity of $J(t)$ and $G(t)$
are  related respectively   to the physical phenomena
of  strain {\it creep} and stress {\it relaxation},
which are experimentally observed.
Later on, we shall outline more restrictive
mathematical conditions that  the material functions must usually
satisfy  to agree with the most common experimental observations.
\vsp
By using the Boltzmann superposition principle,
the general stress--strain relation
can be expressed in terms of one material function [$J(t)$ or $G(t)$]
through a linear hereditary integral of Stieltjes
type, namely
$$
 \epsilon(t) = \int_{-\infty}^t\!\! J(t-\tau)\,d\sigma(\tau)\,,
  \q {\rm or} \q \sigma (t) =
\int _{-\infty}^t  \!\!G(t-\tau)\, d\epsilon (\tau) \,.
 \eqno(1.1)$$
Usually,  the viscoelastic body is quiescent  for
all times prior to some starting instant that we assume as $t=0\,.$
Thus,
under the assumption of causal histories, differentiable
for $t \in \RR^+\,,$
the representations (1.1) reduce to 
$${\epsilon (t)= \int_{0^-}^t  \!\! J(t-\tau)\, d\sigma (\tau )
 =  \sigma (0^+)\, J(t) + \int_0^t  \!\! J(t-\tau)\,
 \dot\sigma (\tau ) \, d\tau} \,,
  \eqno(1.2a)$$
$${\sigma  (t) =
\int _{0^-}^t  \!\!G(t-\tau )\, d\epsilon  (\tau )   =
  \epsilon(0^+)\, G(t) + \int_0^t  \!\! G(t-\tau)\,
 \dot \epsilon	(\tau ) \, d\tau}\,,
  \eqno(1.2b)$$
where the superposed dot denotes time-differentiation.
The lower limits of integration in Eqs (1.2) are written
as $0^-$ to account for the possibility that $\sigma(t)$ and/or
$\epsilon (t) $  exhibit jump discontinuities at $t=0$,
and therefore their derivatives $\dot \sigma (t)$ and
$\dot \epsilon (t)$ involve a delta function $\delta (t)\,. $
Another form of the constitutive equations can be obtained from Eqs (1.2)
integrating by parts:
$${\epsilon (t)=
   J_g\, \sigma (t) + \int_0^t	\!\! \dot J(t-\tau)\,
 \sigma (\tau ) \, d\tau}\,,
  \eqno(1.3a)$$
$${\sigma  (t) =
   G_g \, \epsilon (t) + \int_0^t  \!\! \dot G(t-\tau)\,
  \epsilon  (\tau ) \, d\tau}\,.
  \eqno(1.3b)$$
Here we have assumed  $J_g>0$ and $J_g < \infty \,,$ see (1.7).
The causal functions $\dot J(t)$  and $\dot G(t)$ are referred to as
the {\it rate of creep (compliance)} and
 the {\it rate of relaxation (modulus)}, respectively;
they play the role of memory functions	in the constitutive
equations (1.3).
Being of convolution type,  equations (1.2) and (1.3) can be conveniently
treated by the technique of Laplace transforms to yield
$$ { \epsilons = s\, \Js \, \sigmas\,,\qq
  \sigmas = s\, \Gs \, \epsilons}\,. \eqno(1.4)$$
 Since the creep and relaxation integral formulations must agree
with one another, there must be a one-to-one correspondence
between the relaxation modulus and the creep compliance.
The basic relation  between $J(t)$ and $G(t)$ is found noticing
the following {\it reciprocity relation} in the Laplace domain,
deduced from Eqs (1.4),
  $$ {s\, \Js  = \rec{ s\, \Gs} \iff
    \Js \, \Gs = \rec{s^2}}
 \,. \eqno(1.5)$$
Then, inverting the R.H.S. of (1.5), we obtain
$$ {J(t)   \,*\, G(t) := \int_0^t \!\! J(t-\tau)\,G(\tau)\,d\tau
	  = t}\,. \eqno(1.6)$$
\vfill\eject
\noindent
Furthermore, in view of 
the limiting theorems for the Laplace transform
we can deduce from the L.H.S of (1.5) that
$${J_g = \rec {G_g} \,, \qq J_e = \rec{G_e}} \,, \eqno(1.7)$$
with the convention that $0$ and $+\infty$ are reciprocal to each other.
These remarkable relations allow us to classify the viscoelastic bodies
according to their instantaneous and equilibrium responses.
In fact,  we easily recognize four
possibilities  for the limiting values	of the creep
compliance and relaxation modulus, as listed in Table I.
$$  \vcenter{
\vbox{
\offinterlineskip
\halign{
&
\vrule#
&
\strut
\quad
\hfil
#
\quad
\cr
\noalign{\hrule}
      height 2 pt
 & \omit  && \omit && \omit && \omit && \omit & \cr
 & $Type$ && $J_g$ && $J_e$ && $G_g$ && $G_e$	& \cr
      height 2 pt
 & \omit  && \omit && \omit && \omit && \omit & \cr
\noalign{\hrule}
      height 2 pt
 & \omit  && \omit && \omit && \omit && \omit & \cr
 & $ I $  && $ \;>0\; $&& $<\infty $  && $ <\infty $ && $\;>0\;$ & \cr
 & $ II$  && $ \;>0\; $&& $=\infty $  && $ <\infty $ && $\;=0\;$ & \cr
 & $III$  && $ \;=0\; $&& $<\infty $  && $ =\infty $ && $\;>0\;$ & \cr
 & $ IV$  && $ \;=0\; $&& $=\infty $  && $ =\infty $ && $\;=0\;$ & \cr
      height 2 pt
 & \omit  && \omit && \omit && \omit && \omit & \cr
\noalign{\hrule}
  }
  }
  }
$$
\vsh
\cen{{\bf Table I}: The four types of viscoelasticity}
\vsp
From a mathematical point of view
the material functions turn out to be of the following form [1]
$$ { \l\{ \eqalign{
   J(t) &= J_g + \chi_+ \, \int_{0}^{\infty}
  R_\epsilon (\tau)\, \l( 1-\e^{\ds-t/\tau}\r)\, d\tau
	+ J_+\, t \,,	\cr
   G(t) & = G_e +  \chi_- \, \int_{0}^{\infty}
  R_\sigma  (\tau)\, \e^{\ds-t/\tau}\, d\tau
	+ G_-\, \delta (t)\,.	 \cr }\r.}\eqno(1.8) $$
where all the coefficients and functions are non negative. The
function   $  R_\epsilon(\tau) $ is
referred to as the {\it retardation spectrum} while
$  R_\sigma (\tau) $   as the  {\it relaxation	spectrum}.
For the sake of convenience we shall denote by $R_*(\tau)$
anyone of the two   spectra.
The spectra must  necessarily be locally summable
in $\RR^+\,; $ if they are summable, the supplementary normalization
condition   $ \int_{0}^{\infty}  R_* (\tau)\, d\tau  = 1\,$
is required for the sake of convenience.
We devote particular attention to the
integral contributions	to the material functions (1.8),
 \ie
$$   \l\{ \eqalign{
 \Psi(t) &:=   \chi_+ \,
   \int_{0}^{\infty}
  R_\epsilon (\tau)\, \l( 1-\e^{\ds-t/\tau}\r)\, d\tau	\;
 \Longrightarrow \;  (-1)^n \, {d^n\Psi\over dt^n} < 0\,,\q n \in \NN\,,
  \cr
  \Phi(t) &:=
 \chi_- \,
  \int_{0}^{\infty}
  R_\sigma (\tau)\, \e^{\ds-t/\tau}\, d\tau \;
   \Longrightarrow  \; (-1)^n \, {d^n\Phi\over dt^n} > 0\,,\q n \in \NN\,.
 \cr}
   \r.\eqno(1.9)$$
\noindent
The positive functions $\Psi(t)$ and $\Phi(t)$ are  simply  referred to as
the {\it creep} and {\it relaxation} functions, respectively.
According to  standard definitions, see \eg  [6],  
the alternating sign properties outlined in the R.H.S. of (1.9)
imply
that the creep function is of {\it Bernstein type},
and  the relaxation function  is  {\it completely monotone}.
In particular, we recognize that $\Psi(t)$ is an increasing
function with $\Psi(0) =0$ and
$\Psi(+\infty) = \chi_+ \,$ or $\, +\infty\,, $
while $\Phi(t)$ is a decreasing
function with $\Phi(0) = \chi_-  \,$ or $\, +\infty$ and
$\Phi(+\infty) = 0\,. $
\vfill\eject 
\line{1.2 {\it The Mechanical Models}\hfill}
 \vsp
    To get some feeling for linear viscoelastic behaviour,
it is useful to consider the simpler behaviour of analog
{\it mechanical  models}. They are
 constructed from linear springs and dashpots,
disposed singly and in branches of two (in series or in parallel),
as it is indicated in Fig. 1-1.
\cen{\epsffile{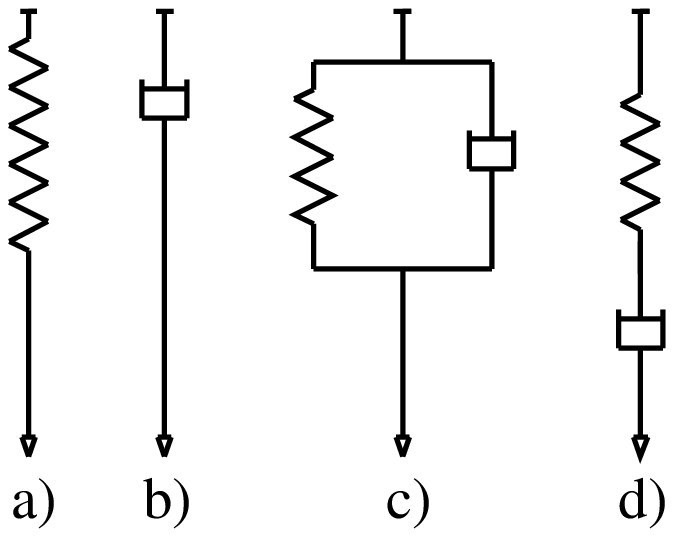}}
\cen{{\bf Fig. 1-1}}
\vsh
\cen{{
The elements of the mechanical models:
 a) Hooke, b) Newton, c) Voigt, d) Maxwell}}
\vs\vsp 
As analog of stress and strain, we use the total extending force
and the total extension.
We note that when two  elements are combined in series [in parallel],
their compliances [moduli] are additive. This can be stated
as  a combination rule: {\it creep compliances add in  series,
while relaxation moduli add in parallel}.
\vsp
The mechanical models play an important role in the literature which is
justified by the historical development.
In fact, the early theories were established with the aid of these
models, which are still helpful to visualise properties and laws of the
general theory, using the combination rule.
\vsp
Now, it is worthwhile to consider the simplest mechanical models
and provide  their governing  stress-strain
relations along with the related material functions.
We point out that the technique of Laplace transform allows one to
easily obtain the requested material functions from the
governing equations.
\vsp
The spring, see Fig. 1-1a), is the elastic (or storage) element, as for it
the force is proportional to the  extension;
it represents a perfect elastic body obeying the Hooke
law (ideal solid).
This model is thus referred to as  the {\it Hooke} model.
We have
    $${
 \sigma(t)  = m\, \epsilon (t)
\qq Hooke \qq
 \left \{ \eqalign{
 J(t) &= 1/m \cr
 G(t) &= m   \cr} \right.} \eqno(1.10)$$
\vfill\eject
\vsp
The dashpot, see  Fig. 1-1b), is the viscous (or dissipative) element,
the force being proportional to the rate of extension;
it represents a perfectly viscous body obeying the Newton law (perfect
liquid).
This model is thus referred to as  the {\it Newton} model.
We have
    $$ {
\sigma(t)  = b\, {d\epsilon\over dt}
 \qq	Newton \qq
   \left\{ \eqalign{
 J(t) &= t/b\cr
 G(t) &= b \, \delta (t) \cr} \right.}\eqno(1.11)$$
We note that
 the {\it Hooke} and {\it Newton} models represent the limiting cases of
viscoelastic  bodies of type $I$ and $IV$, respectively.
\vsp
A branch constituted by a spring in parallel with a dashpot is known as
  the	{\it Voigt}  model, see Fig. 1-1c). We have
$$ {
\sigma(t)  = m\, \epsilon (t) +b\, {d\epsilon\over dt}
 \qq Voigt \qq \left\{ \eqalign{
   J(t) &= \rec{m} \l[ 1-\e^{\ds - t/\tau _\epsilon }\r]\cr
 G(t) &= m  +  b \,  \delta(t) \cr}\right.}
  \eqno(1.12)$$
where $\tau _\epsilon  = b/m\, $ is referred to as
the {\it  retardation time}.
\vsp
A branch constituted by a spring in series with a dashpot is known as
 the  {\it  Maxwell} model, see Fig. 1-1d). We have
$$ {
\sigma(t) +a \, {d\sigma \over dt}    = b\, {d\epsilon \over dt}
 \qq
 Maxwell \qq \left\{ \eqalign{
 J(t) &=  {a\over b} +{t\over b}   \cr
 G(t) &= {b \over a}\,\e^{\ds -t/\tau_\sigma }\cr}\right. }
   \eqno(1.13)$$
where $\tau _\sigma   = a\,$ is is referred to as  the
{\it the relaxation time}.
\vsp
 The {\it Voigt} and the {\it Maxwell} models are thus the simplest
viscoelastic bodies of type   $III$ and $II$, respectively.
The {\it Voigt} model exhibits	an exponential	(reversible)
strain creep but no stress relaxation; it is also referred to
as the retardation element.
The {\it Maxwell} model exhibits  an exponential (reversible) stress
relaxation and a linear (non reversible) strain creep; it is
also referred to as the relaxation element.
\vsp
Adding a spring either in series to a Voigt model, see	Fig. 1-2a), or
in parallel to a Maxwell model, see Fig. 1-2b), means,
 according to the combination rule,
to add a positive constant both to the	Voigt-like creep compliance
and to the  Maxwell-like relaxation modulus so that we obtain $J_g >0$
and $G_e >0\,. $
Such a model was introduced by Zener [7] with the denomination of
{\it Standard Linear Solid} ($S.L.S.$).
 We have
$$ {
\l[1 +a \, {d \over dt}\r] \sigma(t) =
 \l [ m+ b\, {d \over dt}\r] \epsilon (t) \q SLS \q
 \left\{ \eqalign{
 J(t) &=  J_g + \chi_+\,  \l[ 1-\e^{\ds - t/\tau_\epsilon}\r]  \cr
 G(t) &= G_e + \chi_- \,\e^{\ds -t/\tau_\sigma}\cr}\right. }
   \eqno(1.14)$$
\vfill\eject
\noindent
$$ {\l\{\eqalign{
 J_g &= {a\over b}\,, \q
 \chi_+ ={1\over m}- {a\over b}\,,\q
       \tau_\epsilon ={b\over m}\,,\cr
 G_e &= m\,, \q
   \chi_- = {b\over a}- m \,, \q\;
       \tau_\sigma = a\,. \cr}
    \r.} \eqno(1.15)$$
We point out that the condition
$ 0< m<b/ a$  ensures
that  $\chi_+\, ,\,\chi_- $ are positive and hence
$0< J_g <  J_e < \infty \,,$ $\,0 <G_e <G_g < \infty\,$
and
$\, 0 < \tau _\sigma <\tau_\epsilon < \infty \,.$
The $S.L.S.$ is  the simplest (3-parameter)  viscoelastic body
of type $I\,. $
On the other hand, adding a dashpot either in series to a Voigt model,
see  Fig. 1-2c), or
in parallel to a Maxwell model, see Fig. 1-2d), we
obtain the simplest (3-parameter)  viscoelastic body
of type $IV\,. $
\cen{\epsffile{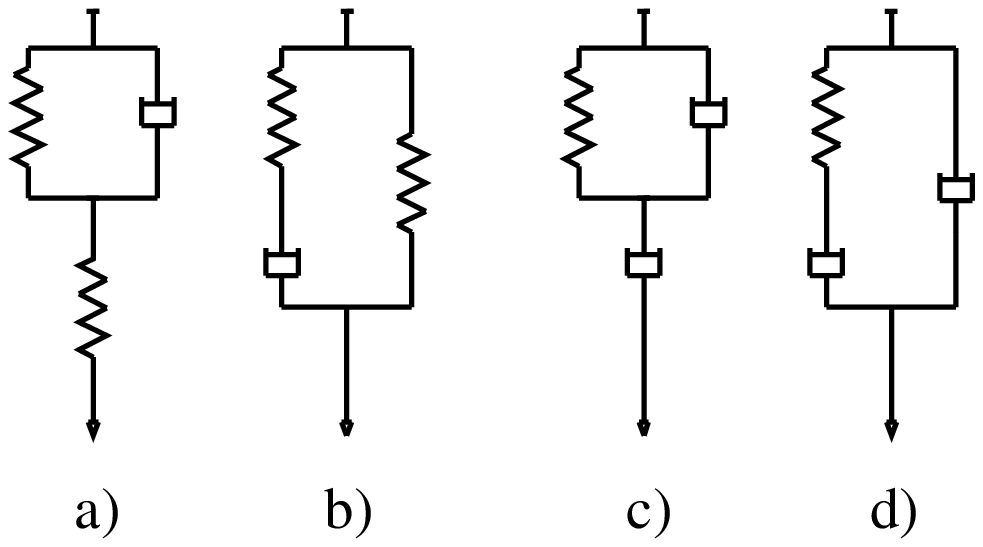}}
\cen{{\bf Fig. 1-2}}
\vsh
\cen{{a) spring in series with Voigt, b) spring in parallel with Maxwell;}}
\cen{{c) dashpot in series with Voigt, d) dashpot in parallel with
     Maxwell.}}
 \vsp
Based on the combination rule, we can construct
models whose  material functions    are of the following type
$$ { \l\{ \eqalign{
   J(t) &= J_g +\sum_{n} J_n \l[ 1-\e^{\ds-t/\tau_{\epsilon,n}}\r]
	+ J_+\, t \,,  \cr
   G(t) & = G_e +  \sum_{n} G_n \,\e^{\ds - t/\tau_{\sigma,n}}
	+ G_-\, \delta (t)\,,	\cr }\r.}\eqno(1.16) $$
where all the coefficient are non-negative.
These functions must be interrelated
because of the
{\it reciprocity relation} (1.5) in the Laplace domain.
Appealing to the theory of  Laplace transforms [2],
it turns out  that
stress-strain  relation  must be   a linear differential
 equation with constant (positive) coefficients of the following form
$$ {
\l[ 1+ \sum_{k=1}^p \,a_k\,{d^k\over dt ^k}\r] \, \sigma (t) =
\l[ m+ \sum_{k=1}^q \,b_k\,{d^k\over dt ^k}\r] \, \epsilon (t)\,, }
\q p=q \q \hbox{or} \q	p=q+1\,.
  \eqno(1.17)$$
Eq. (1.17) is referred to as the
{\it operator equation} for the mechanical models.
\vfill\eject
\line{1.3 {\it The Fractional Viscoelastic Models}\hfill}
 \vsp
Let us now consider a {\it creep compliance} of the form
$$ {J(t) = \Psi(t) = a \, {t^\alpha   \over \Gamma (1+\alpha  )}
	\,, \q a >0\,, \q 0<\alpha  <1\,, }\eqno(1.18) $$
where $\Gamma$ denotes the Gamma function.
Such behaviour is found to be of some interest in
creep experiments; usually it is referred to as
{\it power-law} creep.
This law appears compatible with the mathematical theory presented
in the previous sub-section, in that there exists a corresponding
{\it retardation spectrum}, locally summable, which reads
$$
    R_\epsilon	 (\tau )=    {\sin \pi \alpha  \over \pi }\,
   \rec{\tau  ^{1-\alpha  }}\,. \eqno(1.19)$$
For such a model the
{\it relaxation modulus} can be derived from the reciprocity relation (1.5)
and  reads
$$ {G(t) = \Phi(t) = b	\, {t^{-\alpha }  \over \Gamma (1-\alpha  )}
	\,, \q b = 1/a >0\,.}\eqno(1.20) $$
However, the corresponding {\it relaxation spectrum} does not
exist in the ordinary sense, in that it would be
$$ R_\sigma (\tau )=	{\sin \pi \alpha  \over \pi }\,
   \rec{\tau^{1+\alpha }}\,,\eqno(1.21)$$
and thus not locally summable.
The stress-strain relation in the {\it creep representation},
 obtained from (1.1) and  (1.18) is therefore
$$ \epsilon (t) = {a \over \Gamma (1+\alpha  )}\,
  \int_{-\infty}^t (t-\tau )^\alpha  \, d\sigma \,.\eqno(1.22) $$
Writing $d\sigma = \dot \sigma (\tau)\, d\tau $ and
integrating by parts,  we finally obtain
$$ {\epsilon (t) = {a \over \Gamma (\alpha  )}\,
  \int_{-\infty}^t (t-\tau )^{\alpha -1} \, \sigma(\tau )\, d\tau
  = a \, J_{-\infty}^\alpha  \, \l[\sigma (t)\r] \,,} \eqno(1.23)$$
where $ J_{-\infty}^\alpha $ denotes the {\it fractional integral} of
order $\alpha $ with starting point $-\infty\,, $ see Gorenflo
\& Mainardi [8].
\vsp
In the {\it relaxation representation} the stress-strain relation
can be	obtained from (1.1) and (1.20). Writing
$d\epsilon  =\dot \epsilon (\tau)\, d\tau\,, $	we obtain
$$ {\sigma  (t) = {b  \over \Gamma (1-\alpha  )}\,
  \int_{-\infty}^t (t-\tau )^{-\alpha } \,
 \l[{d\epsilon	(\tau )\over d\tau }\r]\, d\tau
  = b \, {d^\alpha \epsilon (t)\over dt^\alpha	}\,,}
\eqno(1.24)$$
where
$$ {d^\alpha \over dt^\alpha  } =
D_{*\,-\infty}^\alpha  = J_{-\infty}^{1-\alpha }
    \, {d \over dt} \eqno(1.25)$$
denotes the {\it Caputo  fractional derivative} of order $\alpha  $
with starting point $-\infty \,, $ see Gorenflo and Mainardi [8].
\vfill\eject
  Of course, for causal histories, the starting
point  of the integrals in (1.22-25) is $0\,, $
so that we must consider the operators $J^\alpha $ and $D_*^\alpha\,. $
Since in the limit as $\alpha  \to 1$  the fractional integral
 and derivative tend to the ordinary integral and derivative,
 respectively, we note that the classical Newton model
can be recovered from (1.23) and (1.24) by setting $\alpha =1\,. $
\vsp
In textbooks on rheology the  relation (1.24), when
expressed with the fractional derivative,
   is usually referred to as the {\it Scott-Blair  stress-strain law} from
the name of the scientist [9], who in earlier times proposed
such a constitutive equation to introduce a material
property that is intermediate between the elastic modulus (Hooke
solid) and the coefficient of viscosity (Newton fluid).
\vsp
The use of {\it fractional calculus} in linear viscoelasticity
leads to a generalization of the classical mechanical models
 in that the basic   Newton element (dashpot)  is
substituted by the more general Scott-Blair element.
In fact,   we can construct  the class of these generalized models
from Hooke and Scott-Blair elements,
disposed singly and in branches of two (in series or in parallel).
The  material functions are obtained
using the combination rule; their determination
is made easy if we take into account the following
{\it correspondence  principle} between the classical and fractional
mechanical models, as stated by Caputo \& Mainardi [3],
\def\E{{\rm E}}
$${ (0<\alpha  <1)\q \l\{
 \eqalign{
 \,&t	\to {\;t^\alpha  \over \Gamma (1+\alpha  )}\,, \cr
 \,&\delta (t) \to {\;t^{-\alpha } \over \Gamma (1-\alpha )}\,, \cr
 \,& \e^{\ds -t/\tau} \to \E_\alpha  [-(t/\tau)^\alpha ]\,, \cr }
    \r. }  \eqno(1.26)$$
where 
$\,\E_\alpha \,$ denotes the Mittag-Leffler function of order
$\alpha \,,$ heavily used in [8].
\vsp
We verify the {\it correspondence  principle} by considering
the {\it fractional S.L.S.},
formerly introduced by Caputo \& Mainardi [10]
in 1971. Such model is based on the
following  operator equation of fractional order,
which generalises the operator equation (1.14)	for the {\it S.L.S.},
$$ {
\l[ 1+a\, {d^{\,\alpha }  \over dt^{\,\alpha }}\r] \,\sigma (t) =
 \l[ m + b\, {d^{\,\alpha } \over dt^{\,\alpha }} \r] \, \epsilon(t) \,,
	\q 0 <\alpha  \le 1\,. }
\eqno(1.27)  $$
This equation is better analysed
in the Laplace domain where we obtain
$$ {
\l( 1+a\,s^{\,\alpha } \r) \,\sigmas =
 \l( m + b\, s^{\,\alpha } \r) \, \epsilons
   \, \iff \,
  s\Js	=      \rec{s\Gs}= {1+a\,s^{\,\alpha }\over m +b\,s^{\,\alpha }}\,.}
\eqno(1.28)  $$
From the fractional operator equation
we can obtain as particular cases, besides the trivial elastic
model ($a=b=0$) and the {fractional Newton} or {Scott-Blair} model
($a=m=0\,,\, b=\beta$) already considered,
the {fractional  Voigt} model ($a=0$) and
the {fractional Maxwell} model ($m=0$).
\vfill\eject
\vsp
Working in the Laplace domain  and then inverting,
we obtain  for the  fractional Voigt and Maxwell models
$$
\sigma(t) = m\, \epsilon (t) +b\, {d^\alpha  \epsilon\over dt^\alpha  }
 \q Fractional \;Voigt \q \left\{ \eqalign{
   J(t) &= \rec{m} \l\{1-\E_\alpha \l[-(t/\tau_\epsilon)^\alpha  \r]\r\}\cr
 G(t) &= m +  b \,  {\;t^{-\alpha  }\over \Gamma(1-\alpha )} \cr}\right.
  \eqno(1.29)$$
$${
\sigma(t) +a \, {d^\alpha  \sigma\over dt^\alpha }
   = b\, {d^\alpha  \epsilon \over dt^\alpha }
 \q
 Fractional \; Maxwell \q \left\{ \eqalign{
 J(t) &=  {a\over b} +{1\over b} \, {\;t^\alpha  \over \Gamma(1+\alpha	)}  \cr
 G(t) &= {b \over a}\,\E_\alpha  \l[ -(t/\tau_\sigma)^\alpha  \r]\cr}
  \right. }   \eqno(1.30)$$
where $(\tau_\epsilon)^\alpha	= b/m\, $  and
       $(\tau_\sigma)^\alpha   = a\,.$
\vsp
Having recognized with (1.29-30)  the validity of the
Caputo-Mainardi {\it correspondence principle} for the basic models,
we are allowed	to use this principle to obtain the material functions
of higher models, including the fractional S.L.S.,  along with
the corresponding operator equations of fractional order.
Thus, by generalizing (1.16), we  obtain
$$ { \l\{ \eqalign{
  J(t) &= J_g+\sum_{n} J_n\l\{1-\E_\alpha \l[-(t/\tau_{\epsilon,n})^\alpha \r]\r\}
    + J_+\,{\;t^{\alpha } \over \Gamma(1+\alpha )}  \,,  \cr
  G(t) & = G_e +\sum_{n} G_n \,\E_\alpha \l[-(t/\tau_{\sigma,n})^\alpha \r]
	+ G_-\,{\;t^{-\alpha } \over \Gamma(1-\alpha )}\,,  \cr }
   \r.}\eqno(1.31) $$
where all the coefficients are non negative. Extending the procedures
of the classical mechanical models, we will get the
{\it fractional operator equation} in the form which
properly generalises (1.17), \ie
$$ {
\l[1+\sum_{k=1}^p\,a_k\,{d^{\,\alpha _k}\over dt^{\,\alpha _k}}\r] \,\sigma (t) =
\l[m+\sum_{k=1}^q\,b_k\,{d^{\,\alpha _k}\over dt^{\,\alpha _k}}\r] \,\epsilon (t)
 \,, \q \alpha	_k = k + \alpha  -1\,. }   \eqno(1.32)$$
 \vsp
We conclude this section pointing out
the presence of the Mittag-Leffler function in (1.31).
In fact,
 the {\it creep} and {\it relaxation} functions
for the fractional models   contain
contributions of type
$${
\l\{ \eqalign{
\Psi(t) &= \chi_+\,\l\{1-\E_\alpha \l[-(t/\tau_\epsilon)^\alpha  \r]\r\}
	 =  \chi_+ \,
   \int_{0}^{\infty} \!\!\!
  R_\epsilon (\tau)\, \l( 1-\e^{\ds-t/\tau}\r)\, d\tau \,, \cr
\Phi(t) &= \chi_- \,\E_\alpha  \l[ -(t/\tau_\sigma)^\alpha  \r]
    = \chi_- \,     \int_{0}^{\infty} \!\!\!
    R_\sigma(\tau)\, \e^{\ds-t/\tau}\, d\tau \,. \cr	}
   \r.}\eqno(1.33)$$
Denoting  as usual by $*$
 the suffix $ \epsilon$ or $ \sigma$, the  analytical expressions of
the retardation and
relaxation spectra turn out to be identical, namely
$$ R_*(\tau) =
    \rec{\pi\,\tau  }\,
    {\sin\, \alpha  \pi \over (\tau /\tau _*)^\alpha  +
   (\tau /\tau _*)^{-\alpha } + 2\, \cos \, \alpha  \pi } \,.
 \eqno (1.34)$$
 \vfill\eject
\noindent
This result  can be deduced from the spectral  representation of the
Mittag-Leffler function $\E_\alpha  \l[ -(t/\tau_*)^\alpha  \r]$,
as shown by Caputo and Mainardi [3], and recently  by
 Gorenflo \& Mainardi [8] in the framework of their
analysis of  the fractional relaxation equation. 
\vsp
We can have a better insight of the spectral function  $R_*(\tau )$
and of the relaxation function $ E_\alpha[- (t/\tau _*)^\alpha] $
by showing the corresponding plots for a few values of $\alpha \,. $
Assuming $\tau_*=1\,, $ we could simply refer to the  plots
reported in [8] by  Fig. 1a and Fig. 2a, but, for the sake of
convenience,
we prefer to exhibit them again in Fig. 1-3 and Fig. 1-4, hereafter.
\vskip 1.1truecm
 \cen{\epsfxsize=7.5truecm \epsfysize=5.0truecm \epsffile{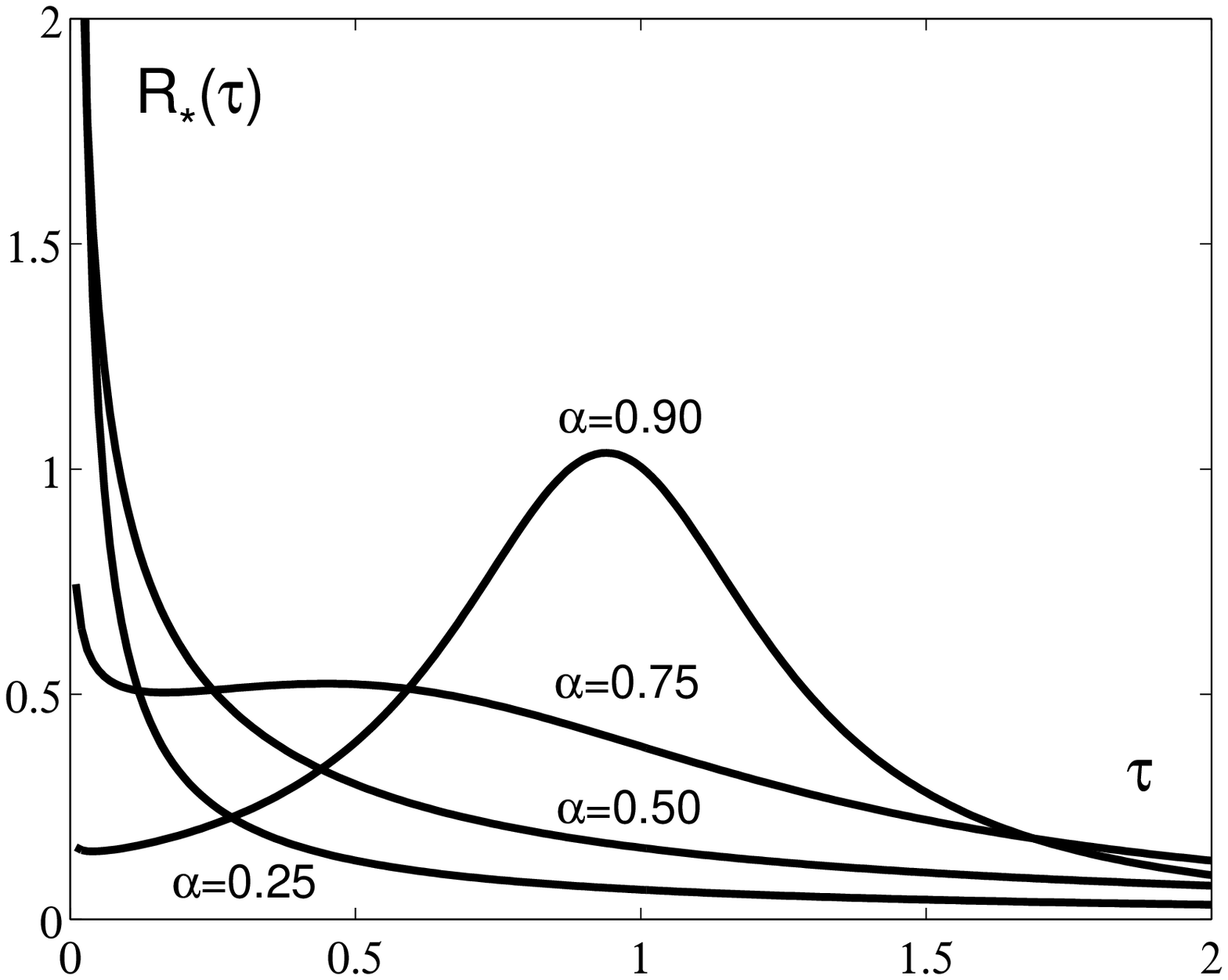}}
\cen{{\bf Fig. 1-3}}
\vsh
\cen{{Spectral function $R_*(\tau )$
for $\alpha  =0.25\,,\,0.50\,,\,0.75\,,\,0.90\,.$}}
\vsp
From the plots of $R_*(\tau)$ in Fig. 1-3 we can easily recognize the
effect of the variation of $\alpha $  on the character of the
spectral function;
for $\alpha \to 1$ the spectrum becomes sharper and sharper until
for $\alpha  =1 $   it reduces to be discrete
with a single retardation/relaxation time.
We also recognize that $R_*(\tau )$ is a decreasing function
of $\tau $ for $0<\alpha   <\alpha  _*$ where $\alpha	_* \approx 0.736$
is the solution   of the  equation
$ \alpha   = \sin \, \alpha   \pi \,; $
subsequently, with increasing $\alpha	\,, $	it first exhibits a minimum
and  then a maximum before tending to the impulsive  function
$\delta (\tau  -\tau _*)$ as $\alpha   \to 1\,. $
Recalling  the analysis of the fractional relaxation equation
by Gorenflo and Mainardi [8],
we  recognize that, compared to the 
exponential  obtained for $\alpha   =1\,, $
the {\it fractional} relaxation function
exhibits very different behaviours,
as can be seen from the plots of $E_\alpha(-t^\alpha)$	in Fig. 1-4.
In particular, we point out  the leading
asymptotic behaviours at small and  large times,
$$  \E_\alpha	(-t^\alpha  ) \sim \cases{\ds
    1- {\;t^\alpha  / \Gamma(1+\alpha )}\,, &as $\;t\to 0^+\,,$\cr
\ds {\;t^{-\alpha  }/ \Gamma(1-\alpha )}\,,& as $\;t\to +\infty\,.$\cr
   }\eqno(1.35) $$
Compared to the solution $ {\rm exp} (-t)$ for the classical models
($\alpha   =1$),
the   solution $\E_\alpha   (-t^\alpha	)$
for the fractional models ($0<\alpha   <1$)
exhibits initially  a much faster decay  (the derivative
tends to $-\infty$ in comparison with $-1$), and
for large times a much
slower decay (algebraic decay in comparison with exponential decay).
In view of its final slow decay,
the phenomenon of fractional relaxation is usually referred to
as a {\it super-slow process}.
$\null$
\vskip 1.1truecm
\cen{\epsfxsize=7.5truecm \epsfysize=5.0truecm \epsffile{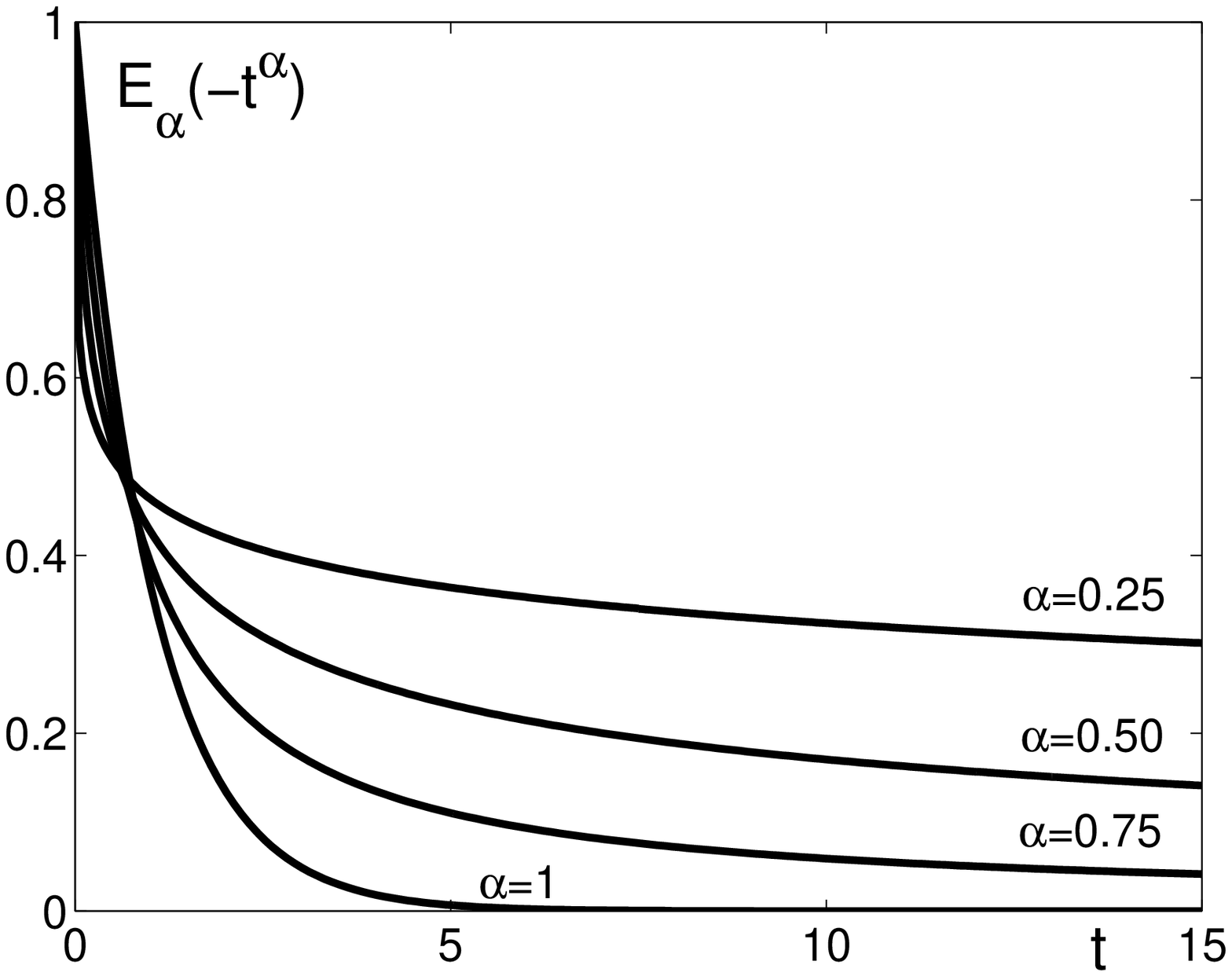}}
\cen{{\bf Fig. 1-4}}
\vsh
\cen{{
Relaxation function $E_\alpha (-t^\alpha)$
for  $\alpha  =0.25\,,\,0.50\,,\,0.75\,,\, 1\,.$}}
\vvs
\line{1.4 {\it Bibliographical remarks} \hfill} \vsp
A number of authors have, {\it implicitly} or {\it explicitly}, used
{\it fractional calculus} as an empirical method of describing the
properties of visco\-elastic  materials.
\vsp
In the first half of this century
Gemant [11-12] and, later, Scott-Blair [9, 13] were early contributors in
the use of fractional calculus to study phenomenological
constitutive equations for viscoelastic media.
\vsp
Inde\-pen\-dent\-ly,  in the former Soviet Union,
Rabotnov [14-15]
introduced  his theory of hereditary solid mechanics with weakly singular
kernels, that {\it implicitly} requires  fractional  derivatives.
This theory was developed also by other soviet scientists including
Meshkov and Rossikhin, see \eg	[16], and Lokshin and Suvorova,
see \eg [17].
\vsp
In 1971, extending earlier work by Caputo [18-20],
Caputo and Mainardi [3,10] suggested that derivatives of
fractional order could be successfully used
to model the dissipation in seismology and in metallurgy.
Since then up to nowadays,  applications  of {fractional calculus}
in rheology have been considered by several authors.
Without claim of being exhaustive, we now quote
some  papers of which the author became aware  during the last 25 years.
In addition to
Caputo [21-24] and Mainardi [25-26] we like to refer to
Smith and de Vries [27],
Scarpi [28], Stiassnie [29],  Bagley and Torvik [30-33],
Rogers [34],  Koeller [35-36],	Koh and Kelly [37],
Friedrich [38],
Nonnenmacher and Gl\"ockle [39-40], 
Makris and Constantinou [41],
Heymans and Bauwens [42],  Schiessel \& al [43],
Gaul \& al [44], Beyer and Kempfle  [45], 
Fenander [46], Pritz [47],
Rossikhin \& al [48-49], and Lion [50].
\vfill\eject

%% file: fmnew22.tex
\pageno = 303
\line{2. THE BASSET PROBLEM VIA FRACTIONAL CALCULUS \hfill} \vsn
\line{2.1 {\it Introduction}\hfill}\vsp
The dynamics of a sphere immersed in an incompressible viscous
fluid represents a classical problem, which has many applications
in flows of geophysical and engineering interest.
Usually,
the low Reynolds number  limit (slow motion approximation)
is assumed so that
the Navier-Stokes equations describing the fluid motion may be
linearised. 
 \vsp
The particular but relevant situation of a sphere subjected to gravity
was first considered  independently by Boussinesq [51] in 1885 and
by Basset [52] in 1888, who introduced
a special hydrodynamic force, related to the
history of the relative acceleration of the sphere,
which is nowadays referred to as {\it Basset force}.
The relevance of these studies	was  in that,
up to then,  only steady motions or small oscillations of bodies
in a viscous liquid had been considered starting from
 Stokes'  celebrated memoir on pendulums [53], in 1851.
The subject matter was	considered with more details
in 1907 by Picciati [54] and Boggio [55], in some notes
presented by the great Italian scientist Levi-Civita.
The whole was summarised by  Basset himself in a later paper [56],
and, in more recent times, by Hughes and Gilliand [57].
\vsp
Nowadays the  dynamics of impurities in unsteady flows
is quite relevant as shown by several  publications,
whose aim is to provide  more general	expressions
for the hydrodynamic forces, including the Basset force,
in order to fit experimental data and numerical simulations,
see \eg [58-66].
\vsp
In the next section we shall recall the general equation  of motion
for a spherical particle, in a viscous fluid,
pointing out the different force contributions
due to effects of inertia, viscous drag and buoyancy.
In particular, the so-called Basset force
will be interpreted in terms of
a fractional derivative
of order $1/2\, $ of the particle velocity relative
to the fluid.
 Based	on our recent works [67-68],
we shall introduce the {\it generalized Basset force},
which is expressed in terms of a fractional derivative of any order
$\alpha $ ranging in the interval $0<\alpha <1\,.$
This generalization, suggested by a mathematical
speculation, is expected to provide a phenomenological insight
for the experimental data.
\vsp
In   section  \S 2.3
we  shall consider  the simplified problem, originally
investigated by Basset,
where the  fluid is quiescent  and
the particle  moves under the action of gravity,
starting at $t=0$ with a
certain vertical velocity.  For the sake of generality,   we prefer
to consider the problem with the generalized Basset force
and will provide the solution for the particle velocity
in terms of Mittag-Leffler -type  functions.
The most evident effect of this
generalization will be to modify the long-time behaviour of the
solution, changing its algebraic decay from $t^{-1/2}$ to
$t^{-\alpha }\,. $ This effect can be of some interest
for a better fit of experimental data.
\vfill\eject
\def\x{{\bf x}}
\def\R{{\bf X}}
\def\X{{\bf X}}
\def\V{{\bf V}}
\def\u{{\bf u}}
\def\g{{\bf g}}
\def\Fb{{\bf F}_B}    \def\Fs{{\bf F}_S}
\def\Fi{{\bf F}_i}
\def\Fd{{\bf F}_d}
\def\Fg{{\bf F}_g}
\def\d{\partial}     \def\dt{\partial t}
\line{2.2 {\it The Equation of Motion}\hfill}\vsp
Let us consider a small rigid sphere of radius $r_0$,  mass $m _p$,
density $\rho _p\,, $
initially centred in $\X(t)$ and moving with velocity $\V(t)$
in a homogeneous fluid, of density $\rho _f$ and kinematic
viscosity $\nu \,,$
 characterized by a flow field $\u(\x,t)\,. $
In general the equation of motion is required to take into account
effects due to inertia, viscous drag and buoyancy, so  it can be
written as
 $$  m_p {d\V\over dt} =   \Fi + \Fd +\Fg  \,, \eqno(2.1)$$
where the forces on the R.H.S.	 
correspond in turn to the above effects.
According to   Maxey and  Riley [60] these forces read,
adopting  our notation,
$$ \Fi = m_f \, \left.{D{\u} \over Dt}\right\vert_{\R(t)}
-{1\over 2} \, m_f \,
 \left(
{d\V\over dt} -
\left.{D{\u} \over Dt}\right\vert_{\X(t)}  \right)
 \,, \eqno(2.2)    $$
$$ \Fd = -  {1\over \mu }\, 
 \left\{
   \l[\V(t) - \u(\X(t),t)\r] + \sqrt{{\tau_0\over \pi}}
    \,	 \int_{-\infty}^t
{d \left[\V(\tau)-\u(\X(\tau),\tau)\right]/d\tau \over
  \sqrt{t-\tau}}\,d\tau  \right\}\,, \eqno(2.3)$$
$$ \Fg = (m_p -m_f) \, \g\,, \eqno(2.4)$$
where $m_f= (4/3) \pi r_0^3 \rho _f$ denotes the mass of the fluid
displaced  by the spherical particle,
and
 $$ \tau _0 := {r_0^2\over  \nu}\,, \eqno(2.5)$$
$$   {1\over  \mu } :=	6\pi\, r_0 \, \nu \rho_f =
	     {9\over 2} \, m_f\, \tau_0^{-1}
\,.\eqno(2.6)  $$
The time constant   $\tau _0$  represents
a sort of time scale induced by viscosity, whereas
the constant $\mu $ is usually referred to as the {\it mobility
coefficient}.
\vsp
In (2.2) we note  two different time
derivatives,   $D/Dt\,,$  $\, d/dt\,,$ which represent the
 time derivatives following a fluid element
and the moving sphere, respectively, so
$$ \left.{D{\u} \over Dt}\right\vert_{\X(t)}
     =	 \l[
 {\d \u\over \dt} + (\u \cdot \nabla) \, \u(\x,t)\r] \,,
\quad
 {d\over dt}\u[\X(t),t] =
  \l[
 {\d \u\over \dt} + (\V \cdot \nabla) \, \u(\x,t)\r] \,,
$$
where the brackets are computed at ${\x=\X(t)}\,. $
\vsp
The  terms on the R.H.S. of (2.2) correspond in turn to the effects of
pressure gradient of the undisturbed flow and of added mass,
whereas those of (2.3) represent respectively
the well-known	viscous Stokes drag,
that we shall denote by $\Fs\,, $
and to the augmented viscous Basset drag denoted by $\Fb\,. $
Using the characteristic time $\tau _0$, the Stokes and Basset forces
read respectively
 $$ F_S =  
    -\, {9\over 2}\, m_f\, {\tau_0 }^{-1} \,
  [\V(t) -\u(\X(t),t)] \,, \eqno (2.7)$$
  $$ F_B =   
    -\, {9\over 2}\, m_f\, \tau_0 ^{-1/2} \,
  \l\{ \rec{\sqrt{\pi}}\, \int_{-\infty}^t
  {d[\V(\tau) -\u(\X(\tau),\tau)]/d\tau \over \sqrt{t-\tau}}\,d\tau \r\}
 \,. \eqno (2.8)
$$
\vfill\eject
We thus recognize that the time constant $\tau_0 $  provides the natural
time scale for the diffusive processes related to the fluid viscosity,
and that the integral expression in brackets at the R.H.S. of (2.8)
just represents the
{\it Caputo  fractional derivative} of order $1/2\,, $
with starting point $-\infty \,,$  of the particle
velocity relative to the fluid $\null^{*}$.    
\footnote{}{\baselineskip=12truept   \note 
\noindent $\null^{*}\,$
Presumably, the first scientist who  has pointed out the relationship
between the Basset  force and the  fractional calculus has been
Tatom [69] in 1988. However, Tatom has limited himself to  note this
fact, without treating any related problem by the methods of fractional
calculus.}
\vsp
We  now
introduce the {\it generalized Basset force} by the definition
$$ F_B^\alpha = -\,{9\over 2} \, m_f \tau_0 ^{\alpha-1} \,
	 {d^{\alpha}  \over dt^{\alpha }}
  [\V(t) -\u(\X(t),t)]\,, \q 0<\alpha <1\,, \eqno(2.9)$$
where  the fractional derivative of order $\alpha $ is in
Caputo's  sense,
in agreement with the notation introduced in \S 1.3 for
the fractional viscoelastic models, see (1.25).
\vsp
Introducing the so-called {\it effective mass}
$$ m_e :=   m_p + {\rec{2} \, m_f}\,, \eqno(2.10)$$
and allowing for the {\it generalized Basset force}  in (2.3),
we can re-write the equation of motion (2.1-4) in the 
more compact  and significant form,
$$ m_e	{d\V\over dt} = {3\over 2} \, m_f\, {D \u \over Dt}
      -\,{9\over 2} \, m_f \,\l [
   \rec{\tau_0}  + \rec{{\tau _0}^{1-\alpha}}\,
 {d^{\alpha }\over dt^{\alpha }}\r]\,(\V-\u)  +
   (m_p-m_f) \, \g\,, \eqno(2.11)$$
that we refer to  as the {\it generalized equation of motion}.
Of course, if in (2.11) we put $\alpha =1/2\,, $
we recover  the  {\it basic equation of motion} with the
original Basset force.
\vvs
\line{2.3 {\it The (Generalized) Basset Problem}\hfill}\vsp
Let us now assume that the fluid is
quiescent, namely $\u(\x,t) = 0\,,\; \forall \, \x,t\,, $ and the
the particle starts to move under the action of gravity, from a given
instant $t_0 =0$ with a certain velocity $V(0^+)= V_0\,, $
in the vertical direction.
This was the problem  considered  by Basset [52], that
was first solved by Boggio  [55], in a cumbersome way, in terms of
Gauss and Fresnel integrals.
\vsp
Introducing the non-dimensional quantities (related to the densities
$\rho _f\,,\,\rho _p $	of the fluid  and  particle),
$$ \chi  := {\rho _p\over \rho _f}\,,\quad
   \beta :={9\rho_f\over 2\rho_p+\rho _f}=   {9\over{1+2\chi}}\,,
 \eqno(2.12)$$
we find it convenient to define a new characteristic time
$$     \sigma_e := \mu \, m_e = {\tau_0 / \beta}  \,, \eqno (2.13)$$
see (2.5), (2.10), (2.12),
and  a characteristic velocity (related to the gravity),
$$  V_S   = ({2/9})\,(\chi -1)\,  g \,\tau_0 \,.  \eqno  (2.14)$$
\vfill\eject
Then we can eliminate the mass factors and the gravity acceleration
in (2.11) and obtain   the
  {\it equation of motion}
in the	form
     $$  {dV \over dt}=
    - {1\over \sigma_e} \, \left[
   1  + {\tau _0}^{\alpha}\,
 {d^{\alpha }\over dt^{\alpha }}\r]\,\V  +
   {1\over \sigma _e} \, V_S
 \,.\eqno(2.15)$$
\vsp
If the Basset term were absent, we  obtain  the  classical
Stokes solution
$$V(t)= 
 V_S +	(V_0 -V_S) \,{\rm e}^{\ds -t/\sigma_e}\,,
\eqno (2.16)$$
where	$\sigma _e$
represents the characteristic  time of the motion,
and $V_S$ the final value assumed by the velocity.
Later we shall show  that  in the presence of the Basset term
the same final value is  still attained by the solution $V(t)$,
but  with  an algebraic rate, which is much slower than the
exponential one found in (2.16).
\vsp
In order to investigate the effect of the (generalized) Basset term,
we compare the exact solution of (2.15) with
the Stokes solution (2.16); for this aim  we find it convenient to scale
times and velocities in  (2.15)
 with  $\{\sigma_e \,, \, V_S\}$,
{\it i.e.}\ to	refer
to the	non dimensional quantities
$  t' = t/\sigma_e \,, \;V' = V /V_S \,, \; V'_0 = V_0/V_S\,. $
The resulting equation of motion reads (suppressing the apices)
$$ \left[
{d\over dt}+ a\,{d^{\alpha}\over dt^{\alpha }}+ 1 \right] \,
   V(t)=1\,, \quad V(0^+) =V_0\,,
 \q a = \beta ^{\alpha}>0 \,,\q 0<\alpha <1 \,.
  \eqno (2.17)$$
This is the  {\it composite fractional
relaxation equation} treated by Gorenflo and  Mainardi [8]
in \S 4.1  by using the Laplace transform method.
Recalling that in an obvious notation we have
\def\V{\widetilde V}  
\def\N{\widetilde N}  
\def\M{\widetilde M}  
\def\ss{{s}^{1/2}} 
$$ V(t) \,\div \, \V(s)\,,    \q
  {d^\alpha\over dt^\alpha } V(t) \, \div \, s^\alpha \, \V(s)-
    s^{\alpha -1}\, V_0\,, \;\; 0<\alpha \le 1\,, \eqno(2.18)$$
the transformed solution of (2.17) reads
$$   \V(s) =  \M(s)\,V_0 + {1\over s}\, \N(s)\,,  \eqno (2.19)	 $$
where
$$\M(s) =  {1 +a\,s^{\alpha -1}\over s +a\,s^\alpha +1}\,, \q
   \N(s) = {1 \over s +a\,s^\alpha +1}\,.
   \eqno (2.20) $$
Noting that
$$ {1\over s}\, \N(s) = {1\over s}- \M(s)  \, \div\,
    \int_0^t N(\tau )\, d\tau  = 1 - M(t) \,
    \Longleftrightarrow \, N(t) =  - M'(t)\,,  \eqno(2.21)$$
the actual solution of (2.17)  turns out to be
$$ V(t) = 1 + (V_0-1)\, M(t) \,, \eqno(2.22)$$
which is "similar" to the Stokes solution  (2.16)
if we consider the substitution of $\e^{-t}$ with the function $M(t)\,. $
\vfill\eject
\vsp
In [67-68] Mainardi, Pironi and  Tampieri have used  a factorisation
method to invert $\N(s)$  and henceforth $\M(s)\,, $
using a procedure indicated by Miller and  Ross [69],
which is valid when $\alpha $ is a rational number, say $\alpha =p/q$,
where $p,q \in \NN\,, \, p<q\,. $  In this way
the actual solution  can be finally expressed as a
linear combination of certain incomplete gamma functions.
This algebraic method
is of course convenient for the ordinary Basset problem ($\alpha =1/2$),
but becomes cumbersome for $q>2\,. $
\vsp
Here,  following the analysis in [8], we prefer  to
adopt the general method of inversion based on the complex Bromwich
formula. 
By this way we are free from the restriction of being $\alpha $ a
rational number and, furthermore,  we are able to provide an integral
representation of the solution,  convenient for  numerical computation,
which allows us to recognize the
monotonicity properties of the solution without need of plotting.
\vsp
We now resume the relevant results  from [8] using the present notation.
The integral representation for $M(t)$ turns
out to be
$$ M(t) = \int_0^{\infty} \e^{\ds\,-rt} K(r)\, dr\,,\eqno(2.23) $$
where
$$ K(r) =
  \rec{\pi}\,
   { a\,r^{\alpha -1}\, \sin \,(\alpha \pi)\over
 (1-r)^2 + a^2\, r^{2\alpha} + 2\, (1-r)\,a\,r^{\alpha} \, \cos \,
(\alpha \pi)} >0\,. \eqno (2.24)$$
Thus   $M(t)$ is a {\it completely monotone}  function
[with spectrum $K(r)$],
 which is decreasing from 1 towards 0 as $t$
runs from  $0$ to $\infty\,. $
The behaviour of $M(t)$ as $t\to 0^+$ and
$t \to \infty$ can be inspected by means of a proper asymptotic analysis,
as follows.
\vsp
The behaviour as  $t\to 0^+$   can be determined
from the behaviour of the Laplace transform
$ \widetilde M(s) = s^{-1} - s^{-2} + O\l(s^{-3+\alpha}\r) \,,	$
as Re$\,\{s\} \to +\infty\,. $
We obtain
$$     M(t) = 1 - t + O\l(t^{2-\alpha}\r)\,,
 \q {\rm as}\q	t \to 0^+ \,.
 \eqno(2.25)$$
\vsp
The spectral representation (2.23-24) is suitable to obtain the
asymptotic behaviour of $M(t)$
as $ t \to +\infty\,, $ by using the Watson lemma.
In fact, expanding the spectrum $K(r)$ for small $r$ and
taking the dominant
term in the corresponding asymptotic series,  we obtain
$$ M(t) \sim
      a\,  {t^{-\alpha} \over \Gamma(1-\alpha)}
 = a\, {\sin \,(\alpha \pi)\over \pi} \,
  \int_0^\infty \! \e^{\,\ds -rt}\,	r^{\alpha -1}\, dr\,,\q
  {\rm as} \q t \to \infty\,. \eqno(2.26)$$
Furthermore, we recognize that
$ 1 > M(t) >  \e^{\ds\, -t} >0\,, \; 0<t<\infty\,,$
namely, the decreasing plot of $M(t)$ remains above that of the
exponential, as $t$ runs from $0$ to $\infty\,. $
Although both the two functions  tend monotonically to $0\,, $
the difference between the two plots
increases with $t$: at the initial point $t=0\,,$  both the curves
assume the unitary value and  decrease with the same initial rate,
but as $t\to \infty$ they exhibit  very different decays,
algebraic (slow) against  exponential (fast).
\vfill\eject
For the ordinary Basset problem it is convenient
to report the result
obtained by the factorisation  method  [67-68].
In this case we must note that $a = \sqrt{\beta}\,, $
see (2.17),  ranges from
$0$ to $3$ since from (2.12) we recognize that
 $\beta $ runs from $0$ ($\chi	= \infty\,, $ infinitely heavy particle)
to $9$ ($\chi  = 0\,, $ infinitely light particle).
\vsp
The actual solution is	obtained expanding
 $\M(s)$ into  partial fractions and then inverting.
Considering the two roots $\lambda _\pm$ of   the polynomial
$P(z) \equiv z^2 + a \,z  +1\,,$ with $z=s^{1/2}$
we must treat separately the following two cases
$$ i)\;\; 0<a<2\,, \q{\rm or}\q  2<a<3\,, \q {\rm and} \q
   ii)\;\; a=2\,, $$
which correspond to two distinct roots ($\lambda _+ \ne \lambda _-$),
or two coincident roots ($\lambda _+ \equiv \lambda _-= -1$),
respectively.
We obtain
\vsn   $ i)\q	a \ne 2 \, \Longleftrightarrow\,
  \beta \ne 4 \,, \; \chi  \ne 5/8\,,$
$$  \M(s) = {1+a\, s^{-1/2}\over s+ a \,\ss +1}
 = {A_-\over \ss\,(\ss -\lambda_+ )} +
    {A_+\over \ss\,(\ss -\lambda_- )} \,,
\eqno (2.27)  $$
with
$$\lambda_\pm = {-a \pm (a^2 -4)^{1/2} \over 2}={1\over\lambda_\mp}
     \,, \quad
   A_{\pm} = {\pm}\, { \lambda_\pm  \over \lambda_+  - \lambda_- }\,;
 \eqno(2.28)
$$
\vsn   $ii) \q a = 2 \, \Longleftrightarrow\,
   \beta = 4 \,, \; \chi  = 5/8\,,$
$$  \M(s) = {1+2\, s^{-1/2}\over s+ 2 \,\ss +1}  =
{1\over(\ss +1)^2} +   {2\over \ss\,(\ss +1)^2} \,.
\eqno (2.29)$$
The Laplace inversion of $ (2.27-29)$ can be expressed
in terms of  Mittag-Leffler functions
of order $1/2\,, $
$E_{1/2} (\lambda\sqrt{t}) = {\rm exp}(\lambda^2t)\,
 {\rm erfc} (-\lambda\sqrt{t})\,, $ as shown in the Appendix of [8].
We obtain
$$M(t) = \l\{ \eqalign{
i)\,&\;
A_-\,E_{1/2}
\,(\lambda_+\,\sqrt{t})+A_+\,E_{1/2}\,(\lambda_-\,\sqrt{t})\,,\cr
ii)\,&\;
    (1-2t)\, E_{1/2}\,(-\sqrt{t})  + 2\, \sqrt{t/\pi}\,. \cr}
\r.
   \eqno (2.30)$$
We recall that the  analytical solution  to the classical
Basset problem was
formerly provided by Boggio [55] in 1907 with a different
(cumbersome) method.
One can show that our solution	(2.30),
derived by the tools of the Laplace transform  and fractional calculus,
coincides  with Boggio's solution.
Also Boggio arrived at the analysis of the two roots
$\lambda _\pm$ but his expression
of the solution in the case of two conjugate complex roots
($\chi	> 5/8$)
given as a sum of Fresnel integrals could induce one
to forecast  unphysical oscillations, in the absence of
numerical tables or plots.
This disturbed Basset
who, when he summarised the state of art about his problem
 in a later paper of 1910 [56], thought there was some physical
deficiency in his own theory.
With our integral representation
of the solution, see  (2.23-24),
we can prove the monotone character of the solution, even
 if the arguments of the exponential and error functions
are complex.
\vfill\eject
\vsp
In order to have some insight about the effects of the
two parameters $\alpha $ and $a$ on the (generalized) Basset problem
we exhibit some (normalized)
plots for the particle velocity $V(t)$, 
corresponding to the  solution of Eq. (2.17),
assuming for simplicity a vanishing initial velocity ($V_0=0\,$).
\vsp
We consider 3 cases for $\alpha\,, $  namely
$\alpha =1/2\,$ (the ordinary Basset problem) and
$\alpha = 1/4\,, \, 3/4\, $  (the generalized Basset problem),
corresponding to Figs 2-1, 2-2, 2-3, respectively.
For each  $\alpha $ we consider  four values of
$a$ corresponding to $\chi :=  \rho_p/\rho_f = 0.5, 2, 10, 100\,. $
For each couple $\{\alpha \,,\, \chi\}$
we compare the Basset solution	(in continuous line)
with its asymptotic expression
(in dashed-dotted line)
for large times and the Stokes solution (dashed line).
We remind that the Stokes solution is the solution of Eq. (2.17)
with $a=0$ and hence is independent of $\alpha\,.$
\vsp
From these figures we can recognize
the retarding effect of the (generalized) Basset force, which
is more relevant for lighter particles,
in reaching the final value of the velocity. This effect is
of course due to the algebraic decay of the function $M(t)$,
see (2.26),
which is much slower  than the exponential decay of the Stokes solution.
$\null$
\vskip 0.5truecm
$\null$
\cen{\epsfxsize=12.0truecm \epsfysize=9.0truecm \epsffile{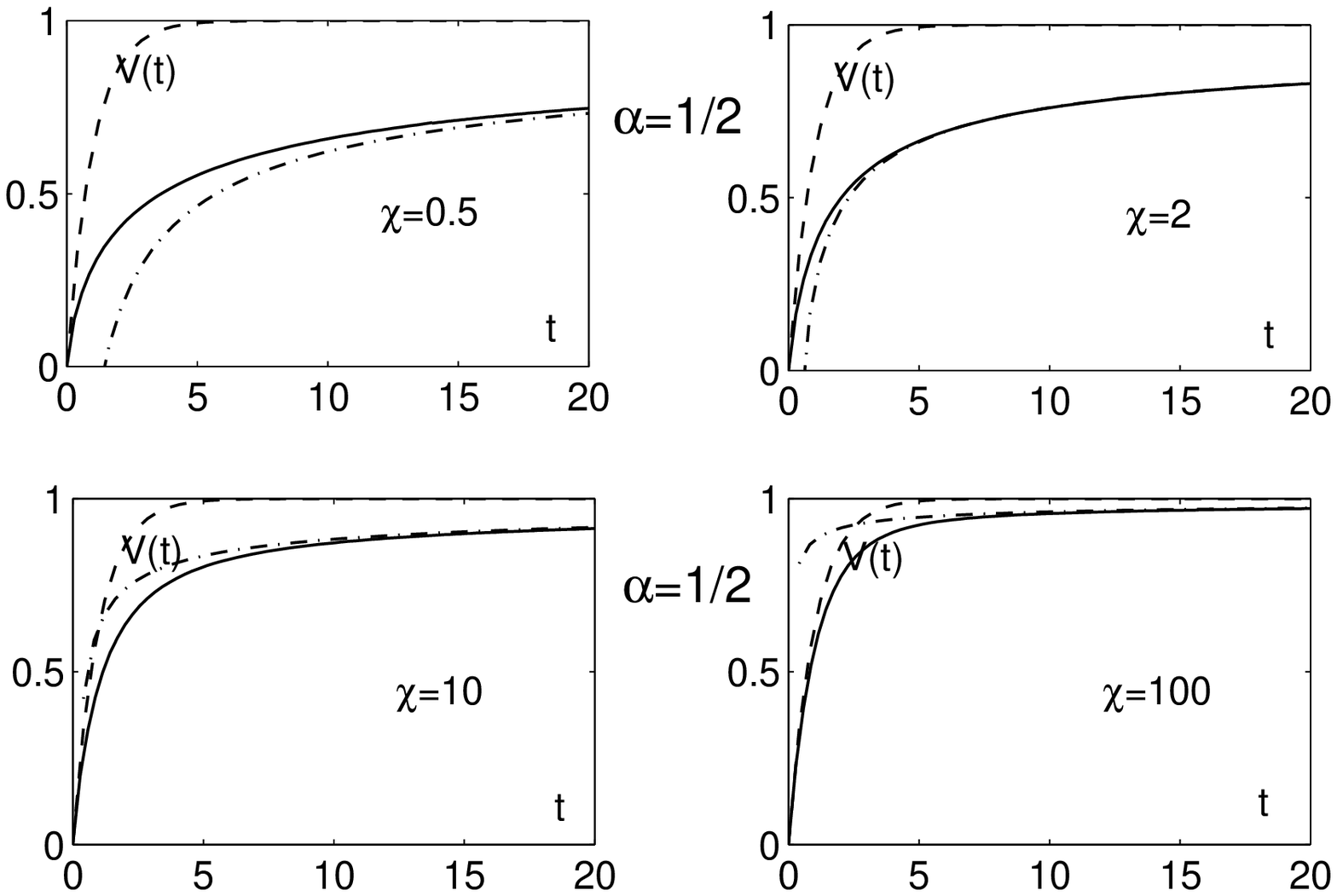}}
\vsh
\cen{{\bf Fig. 2-1}}
\vsh
\cen{
The normalized velocity $V(t)$ for $\alpha =1/2$ and
  $\chi = 0.5\,,\,  2\,, 10\,,\, 100\,:$}
\cen{Basset exact  ----- $\,$;	
     Basset asymptotic  $-\, \cdot\, -\, \cdot\, -\,;$
     Stokes $ - - -\,.$}
\vfill\eject
\cen{\epsfxsize=12.0truecm \epsfysize=9.0truecm \epsffile{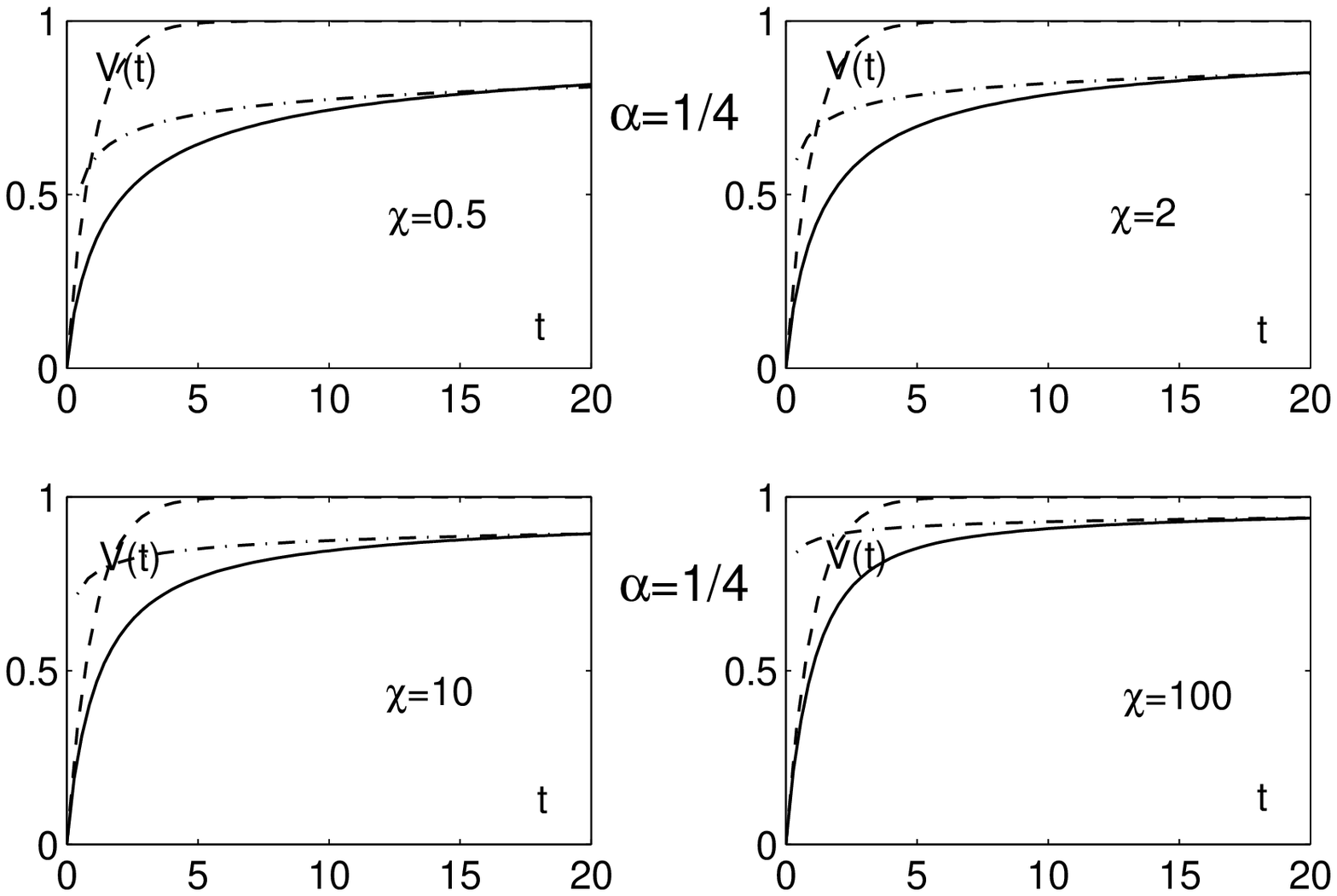}}
\cen{{\bf Fig. 2-2}}
\vsh
\cen{The normalized velocity $V(t)$ for $\alpha =1/4$ and
 $\chi = 0.5\,,\,  2\,, 10\,,\, 100\,:$}
\cen{Basset exact  ----- $\,$;
    Basset asymptotic  $-\, \cdot\, -\, \cdot\, -\,;$	
    Stokes  $ - - -\,.$}
\cen{\epsfxsize=12.0truecm \epsfysize=9.0truecm \epsffile{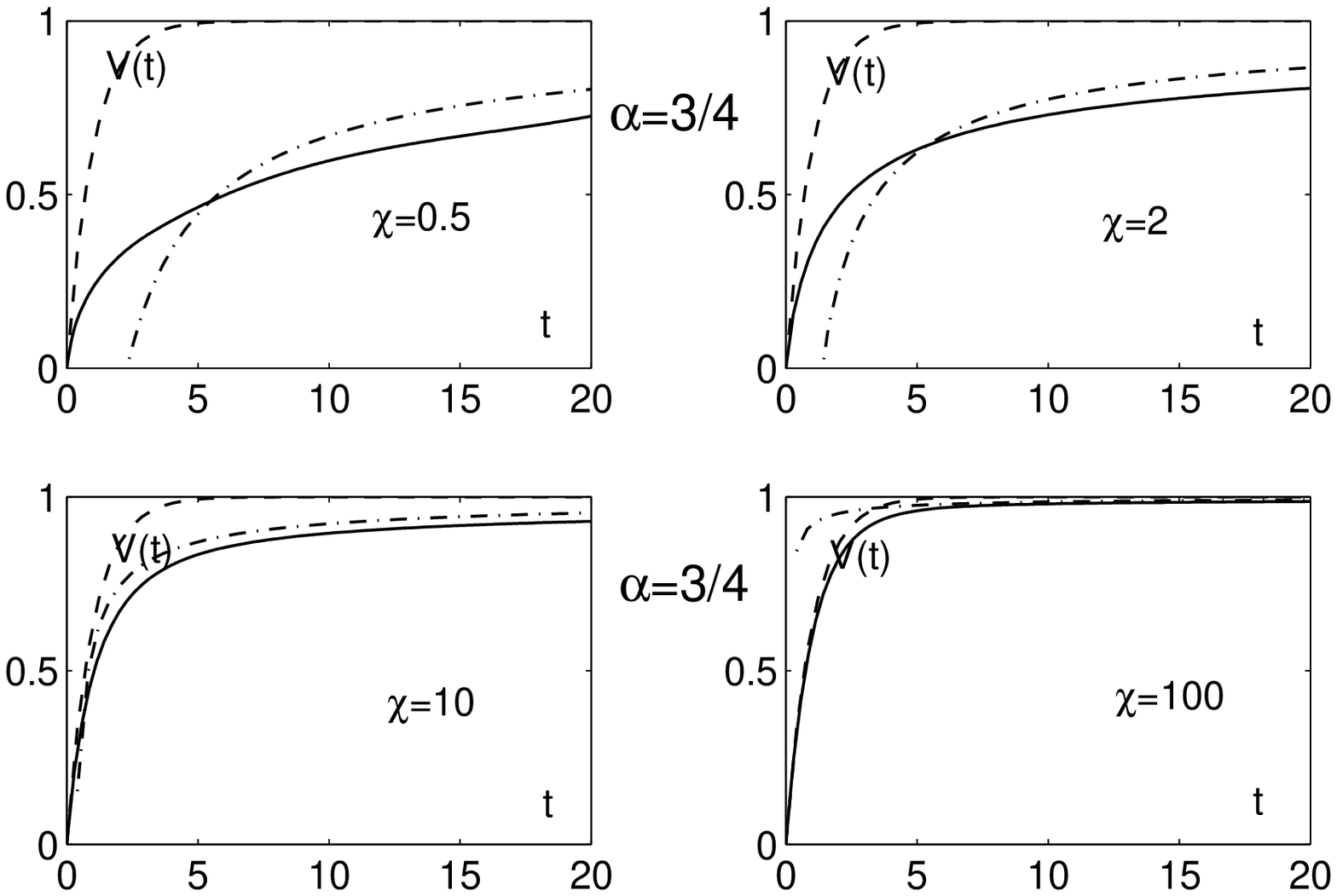}}
\cen{{\bf Fig. 2.3}}
\vsh
\cen{The normalized velocity $V(t)$ for $\alpha =3/4$ and
$\chi = 0.5\,,\,  2\,, 10\,,\, 100\,:$}
\cen{Basset exact  ----- $\,$;	
     Basset asymptotic  $-\, \cdot\, -\, \cdot\, -\,;$
    Stokes $ - - -\,.$}
\vfill\eject

%% file: fmnew23.tex
\pageno=311
\def\V{\barr V}  
\def\M{\barr M}  
\def\ss{{s}^{1/2}} 
\line{3. BROWNIAN MOTION AND FRACTIONAL CALCULUS\hfill} \vs
\line{3.1 {\it Introduction}\hfill} \vsp
According to the classical approach started by Langevin
normal diffusion  and Brownian motion
are associated with the Langevin  equation.
More specifically,  
the classical Langevin equation addresses the dynamics of a Brownian particle
through Newton's law by incorporating the effect of the Stokes fluid
friction and that  of thermal fluctuations in the vicinity of the particle
into a random force,
see \eg Wax [70],
Fox and  Uhlenbeck [71], Fox [72], Kubo {\it et al} [73].
\vsp
Since the pioneering computer experiments by Alder and	Wainwright [74] in
1970, which have shown that the velocity autocorrelation
function for a Brownian particle in a dense fluid goes asymptotically
as $t^{-3/2}$ instead of exponentially as predicted by stochastic
theory, many attempts have been made to reproduce  this result by purely
theoretical arguments, see \eg	[75-97];
in most cases  hydrodynamic models  are adopted.
\vsp
 Recently, a great interest on the subject matter has
been raised because of the possible connection among
{\it long-time correlation} effects, {\it fractional Brownian motion}
and {\it anomalous diffusion}, see \eg [98-102].
We recall that {\it anomalous diffusion} is the phenomenon, usually
met in disordered or fractal media, according to which the
{\it displacement  variance} is no longer linear in time
but proportional to a power $\alpha $ of time with
$ 0<\alpha <1$ ({\it slow diffusion}) or
$1<\alpha <2$ ({\it fast diffusion}), 
 see  Bouchaud and   Georges [99] for a review.
\vsp
We also point out that, in view of the linear-response theory,
 Kubo in 1966 [103] stated a {\it fluctuation-dissipation  theorem}
\footnote{*}{\baselineskip=12truept   \note 
For a critical analysis of 
Kubo's  fluctuation-dissipation theorem see  Felderhof
[104]}
by introducing	a   {\it generalized Langevin equa\-tion} ($GLE$),
with an indefinite memory function as an integral kernel.
In other words, this theorem may be represented
by a stochastic equation describing the fluctuation, which is a
generalization of the classical Langevin equation; in the $GLE$
the friction force becomes retarded or frequency dependent and
the random force is no longer a white noise. As a matter of fact,
the hydrodynamic models
introduced in the literature
appear as particular cases of Kubo's $GLE$.
\vsp
Here, after resuming in \S 3.2 the classical results
derived from the ordinary Langevin equation, in \S 3.3
we shall revisit a  hydrodynamic model
which takes into account, in addition to the Stokes viscous
drag,  the inertial  effect due to the added mass and the
retarding effect due to the Basset memory force.
So doing, we obtain
 a stochastic differential equation  which
 contains a time derivative of order 1/2. This $GLE$
will be referred to as the {\it fractional
Langevin equation}.
\vfill\eject
\vsp
The present approach
is based  on a recent analysis	 carried out
by  the author and collaborators [105-106],
in order  to  model the  Brownian motion
more realistically  than in the classical approach  (based on the
Langevin equation).
\vsp
Using  Kubo's  {\it fluctuation-dissipation theorem}
and   the techniques of  {\it fractional calculus},
we shall provide the analytical expressions
of the autocorrelation functions (both for the random force and  the
particle velocity)  and of the displacement variance.
Consequently, the well-known results of the classical theory
of the Brownian motion	will be properly generalized.
\vsp
In the final section, \S3.4, we shall present and discuss some numerical
results implied by our analysis.
\vvs	 
\line{3.2 {\it The Classical Approach to the Brownian Motion}\hfill}
 \vsp
We assume that the Brownian particle of mass $m_p$ executes
a random motion in one dimension with velocity	    $V= V(t)$	and
displacement $X=X(t)$.
The classical approach to the Brownian motion is based
on  the following stochastic  differential equation
({\it Langevin equation})
  $$  m_p\, {dV\over dt} =  F(t)  + R(t)\,,\eqno (3.1)$$
where $F(t)$ denotes  the {\it frictional force} exerted from
the fluid on the particle
and $R(t)$ denotes the {\it random force} arising from rapid
thermal fluctuations, subjected to the condition
  $ \langle \, R(t)\, \rangle= 0\,.$
As usual, we have denoted with brackets the average
taken over an ensemble in thermal equilibrium.
Therefore the total force has been divided into a mean force $F$
and a fluctuating force $R$. The fact that $F(t)$ is independent
of the fluid variables is due to the boundary condition that the
fluid velocity be equal to the particle velocity, $V(t)\,, $
at the surface of the particle.
\vsp
Assuming for the mean force  the familiar Stokes  approximation
for a drag of spherical particle of radius $r_0\,,$  we
obtain the classical formula
$$ F = -\rec{\mu } \, V(t) \,, \q
    \rec{\mu }= 6 \pi \, r_0 \,\rho_f \,\nu\, \,, \eqno(3.2)$$
where $\mu$ denotes the {\it mobility coefficient} and
$\rho _f$ and $\nu $ are the density and the kinematic viscosity
of the fluid, respectively.  In this approximation
the time derivative of the fluid velocity field has been neglected.
If we introduce the friction characteristic time
$ \sigma_p   :=   \mu\, m_p \,,$
the Langevin equation (3.1) explicitly reads
   $$  {dV\over dt} =
 -\rec{\sigma_p }\,  V(t) +  \rec{m_p} \, R(t)\,.\eqno (3.3)$$
\vfill\eject
\vsp
The stochastic processes $V(t)$ and $R(t)$
are assumed to be    {\it Gaussian-Markovian} and
 {\it stationary}. The stationarity  implies
 that the
{\it autocorrelation functions} $C_V$ and $C_R$
depend only on the time shift, namely
$$ C_V(t_1,t) :=\langle \,V(t_1)\, V(t_1 +t)\, \rangle = C_V(t)\,,
   \eqno(3.4)$$
$$ C_R(t_1,t):=  \langle \,R(t_1)\, R(t_1 +t)\, \rangle = C_R(t)\,,
  \eqno(3.5)$$
for any $t_1$ and $t\,. $
Hereafter we assume $t_1=0$ and $t\ge 0\,. $
 \vsp
Following the classical approach to the  Brownian motion,
we require that the  variance of the velocity at $t=0\,,$
$ \, C_V(0) = \langle \, V^2(0)\, \rangle\,, $
satisfies the {\it  equipartition law  for the energy distribution},
\ie
$$ m_p \, \langle \, V^2(0)\, \rangle  =  k \, T
    \, \Longleftrightarrow \,
   \sigma _p \, \langle \, V^2(0)\, \rangle  =	\mu \,k \, T
\,, \eqno(3.6)$$
where  $k$ is the Boltzmann constant,  as if
the Brownian particle were kept
for a sufficiently long time in the fluid at (absolute) temperature
$T\,,$	and
that the random force is uncorrelated
to  the  particle velocity at
$t=0\,, $ \ie
$$   \langle \, V(0) \, R(t) \, \rangle = 0\,,\q t\ge 0\,.
 \eqno(3.7)$$
As well known, the previous assumptions  lead to the
 relevant results,
$$
   C_V(t) =  \langle \, V^2(0)\, \rangle  \, {\rm e}^{\dis -t/\sigma_p}
    \,, \q t\ge 0\,,
\eqno(3.8)$$
$$ C_R(t) = {m_p^2 \over \sigma_p } \,
  {\langle \, V^2(0)\, \rangle} \, \delta (t)  \,, \q t\ge 0\,,
\eqno(3.9)$$
where $\delta (t)$ denotes the Dirac distribution.
The result (3.8)  shows that the velocity autocorrelation
function  decays exponentially with characteristic time
$\sigma_p  \,,$ whereas (3.9) means that
 $R(t)$ is  a {\it white noise}.
\vsp
It can be readily shown that the mean squared displacement of
the Brownian  particle
(starting at the origin at $t_0 =0\,$), \ie the {\it displacement
variance}, is given by
$$ \langle\,X^2(t)\, \rangle=2\,  \int_0^t \!\! (t-\tau)\,
C_V(\tau)\,d\tau = 2\, \int_0^t d\tau_1 \int_0^{\tau _1}
  \!\! C_V(\tau )\, d\tau \,, \q t\ge 0\,.  \eqno(3.10)$$
 For this it is sufficient  to recall that
 $X(t)=\int_{0}^t V(t')\,dt'\,,$ and to use the definition (3.4) of
 $C_V(t)$ for $t\ge t_0=0\,.$
As a consequence of (3.8) and (3.10)
we obtain
\def\D{{\cal D}}
$$\langle\,X^{2}(t)\,\rangle= 2\, \sigma_p\,{\langle\, V^2(0)\,\rangle}
 \,   \left[\,t- \sigma_p\left(1-{\rm e}^{\dis - t/\sigma_p}\right )
  \,\right]\,, \q t\ge 0\,,
\eqno(3.11)$$
from which  we recognize that for sufficiently large times
the variance increases linearly with time.
\vfill\eject
\vsp
It is usual to introduce
the {\it diffusion coefficient}     as
  $$ \D:= \lim_{t \to \infty} \,
   {\langle\,X^2(t)\, \rangle \over 2\,  t}   \,. \eqno(3.12)$$
Then from (3.11-12) we obtain  the chain of equalities
$$ \D=\sigma_p\,{\langle\, V^2(0)\,\rangle} =
    \int _0^{\infty} \!\! C_V(t) \, dt
 \,, \eqno(3.13) $$
and, using (3.7),
$$ \D=	  \mu \, k\, T
   \,. \eqno(3.14)$$
The identity (3.14) is known as {\it Einstein relation}.
In particular, we point out the asymptotic behaviour of the
variance for large times,
$$\langle\,X^{2}(t)\,\rangle
    =  2 \D\, t\,
 \l[1- (t/ \sigma_p)^{-1}+ EST \r]\,,
\q {\rm  as}\q	 t\to \infty \,, \eqno(3.15)$$
where $EST$ denote exponentially small terms.
\vvs
\line{3.3 {\it The Hydrodynamic Approach to the Brownian Motion}\hfill}
 \vsp
On the basis of hydrodynamics, the Langevin equation  (3.3) is
not completely correct, since it ignores the effects of the added mass and
Basset history force, which are due to the acceleration of the
particle. This was formerly pointed out  in the early seventies
by a number of authors, just after the cited
computer experiments by Alder and  Wainwright [74].
 \vsp
The added mass effect
requires to substitute the mass of the particle
with the so-called effective mass, $m_e$ introduced in (2.10).
As a consequence, in order to keep unmodified the mobility coefficient
in the Stokes drag, we have to introduce a new friction
characteristic time,  $\sigma_e \,,  $ such that
$$ \mu := {\sigma_p \over m_p} ={ \sigma_e   \over m_e} \,
     \Longleftrightarrow \,
  \sigma_e  := \sigma \, \l(1 +\rec{2\chi  }\r) \,,
   \q {\rm with} \q \chi   := {\rho _p\over \rho _f}\,.
 \eqno(3.16)
$$
The corresponding Langevin equation is obtained form  (3.3)
by replacing
$m_p$ with $m_e$ and $\sigma_p $ with $\sigma _e\,. $
With respect to the classical analysis, it turns out that the  added mass
effect, if it were present alone,  would be  only
to lengthen the time scale ($ \sigma_e	 >\sigma_p \,$)
in the exponentials entering the basic formulas (3.8) and (3.11)
and to decrease the  velocity  variance
$\langle V^2(0)\rangle\,, $
consistently
with  the energy equipartition law  at the same temperature,
$$ m_e \, \langle \, V^2(0)\, \rangle  =  k \, T
    \, \Longleftrightarrow \,
   \sigma _e \, \langle \, V^2(0)\, \rangle  =	\mu \,k \, T
  \,. \eqno(3.17)$$
 Consequently,	the diffusion coefficient  turns out to
be not altered by the added mass effect and
the Einstein relation  still holds.
\vfill\eject
\vsp
In view of
Kubo's {\it fluctuation-dissipation theorem},
an arbitrary retarding effect in the friction force
(in particular that due to
the Basset force) can be taken into account
by introducing	a suitable memory function
$\gamma (t)$  in the Langevin equation.
The consequent $GLE$ 
reads  (in our notation)
 $$   {dV\over dt} = -\int_{0^-}^{t^+}
  \gamma (t-\tau) \, V(\tau ) \, d\tau	 + \rec{m_e}\, R(t)\,,
  \q t\ge 0\,,
\eqno (3.18)$$
where, as usual, the limits of integration  are  extended
to account for the possibility	of  Dirac-type	distributions.
The {\it fluctuation-dissipation  theorem} can be  readily expressed
by the Laplace transforms, see \eg Mainardi and  Pironi [105].
In our notation this theorem leads to
$$ \barr C_{{V}}(s) := \barr{{\langle\,  {V}(0) \, {V}(t)\, \rangle}}
   =  { \langle \, {V}^2(0)\, \rangle \over s + \barr \gamma (s)}
     \,, \eqno(3.19)$$
and
$$
 \barr C_{R}(s) := \barr{{\langle\,  {R}(0) \, {R}(t)\, \rangle}}
=  m_e^2 \,   { \langle \, {V}^2(0)\, \rangle}\,
     \barr \gamma (s)\,. \eqno(3.20)$$
The classical results are easily recovered for
$t\ge 0\,$ by noting that,
in the absence of added mass and retarding effects,
we get
$ \barr\gamma(s)=1/\sigma_p \,\div\, \gamma(t)=\delta(t)/\sigma_p \,.
$
\vsp
Taking into account both the added mass and the Basset history force
(whose expression has been given in the previous section in terms
of a fractional derivative)
the  Langevin equation	(3.3) turns out to be  modified into
$$		  {dV\over dt} =
 -\rec{\sigma_e }\,\l[
  1 + \sqrt{\tau_0 } \,{d^{1/2}\over dt^{1/2}}\r] \, V(t)
+  \rec{m_e} \, R(t)\,, \q \tau _0 := {r_0^2  \over \nu}\,.
\eqno (3.21)$$
Here the fractional derivative is intended
in the Caputo sense with starting point $t_0=0\,,$     \ie
$$ {d^{1/2}\over dt^{1/2}} \, V(t) =
    \rec{\sqrt{\pi}}\, \int_{0}^t
  { dV/d\tau \over \sqrt{t-\tau }}\, d\tau
 \,. \eqno(3.22)$$
We agree to refer to (3.21) as	the {\it fractional Langevin equation}.
\vsp
We easily recognize that our fractional Langevin equation (3.21) can be
considered a particular case of the $GLE$ (3.18)
by noting that
$$ \barr \gamma(s) =   \rec{\sigma_e  }\,
     \l[ 1 + \sqrt{\tau_0 } \, s^{1/2} \r]
	\,  \div \,
\gamma (t) = \rec{\sigma_e  }\,
 \l[ \delta (t) - \sqrt{\tau_0 } \,
   {1 \over 2 \sqrt{\pi}} \,  t^{-3/2} \, \Theta(t)\r]
 \,, \eqno(3.23) $$
where $\Theta(t)$ is the  Heaviside step function.
Therefore the expression for $\gamma (t)$ turns out to be defined
only in the sense of  distributions. Specifically, $\delta (t)$
is the well-known Dirac delta function
and  $t^{-3/2} \, \Theta(t)$ is the linear functional over
test functions, $\,\phi(t)\,, $ such that
$$
 { \langle \, t^{-3/2}\,\Theta(t)\,,\, \phi(t) \rangle}  \,
  = \int _0^\infty {[\phi(t)-\phi(0)]\over t^{3/2}}\, dt\,. $$
For more details on distributions,  see \eg [107] or [108].
\vfill\eject
The significant change	with respect to the classical case results
from the $\,t^{-3/2}\,$ term. Not only does it imply
a non-instantaneous relationship between the force and the
velocity, but also it is a slowly decreasing function so
that the force is effectively related to the velocity over
a large time interval.
The representation of the force in terms of distributions,
as required by the $GLE$, is not
strictly necessary since
we can use  the equivalent fractional form.
\vsp
Let us	consider the {autocorrelation} for the random force. The
inversion of the Laplace transform $\barr C_R (s)$ yields,
by (3.22-23),
$$ C_R (t) =
 {m_e^2\over \sigma_e  } \,
  { \langle \, V^2(0)\, \rangle}  \,
\l[ \delta (t) - \sqrt{\tau_0 } \, {1\over 2 \sqrt{\pi}}
  \, t^{-3/2}
\r]
 \,,\q t\ge 0\,,  \eqno(3.24)$$
to be compared with the classical result (3.9).
Thus, we recognize that,  in the presence of the Basset history force,
the random force can no longer
be  represented uniquely by a white noise;  an
additional "fractional" or "coloured"   noise is  present due to the term
$t^{-3/2}$  which, as  already noted,
is to be interpreted in the  sense
of distributions.
Since the fluctuating force  is no longer uncorrelated
at different times, the fractional Langevin equation
does {\it not} represent a Markovian process. Nevertheless,
it is still Gaussian (since the Gaussian nature
of the driving sources for the fluid is  assumed), and stationary
(in view of the time-shift  invariance).
since
\vsp
Let us now consider  the  autocorrelation for the velocity field.
Inserting (3.23) in (3.21), it turns out as
$$ \barr C_V (s) =
    { {\langle \, V ^2(0)\, \rangle} \over
      s +\l[1+ \sqrt{\tau_0 } \, s^{1/2}\r]/ \sigma_e }
 = { {\langle \, V ^2(0)\, \rangle} \over
      s + \sqrt{\beta /\sigma_e } \, s^{1/2} +1/\sigma_e  }
\,, \eqno(3.25)$$
where
$ \beta :=  {\tau_0 / \sigma_e	}\,, $ see (2.12-13).
We first note that the effect of the
Basset force is expected to be negligible for $\beta \to 0\,$
($\chi := \rho _p/\rho_f \to \infty$),
\ie  for particles
which are  sufficiently
heavy with respect to the fluid.
In this case we can  assume also $\sigma _e \approx \sigma _p$
so the classical results
(3.8), (3.9) and (3.11) turn out to be true.  
\vsp
A first result concerning the asymptotic behaviour of $C_V(t)$ as
$t\to \infty$
can be easily obtained	from (3.25)
by
applying the asymptotic theorem for the Laplace transform as
$s \to	 0\,, $  see \eg Doetsch [109]. In fact, from
$$   \barr C_V (s) \sim        \sigma _e
 \, {\langle \, V ^2(0)\, \rangle} \,
 (1 - \sqrt{\beta \sigma _e}\; s^{1/2})\,, \q s \to 0\,, $$
we get
$$ C_V (t) \sim
    {\langle \, V ^2(0)\, \rangle}\,
 \sqrt{\beta /( 4  \pi)}
 \,   \l(t/ \sigma_e \r)^{-3/2}\,, \q t\to \infty
    \,. \eqno(3.26)$$
The presence of such a long-time tail
is thus in agreement with that	formerly observed in computer
simulations 
by Alder and  Wainwright [74].
\vfill\eject
\vsp
The explicit inversion of the Laplace transform  in (3.25)
can be carried out  in a way similar to that used in the
(deterministic) ordinary Basset problem treated
in the previous Section, see (2.27-30).
For this purpose we need to consider the function
     $$ \barr N(s) =
 {1\over s+ a \,\ss +1}  \,, \q a = \sqrt{\beta}\,, \eqno(3.27)$$
and recognize that
$$ {C_V (t)\over{\langle \, V^2(0)\, \rangle}}	= N(t/\sigma_e)
    \, \div \, \sigma _e\, \barr N( \sigma _e\,s) =
  {1 \over    s + \sqrt{\beta /\sigma_e } \, s^{1/2} +1/\sigma_e  }
\,. \eqno(3.28)$$
Thus, the actual solution is  obtained by expanding
 $\barr N(s)$ into  partial fractions and then inverting.
\vsp
We first obtain
$$ \barr N(s) =
 {1\over s+ a \,\ss +1}= \l\{
\eqalign{
i)\,&\, {A_+\over \ss\,(\ss -\lambda_+ )} +
    {A_-\over\ss (\ss -\lambda_- )} \,, \cr
ii)\,&\, {1 \over (\ss +1)^2}\,, \cr }
   \r. \eqno(3.29)$$
where	$\lambda_\pm$ and $A_\pm$ are given by (2.28), and the
distinction of cases $i)$ and $ii)$ is the same as there.
\vsp
Then, the Laplace inversion of $ (3.29)$ can be expressed
in terms of  Mittag-Leffler functions
of order $1/2\,, $
$E_{1/2} (\lambda\sqrt{t}) = {\rm exp}(\lambda^2t)\,
 {\rm erfc} (-\lambda\sqrt{t})\,, $ as shown in the Appendix of [8].
We obtain,  using (3.28),
$$  {C_V (t)\over{\langle \, V^2(0)\, \rangle}} =
  \l\{
\eqalign{
i)\,&\,   A_+\,  E_{1/2} (\lambda _+\,\sqrt{t/\sigma _e}\,)
   + A_-\,  E_{1/2}(\lambda _-\,\sqrt{t/\sigma _e}\,)
     \,, \cr\cr
ii)\,&\,  (1 +2\,t/\sigma _e)\,   E_{1/2} (-\sqrt{t/\sigma _e}\,)
       - (2/ \sqrt{\pi} )\, \sqrt{t/\sigma _e} \,.	  \cr }
   \r. \eqno(3.30)$$
\vsp
Furthermore, it can be shown
that $N(t)$ is a  {\it completely monotone} function for $t>0\,, $
decreasing  from $1$ to $0\,, $
as $t$ runs from $0$ to $\infty\,. $
\vsp
Let us now consider the displacement variance, which is provided
by the repeated integral of the velocity autocorrelation as
indicated in (3.10).
 From the Laplace transform
  $\barr{{\langle\,X^2(s)\, \rangle}}= 2\, \barr C_V (s)/s^2\,, $
we first derive  the asymptotic behaviour  of
 ${\langle\,X^2(t)\, \rangle}$
as $t\to \infty\,. $  We easily obtain
$$ {\langle\,X^2(t)\, \rangle}	=
2 D\, t \l\{1 -
 2\sqrt{\beta / \pi}\,	(t/\sigma _e) ^{-1/2} +
   O\l[ (t/\sigma _e)^{-1} \r] \r\}
\,,  \q t\to \infty \,, \eqno(3.31)$$
where  $D$   is the diffusion coefficient defined in (3.12-14).
\vfill\eject
\vsp
The explicit expression of the displacement variance can be obtained
by expanding $\barr N(s)/s^2$ into partial fractions and then inverting.
In the case $i)\;\beta \ne 4\,, $ we obtain
$$  \eqalign{
{\langle\,X^2(t)\, \rangle}  =\,&  2 D \,
  \l\{ t -  2\,\sqrt{\beta\, \sigma _e\, t\over \pi}\r.\cr
 +& \,\sigma _e\, \l.
{\lambda _+^3 \,[1- E_{1/2}(\lambda _- \sqrt{t/\sigma _e}\,)]
 -\lambda _-^3 \,[1- E_{1/2}(\lambda _+ \sqrt{t/\sigma _e}\,)]
 \over	(\lambda _+-\lambda _-)}
 \r\}\,. \cr}
\eqno(3.32)$$
 \vsp
Thus,  the   displacement  variance is proved
to maintain, for sufficiently long times,
the linear behaviour  which is typical of normal diffusion
(with the same diffusion coefficient  as in the classical case).
However, the Basset history force,
which is responsible of the algebraic decay of the velocity correlation
function, induces  a retarding effect  in the
establishing   of the linear behaviour of the displacement variance.
As we shall see hereafter,
this retarding effect  is  more evident
when the Brownian particle is  lighter, such as to give rise to
regimes  of effective {\it fast anomalous diffusion}
characterized by the law
 $$ {\langle\,X ^2(t)\, \rangle} \sim 2 \,D_a\,t^{\alpha} \,,
   \q D_a= a\, D\, (\sigma_p) ^{1-\alpha}\,;
 \q  0< a<1\,, \q 1< \alpha < 2\,. \eqno(3.33) $$
\vs
\line{3.4 {\it	Numerical Results and Discussion}\hfill}
\vsp
In order to get a physical insight of the effect of the
Basset history force (coupled with the added mass)
on the classical Brownian motion,
we exhibit the results obtained recently by
Mainardi and  Tampieri [106]
concerning plots of the  velocity autocorrelation (3.30) and the
displacement variance (3.32).
As an example we consider relatively light Brownian particles,
by assuming  $\chi= 0.1\,$ and $\chi= 0.5\,. $
We  take non-dimensional quantities,
by scaling  the time with the  decay constant $\sigma_p$
of the classical Brownian motion and the displacement
with the diffusive scale $(D\,\sigma_p )^{1/2}\,. $
Please note that here we have preferred to scale the time with
$\sigma_p $ more than with $\sigma _e\,, $ since in the classical approach
the added mass effect is neglected!
With these scales the asymptotic equation for the displacement variance
reads ${\langle\,X ^2(t)\, \rangle} \sim 2 \,t\,. $
 \vsp
In Figs 3-1 and 3-2 we plot versus the normalized time 
the velocity autocorrelation
normalized with its initial value
 ${\langle\,V ^2(0)\, \rangle} \,$
and the displacement variance normalized with its asymptotic
value $2\,t\,.$
We compare any function, provided by our full
hydrodynamic approach (added mass and Basset force), in continuous
line, with  the corresponding one,
provided by the classical analysis, in dashed line,
and by	the only effect of the added mass, in dashed-dotted line.
For large times we also exhibit the asymptotic estimations (3.26)  and
(3.31), in dotted line, in order to recognize  their range of validity.
\vfill\eject
$\null$
\cen{{\epsfxsize=7.5truecm \epsfysize=8.0truecm \epsffile{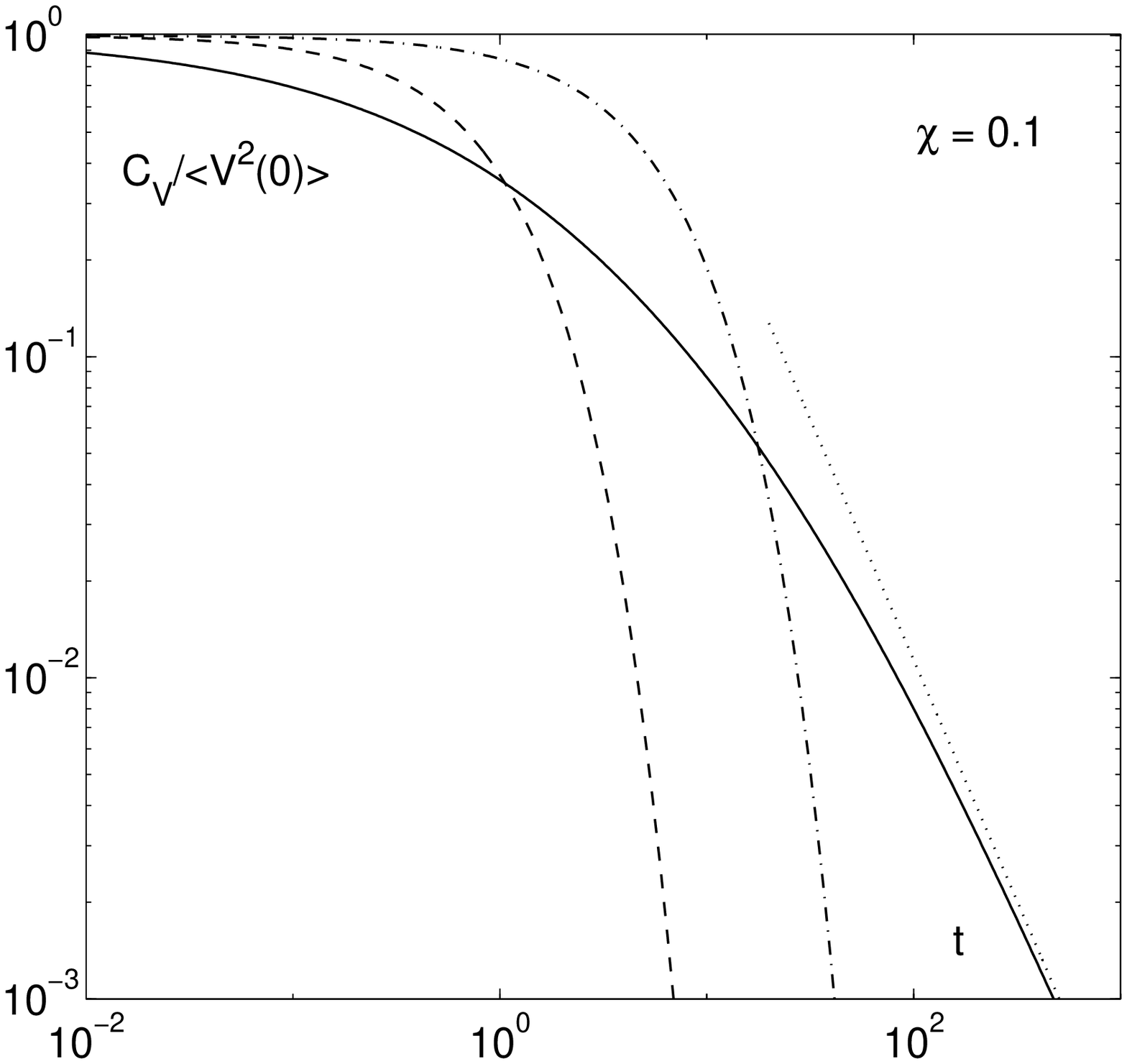}}
 {\epsfxsize=7.5truecm \epsfysize=8.0truecm \epsffile{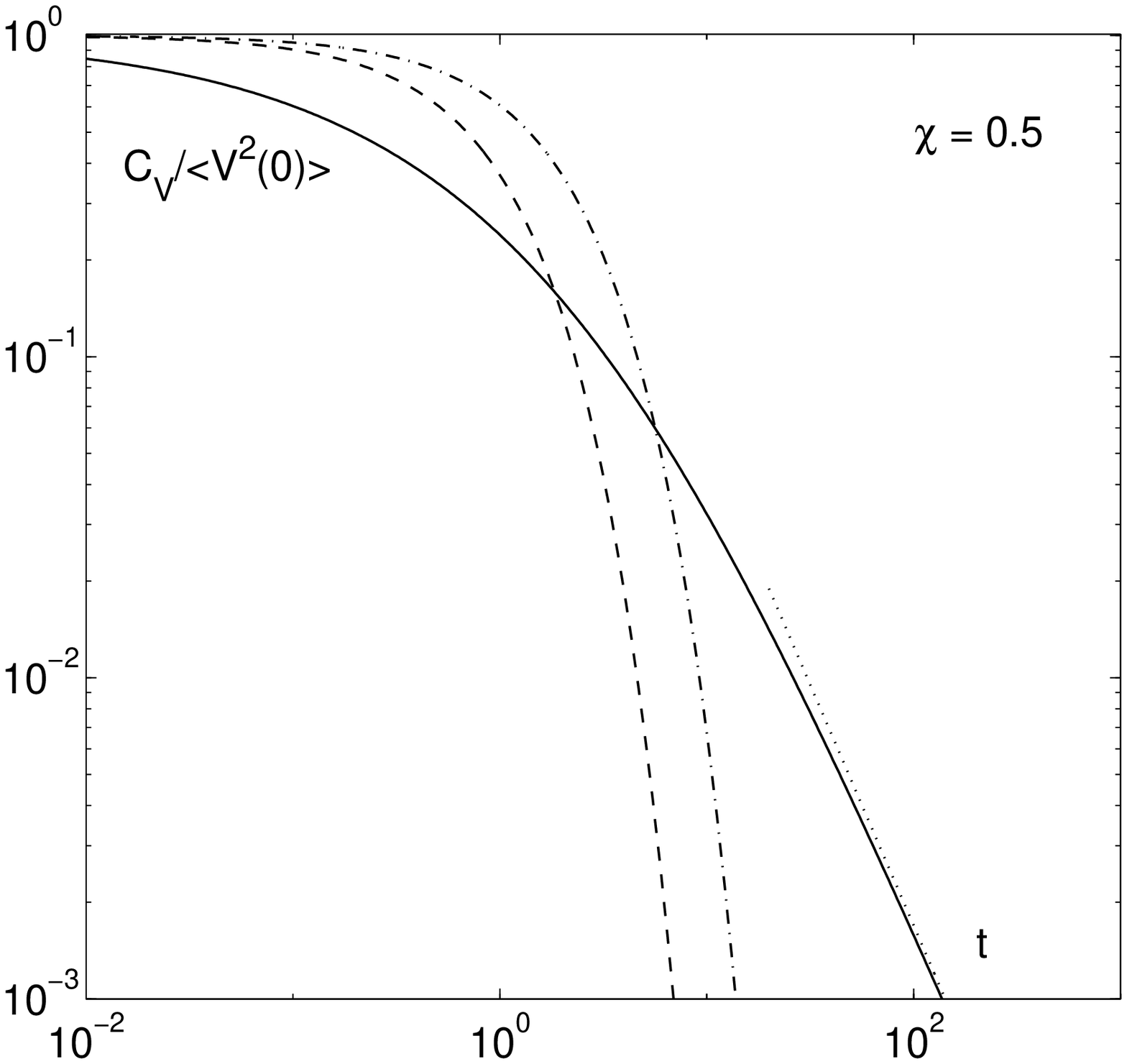}} }
\vskip 0.1 truecm
\cen{{\bf Fig. 3-1}}
\vsh
\cen{Velocity autocorrelation versus time
for $\chi= 0.1$ (left) and for $\chi=0.5$ (right):}
\cen{full hydrodynamic ----- $\,$; 
added mass $- \,\cdot\, - \cdot \, -\,$;
  classical $ - - - \,.$}
$\null$
\vskip 0.1 truecm
\cen{{\epsfxsize=7.5truecm \epsfysize=8.0truecm \epsffile{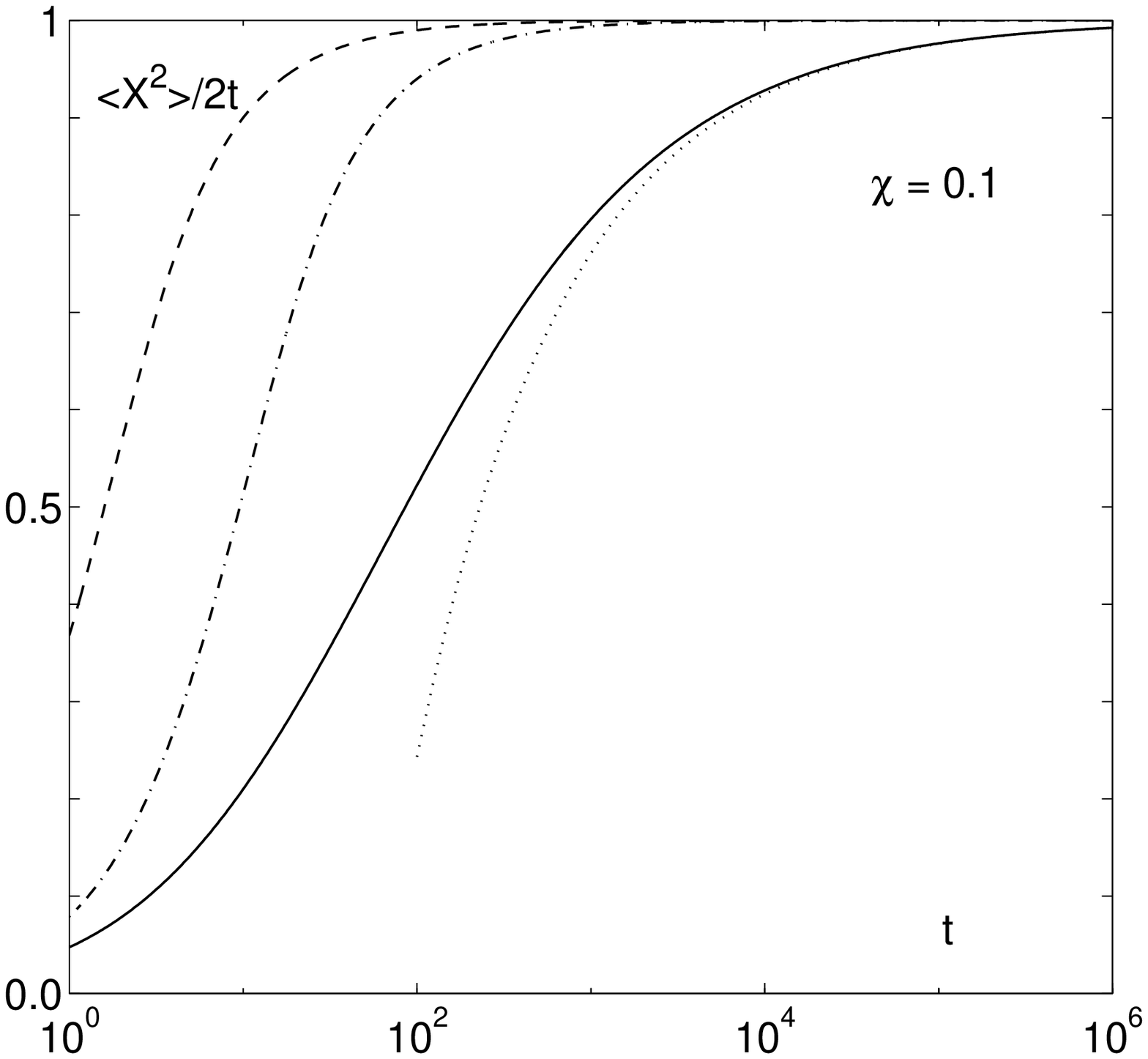}}
 {\epsfxsize=7.5truecm \epsfysize=8.0truecm \epsffile{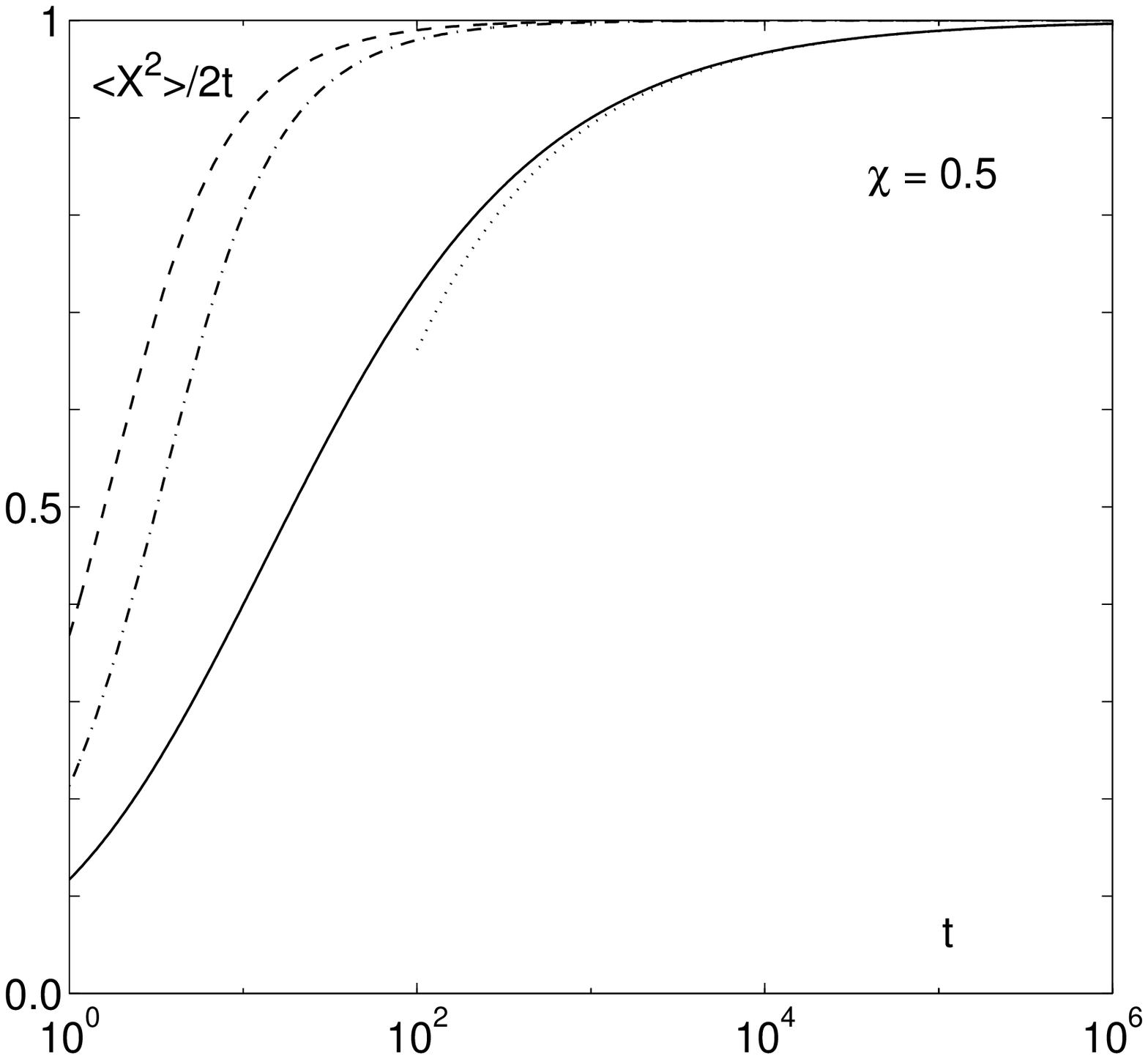}} }
\vskip 0.1 truecm
\cen{{\bf Fig. 3-2}}
\vsh
\cen{Displacement variance versus time
     for $\chi= 0.1$ (left) and for $\chi=0.5$ (right):}
\cen{full hydrodynamic ----- $\,$; 
added mass $- \,\cdot\, - \cdot \, -\,$;
  classical $ - - - \,.$}
\vfill\eject
The retarding effect of the Basset force is   more evident when
the Brownian particle is  lighter, such as to appear  a manifestation
of fast anomalous diffusion. In fact, if we consider
a time interval  (say two decades)
starting when
the classical analysis foresees the establishment of the asymptotic
linear behaviour  for the displacement variance,
 a law of anomalous diffusion
 ${\langle\,X ^2(t)\, \rangle} \sim 2 \,a\,t^{\alpha} \,,
   \; a>0\,, \;  \alpha \ne 1\,,  $
can  well approximate the exact formula (3.32), provided
by the full hydrodynamic model.
By evaluating	the parameters of the anomalous  diffusion,
$a $ and $\alpha \,, $
with a best fit based on the least squared  method,
we find $0<a<1$ and $1<\alpha <2\,. $
We recognize that the effective {\it anomalous diffusion} turns out to be
{\it fast}; in particular,  it is faster as $\chi$ is smaller, with
parameters $a \to 0^+$ and $\alpha \to 2^- $  as $\chi \to 0^+\,.  $
Of course, the normal diffusion is recovered as $\chi \to \infty\,,$
since $a \to 1^-$ and $\alpha \to 1^+\,.  $
In Fig. 3-4 we show the function ${\langle X^2 \rangle}/ 2$
versus time (in the 2-decade range $10^1 \div 10^3$)
 corresponding either to our analysis  
 and to the classical analysis.  
 While the classical curve, in dashed line above,  is  practically
coincident  with the linear one (regime of normal diffusion)
our curve, in continuous line below, is fitted with a power-law curve,
in dashed-dotted  line,  with an exponent $\alpha >1\, $
(regime of fast anomalous diffusion).
\cen{{\epsfxsize=7.5truecm \epsfysize=8.0truecm \epsffile{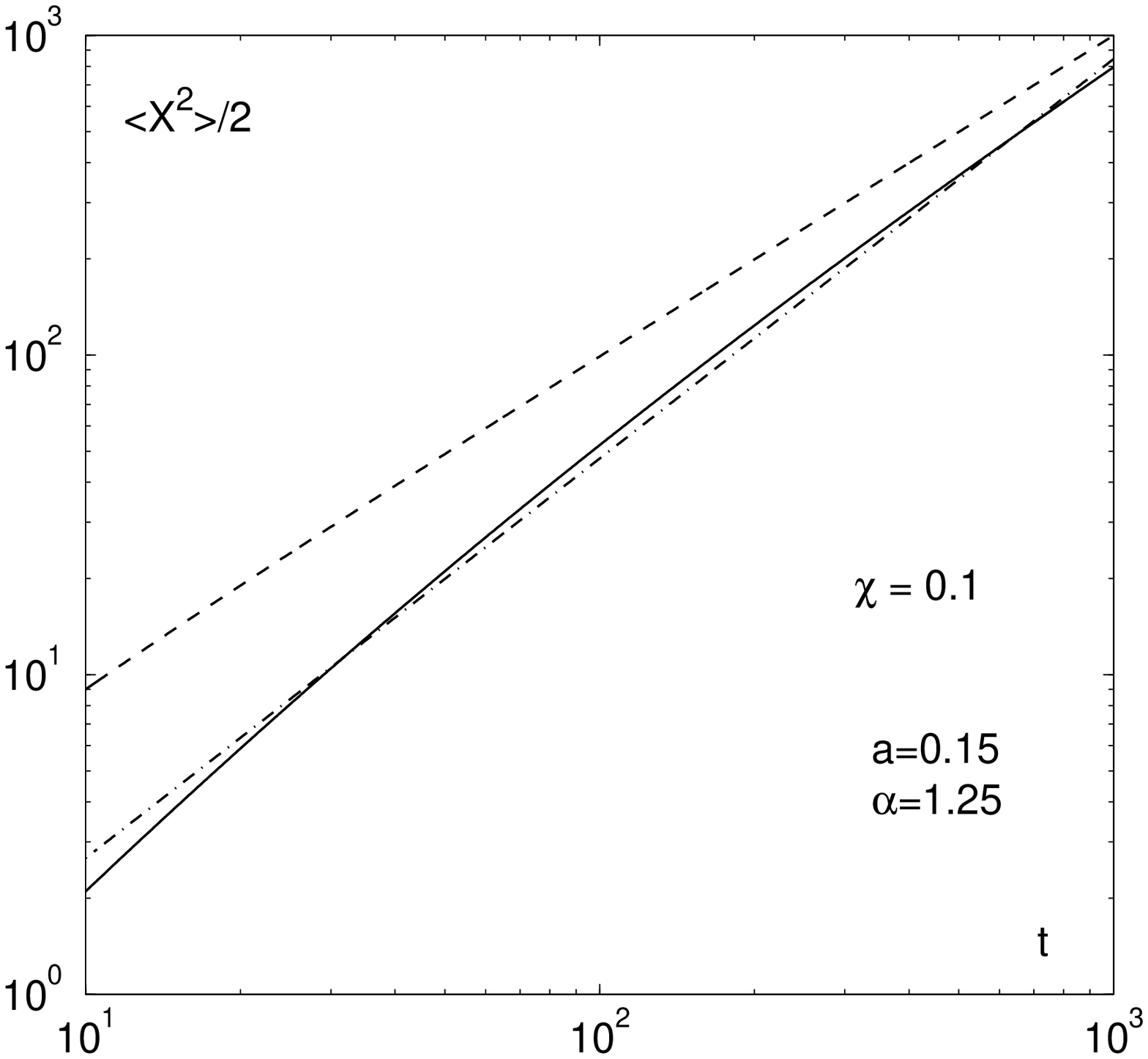}}
 {\epsfxsize=7.5truecm \epsfysize=8.0truecm \epsffile{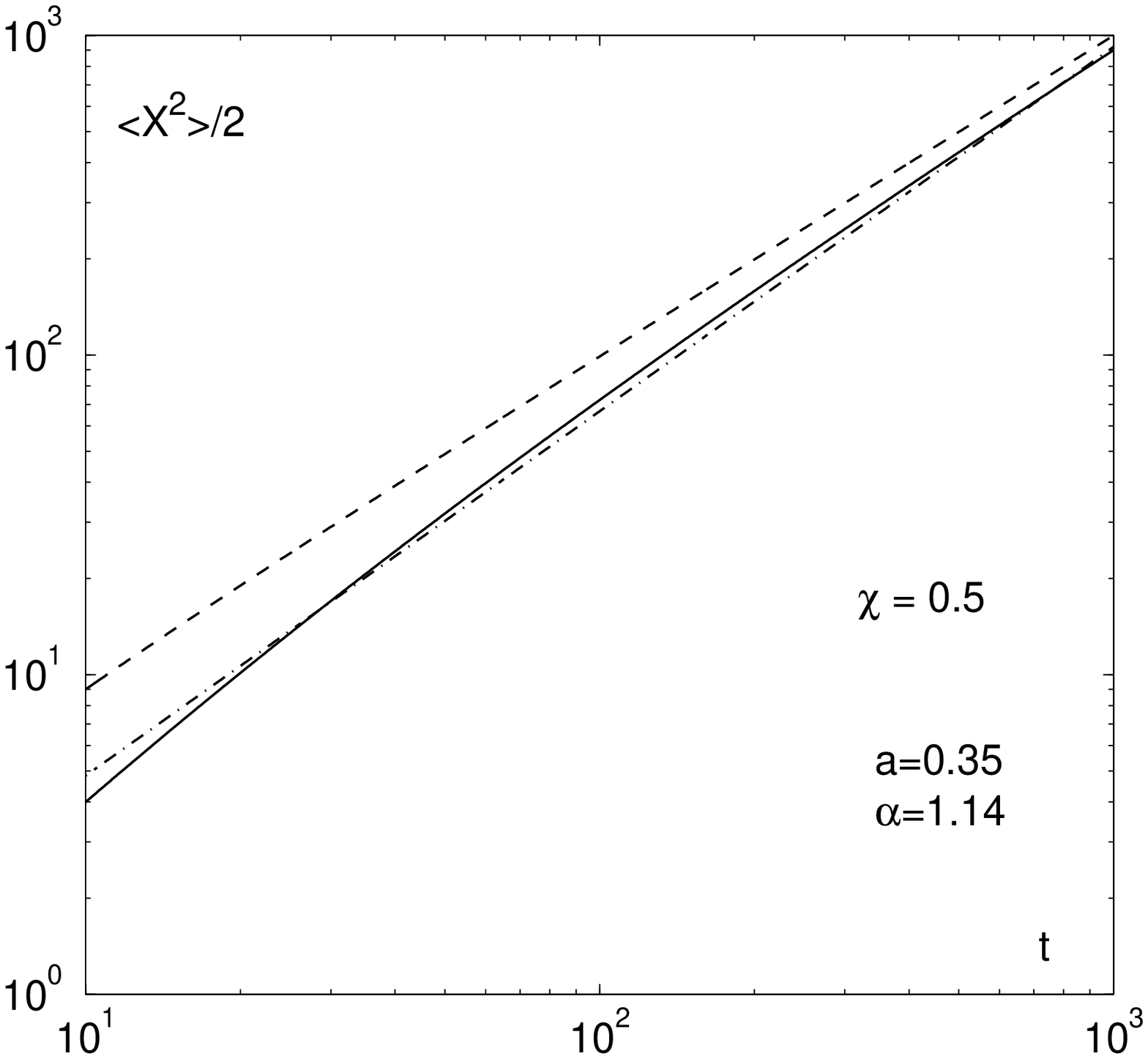}} }
\vskip 0.1 truecm
 \cen{{\bf Fig. 3-3}}
\vsh
\cen{
 The displacement variance at large times
for $\chi  = 0.1\,,\, 0.5\,:$}
\cen{full hydrodynamic	 -----$\,;$  
best-fit  $- \,\cdot\, - \,\cdot \,-\,;$
  classical $---\,.$}
\vsp
From the above analysis we conclude that  if an observer investigates the
time evolution of a cloud of sufficiently light Brownian particles, he
recognises
that the normal diffusion is preceded by a regime of fast anomalous
diffusion, which lasts for long time.
If the observation interval is not sufficiently long, he may be
induced to trust in the occurring of fast anomalous diffusion.
\vfill\eject

%% file: fmnew24.tex
\pageno=321
\def\D{{\cal D}}
\def\Gc{{\cal {G}}_c}	\def\Gcs{\barr{\Gc}} 
\def\Gs{{\cal {G}}_s}	\def\Gss{\barr{\Gs}} 

\def\argsa{(x/ \sqrt{\D})\, s^{\nu }}
\def\arg{ x^2/ (4\D\, t)}
\line{4. THE FRACTIONAL DIFFUSION-WAVE EQUATION \hfill} \vs
\line{4.1 {\it Introduction}\hfill} \vsp
By {\it fractional diffusion-wave} equation we mean the  linear
integro partial differential equation
obtained from the
classical diffusion or wave equation by replacing the first- or
second-order time derivative by a fractional derivative (in the Caputo
sense) of order
$\beta	  $ with $0 <\beta    \le 2$. In our notation it reads
$${\d^{\beta  }  u\over \dt^{\beta   }} =
 \D\,{\d^2 u \over \dx^2} \,,\qq u=u(x,t)\,, \q
0<\beta  \le 2 \,, \q \D>0\,,\eqno(4.1)$$
where
$\D$ denotes a positive constant with the dimensions $L^2\,T^{-\beta  }\,,$
$x$ and $t$ are the space-time variables,
and  $u=u(x,t)$ is the field variable, which is assumed
to be a {\it causal} function of time, \ie
vanishing for $t<0\,. $
From  the Chapter of  Gorenflo and  Mainardi [8], see in \S 1.3
 Eq. (1.17),  we  remind the definition of the Caputo fractional derivative
of order  $\beta  >0 $
for a (sufficiently well-behaved) {\it causal} function $f(t)\,, $
$$
 D_*^\beta  \,f(t) := J^{m-\beta } D^m \, f(t) =
\cases{
    {\ds \rec{\Gamma(m-\beta )}}\,{\ds\int_0^t
 {\ds {f^{(m)}(\tau)\over (t-\tau )^{\beta  +1-m}}} \,d\tau} \,,
  & $\; m-1<\beta  <m\,, $\cr\cr
     {\ds {d^m\over dt^m}} f(t)\,,
    & $\; \beta  =m\,. $\cr\cr }
  $$
Introducing   the causal power function
$$ \Phi_{\lambda}(t) := { t_+^{\lambda	-1}\over  \Gamma(\lambda )}\,,\q
   \lambda   >0\,, $$
where the suffix $+$ is just denoting that the function is
vanishing for $t<0\,, $ and recalling the Laplace transform pair
$\Phi_\lambda (t)\,\div\, s^{-\lambda }\,, $ we easily recognize
that
$$	  D_*^\beta   \,f(t)
     :=  \Phi_{m-\beta }(t) \,*\, f^{(m)}(t)
\, \div \,
	s^\beta \,  \widetilde f(s)
   -\sum_{k=0}^{m-1} s^{\beta  -1-k} \, f^{(k)}(0^+)\,,
\q m-1 < \beta \le m\,. $$
We note  $\Phi_{\lambda}(t) \,*\, \Phi_{\mu }(t) =
  \Phi_{\lambda+\mu }(t)\,.$
In Eq. (4.1) we thus need to distinguish  two cases
 $i)\q 0<\beta	\le 1\,, \q {\rm and}\q ii)\q 1<\beta  \le 2\,, $
for which the equation
assumes the explicit forms as follows :
$$ \Phi_{1-\beta } (t) \,*\, {\d u\over \dt} =
    \rec{\Gamma(1-\beta )}\,
 \int_0^t (t-\tau )^{-\beta }\, \l( {\d u\over \d\tau}\r) \, d\tau
   =   \D\,{\d^2 u \over \dx^2} \,,\qq 0<\beta	\le 1 \,; \eqno(4.2)$$
$$ \Phi_{2-\beta } (t) \,*\, {\d^2 u\over \dt^2} =
    \rec{\Gamma(2-\beta )}\,
 \int_0^t (t-\tau )^{1-\beta } \, \l({\d^2 u\over\d\tau^2}\r)\, d\tau
   =   \D\,{\d^2 u \over \dx^2} \,,\q 1<\beta  \le 2 \,. \eqno(4.3)$$
The equations $(4.2)$ and $(4.3)$ can be properly referred to as  the
{\it time-fractional diffusion} and the {\it time-fractional wave}
equation, respectively.
\vfill\eject
A {\it fractional diffusion} equation akin to (4.2)
has been explicitly introduced in physics by Nigmatullin
[110] to describe  diffusion in special types of porous media,
which exhibit a  fractal geometry.
The author [111] has shown  that
the {\it fractional wave} equation governs the propagation of mechanical
diffusive waves in viscoelastic media which exhibit a simple power-law
creep. This problem of dynamic viscoelasticity,
formerly treated by Pipkin [5] 
but unaware of the interpretation by fractional calculus,
thus provides  an interesting example of the relevance of (4.3)
in physics.
Of course, anytime some  hereditary mechanisms of power-law
type are present in  diffusion or wave phenomena, the
appearance of time fractional derivatives in the evolution
equations  is expected.
\vsp
In a series of papers [112-116] the author has pursued his analysis
on the fractional diffusion-wave equation (4.1),
based on Laplace transforms and special functions of Wright type.
Mathematical aspects of  integro differential equations
akin to  (4.2-3) and based on the use of integral transforms
and special functions
have been treated in some relevant papers
by Wyss [117],
Schneider and  Wyss [118], Schneider [119]
(Mellin transforms and Fox $H$ functions)
and by Fujita [120]  (Fourier transforms and
Mittag-Leffler functions).
More formal  approaches based on semigroup theory in Banach spaces
have been given by Kochubei [121-122]  and El-Sayed [123].
Recently the integro-differential equation
treated by  Fujita has been considered by Engler [124]
in a very interesting paper
in view of the connection between similarity solutions and
stable probability distributions.
\vsp
Hereafter we  present a review	on the fractional
evolution equation (4.1), essentially  based on our works [111-116].
In \S 4.2 we  analyse
the two basic boundary-value problems,
referred to as
the {\it Cauchy problem}  and the {\it Signalling problem},
by   the technique of  the Laplace transform and
we derive  the transform expressions
of the respective fundamental solutions (the {\it Green
functions}).
 \vsp
In \S 4.3 we  carry out the inversion of the relevant Laplace transforms
and we outline a reciprocity relation between the Green functions
themselves in the space-time domain.  In view of this relation
the Green functions can be expressed
in terms  of two interrelated {\it auxiliary functions}  in
the similarity variable $r = |x|/(\sqrt{D}t^{\beta  /2})\,.$
These auxiliary
functions can be analytically continued in the whole complex plane
as entire functions of {\it Wright} type.
 \vsp
In \S 4.4  we show the evolution of the fundamental solutions
of both the Cauchy and Signalling problems for some (rational) values
of the order of time derivation. To
gain more insight into the phenomenon of fractional diffusion we also
exhibit the evolution
of an initial box function in the Cauchy problem.
This allows us to better recognize
the processes of slow diffusion ($0<\beta  <1$) and the   intermediate
processes   between diffusion and wave propagation ($1<\beta  <2$).
\vfill\eject
\vsp
Finally, in the Appendix, we provide the reader
with a review of the main mathematical	properties of
our auxiliary	functions in the framework of the Wright
functions.   The interesting connection with the
stable probability distributions is not treated
but deferred to recent works of ours, there quoted.
\vvs
\line{4.2 {\it Analysis of the Cauchy and Signalling Problems
with the Laplace Transform}\hfill} \vsp
As well known, the two basic boundary-value problems
for the  evolution equations of diffusion
and wave  type	are the {\it Cauchy} and {\it Signalling problems}.
In the	{\it Cauchy problem}, which concerns  the space-time domain
$-\infty <x< + \infty\,, $ $\, t \ge 0\,, $
the data are assigned at $t=0^+$ on
the  whole space axis (initial data).
In the {\it Signalling problem}, which concerns  the space-time domain
$x\ge 0\,, $ $\, t \ge 0\,, $
the data are assigned both at $t=0^+$ on the semi-infinite
space axis $ x >0 $ (initial data) and at $x=0^+$ on the semi-infinite
time axis  $ t>0$ (boundary data); here, as mostly usual,
the initial data are assumed to be vanishing.
Extending
the classical analysis to our fractional equation (4.1), and
denoting by $f(x)$ and $h(t)$ two given, sufficiently well-behaved
functions,    the basic problems are thus formulated as following:
\vsh\pni
{\it a) Cauchy problem}
$$   u(x,0^+)=f(x) \,, \q -\infty <x < +\infty\,; \qq
     u(\mp \infty,t) = 0\,,\q \, t>0\,;  \eqno(4.4a)
$$
\pni
{\it b) Signalling problem}
$$  u(x, 0^+) =0 \,, \q  x>0\,;\qq
    u(0^+,t ) =h(t) \,, \q u(+\infty,t) =0 \,, \q   t >0 \,. \eqno(4.4b)
$$
If $1<\beta    \le 2\,, $ we  would add in (4.4a) and (4.4b)
the initial value of the first-order  time derivative 
of the field variable,
\ie ${\d \over \d t} u(x,0^+)= g(x)\,,$ since in this case
Eq. (4.1) turns out to be
of the second order in time, see   the integro-differential equation
(4.3), and, consequently, two linearly independent solutions are to be
determined.   We limit ourselves to choose
$g(x) \equiv 0\,. $
We easily recognize that the above Cauchy problem for (4.1)
can be expressed
through the integral equations	of fractional order
$$  u(x,t) = \cases{
 f(x)  +
   {\ds {\D\over \Gamma(\beta )}}
   {\ds \int_0^t} \!\!
   \l[{\ds{\d^2 u\over \dx^2}}(x,\tau)\r](t-\tau)^{\beta-1}\,d\tau \,,
  & $0<\beta  \le 1,$\cr\cr
f(x) + t\, g(x) +
  {\ds {\D\over \Gamma(\beta )}}
  {\ds\int_0^t }
 \l[{\ds{\d^2 u\over \dx^2}}(x,\tau)\r](t-\tau)^{\beta-1}\, d\tau\,,
 & $1<\beta \le 2.$ \cr}
\eqno(4.5)$$
We thus note that for $1<\beta\le 2$ the choice $g(x)=0$ ensures
the continuous dependence  of the solution on the parameter $\beta $
also in the transition from $\beta    =1^-$ to	$\beta	  =1^+\,.$
\vsp
In view of our analysis we find it convenient to put
$$  \nu  ={\beta  \over 2}\,,
   \q 0<\nu  <1\,.  \eqno(4.6)$$
\vfill\eject
\vsp
For the {\it Cauchy} and {\it Signalling}  problems we
introduce
the so-called Green functions $\Gc (x,t;\nu  )$ and $\Gs(x,t;\nu  )$,
which represent the respective fundamental solutions,
obtained when $g(x) = \delta (x)$ and $h(t) = \delta (t)\,. $
As a consequence, the solutions of the two basic problems
are obtained by a space or time convolution  according to
$$ \eqalignno{u(x,t;\nu )
&= \int_{-\infty}^{+\infty} \Gc(x-\xi ,t;\nu   ) \, f(\xi)
\, d\xi     \,,  &(4.7a)\cr
 u(x,t;\nu  ) &= \int_{0^-}^{t^+} \Gs(x,t-\tau;\nu   ) \, h(\tau ) \,
d\tau	  \,.  &(4.7b)\cr}$$
It should be noted that  $\Gc(x ,t ;\nu ) =
  \Gc(|x|,t ;\nu )$ since the Green function turns out to be
an even function of $x$.
\vsp
For the {\it standard diffusion} equation ($\nu   =1/2$)
it is well known that
$$   \eqalignno{
\Gc(x,t;1/2) := \Gc^d (x,t)
 &= {1\over 2\sqrt{\pi \D}}\,t^{-1/2}\, \e^{-\ds\arg}\,,
 &(4.8a) \cr\cr
\Gs(x,t;1/2) := \Gs^d (x,t)
 &= {x\over 2\sqrt{\pi \D}}\, t^{-3/2} \, \e^{-\ds\arg}\,.
 &(4.8b) \cr}$$
 \vsp
For the {\it standard wave} equation ($\nu   =1$)
it is well known that, putting $c = \sqrt{\D}\,, $
$$   \eqalignno{
\Gc(x,t;1) := \Gc^w (x,t)
 &={1\over 2} \l[ \delta (x-ct) + \delta (x+ct)\r] \,,
 &(4.9a) \cr\cr
\Gs(x,t;1) := \Gs^w (x,t)
 &= \delta (t-x/c)  \,. &(4.9b) \cr}$$
\vsp
In the general case $0<\nu  \le 1$ the two Green functions will be
determined by using the technique of the Laplace transform. This technique
  allows us  to obtain	the transformed
functions $\Gcs(x,s;\nu  )$, $\Gss (x,s;\nu  )$,
by solving ordinary differential equations of the 2-nd order in $x$
and then,
by inversion,  $\Gc(x,t;\nu  )$ and $\Gs(x,t;\nu  )$.
\vsp
 For the {\it Cauchy Problem} (4.4a)
the application of the Laplace
transform to (4.1)  with
$u(x,t)=\Gc(x,t;\nu )\,, $
 and $ \Gc(x,0^+; \nu  )= f(x) =\delta (x)\,, $
[and ${\d\over \d t} \Gc(x,0^+;\nu ) = 0\,$ if $1/2 <\nu  \le 1\,$]
leads  to the non-homogeneous differential equation
satisfied by the image of the Green function,
$\Gcs(x,s;\nu  )\,, $
$$
 \D\,{d^2 \Gcs	\over dx^2}-  s^{2\nu }  \,\Gcs=
   -\delta (x)\,s^{2\nu  -1}\,, \q -\infty <x<+\infty
    \,.       \eqno(4.10)$$
Because of the singular term $\delta (x)$ we have to consider
the above equation separately in the two intervals $x<0$ and $x >0$,
imposing the   boundary conditions  at $x=\mp \infty\,, $
$\Gc({\mp \infty, t;\nu }) =0\,, $
and the necessary matching conditions
at $x= 0^{\pm}$.
\vfill\eject
\noindent
We obtain
$$  \Gcs(x,s;\nu ) =
    \rec{2\sqrt{\D}\, s^{1-\nu }} \,
 \e^{\ds -(|x|/\sqrt{\D})s^{\nu }}\,, \q -\infty <x<+\infty
\,. \eqno(4.11)$$
In fact,
from (4.10)  $\Gcs$ is expected in the form
 $$ \Gcs(x,s;\nu ) = \cases{
 c_1(s)\,\e^{-\argsa}+c_2(s)\,\e^{+\argsa}\,,& if $\q x >0\,; $  \cr\cr
 c_3(s)\, \e^{- \argsa} + c_4(s) \,\e^{+ \argsa}\,,& if $\q x <0\,. $
       \cr}\eqno(4.12) $$
 Clearly, we must set $c_2(s) = c_3(s) = 0\,, $
in order to ensure that the solution vanishes as $|x| \to \infty\,. $
 We recognize from (4.10) that in $x =0$ the function
 $\Gcs(x,s;\nu )$ is continuous but not its first derivative: we write
 $$ \Gcs(0^+,s;\nu ) - \Gcs(0^-,s;\nu  )
	  =c_1(s)-c_4(s)  =0\,,\eqno(4.13)$$
 and, by integrating (4.10) with respect to $x$ from $x=0^-$ to $x=0^+$,
 $$ {d\over dx}\,\Gcs(0^+,s;\nu ) - {d\over dx}\,\Gcs(0^-,s;\nu )
 =-[c_1(s) +c_4(s)]\, {s^{\nu }\over \sqrt{\D}}
 = -{s^{2\nu  -1}\over \D}\,. \eqno(4.14)$$
 Therefore, using (4.13-14) we obtain
 $ c_1(s) = c_4(s) =1/(2\sqrt{\D} \, s^{1-\nu })\,,$ and
consequently
 the expression  (4.11).
\vsp
 For the {\it Signalling Problem} (4.4b)  the application of the Laplace
transform to (4.1) with
$u(x,t)=\Gs(x,t;\nu )\,, $
 $ \Gs(x,0^+; \nu  ) = 0\,, $
[and ${\d\over \d t} \Gs(x,0^+;\nu ) = 0\,$ if $1/2 <\nu  \le 1\,$],
 leads to   the  homogeneous differential equation
$$   \D\,{d^2  \Gss \over dx^2}
   -s^{2\nu  } \,\Gss	= 0 \,, \q x\ge 0  \,.	     \eqno(4.15)$$
Imposing the boundary conditions at $x=0\,, $
$\Gs(0^+,t;\nu )=h(t) = \delta (t)\,, $
and at $x=+\infty\,, $
$\Gs(+\infty,t;\nu )=0\,, $
we obtain
$$  \Gss(x,s;\nu  )  =
 \e^{\ds -(x/\sqrt{\D})s^{\nu  }}\,, \q x\ge 0\,. \eqno(4.16)$$
In fact,
  from (4.15)  $\Gss$ is expected in the form
 $$ \Gss(x,s;\nu ) =  c_1(s)\,\e^{-\argsa}+c_2(s)\,\e^{+\argsa}\,,
    \q x\ge 0\,. \eqno(4.17)$$
Clearly, we must set $c_2(s) =	0$ to ensure
 that the solution vanishes as $x \to +\infty\,, $
and  consequently we obtain
$c_1(s) = \Gss(0,s;\nu	) = \widetilde \delta (s) =1\,. $
\vfill\eject
\line{{\it 4.3 The Reciprocity Relation and the Auxiliary Functions}
   \hfill}\vsp
From (4.11) and (4.16) we recognize
$$ {d\over ds}\, \Gss = -2\, \nu  \,x \,\Gcs \,, \q x >  0
   \,, \eqno(4.18)$$
which implies for the  original Green functions
 the following {\it reciprocity relation}
$$ 2\nu  \, x\,  \Gc(x,t;\nu  )  = t\, \Gs(x,t;\nu  )
  \,,\q x > 0\,,\q t > 0 \,.\eqno(4.19) $$
\vsp
The above  relation  can be easily verified in the case of standard
diffusion ($\nu  =1/2$) where the explicit expressions
(4.8) of the Green   functions leads to the identity
(for $x > 0\,, \, t > 0$)
    $$ x\, \Gc^d (x,t) = t\, \Gs^d (x,t) = {1\over 2\sqrt{\pi}}
   \, {x \over \sqrt{\D\,t}}\,	\e^{-\ds\arg}=
F^d(r) =  {r\over 2}\,M^d(r)\,,
 \eqno(4.20)$$
where
  $$r={x/(\sqrt{D}\, t^{1/2})} > 0\,,\eqno(4.21)$$
is the well-known {\it similarity variable} and
$$ M^d(r)= \rec{\sqrt{\pi}} \,\e^{-\ds r^2/4}\,.\eqno(4.22)$$
We  can refer to $F^d(r)$ and $M^d(r)$ as
to the {\it auxiliary functions}
for the diffusion equation because each of them
provides the fundamental solutions through (4.20).
We note that $M^d(r)$ satisfies the normalization condition
$ \int_0^{\infty} \! M^d(r)\, dr =1\,.$
\vsp
Now we are going to show how, in the general case $0<\nu  <1\,, $
the inversion of the Laplace transform in
(4.11) or (4.16) leads us to generalize the auxiliary functions
$F^d(r)$ and $M^d(r)$
by introducing
the proper {\it similarity variable} for $x>0\,, \, t>0\,, $
$$r={x/(\sqrt{\D}\, t^{\nu  })} >0\,.\eqno(4.23)$$
\vsp
The new {\it auxiliary functions}, that we denote by
$F(r ;\nu  )$ and $M(r ;\nu )\,,  $
turn out to be expressed in terms of
Bromwich complex integrals as shown hereafter.
\vsp
Applying  in the reciprocity relation (4.19)
 the complex inversion formulas for the transformed Green functions
(4.11) and (4.16), we obtain
$$ 2\nu  x\,\Gc  (x,t;\nu  ) = {2\nu  x\over 2\sqrt{\D}}\,
\rec {2\pi i}\, \int_{Br} \!\!	\e^{\ds st - (x /\sqrt{\D})\, s^{\nu }}
\, {ds \over s^{1-\nu }}\,,\q x>0\,,\q t>0\,,
 \eqno(4.24)$$
and
$$ t\,\Gs  (x,t;\nu  ) = t\,
\rec {2\pi i}\, \int_{Br} \!\!	\e^{\ds st - (x/\sqrt{\D})\, s^{\nu }}
\, ds \,,\q x>0\,,\q t>0\,,
\eqno (4.25)$$
where $Br$ denotes the Bromwich path.
\vsp
In order to express the Bromwich integrals in terms of
the single {\it similarity variable} $r$ defined by (4.23)
let us	change the  integration variable in (4.24-25)
setting  $  \sigma =st\,. $
   From (4.24) we obtain
 $$ 2\nu  x\,  \Gc(x,t;\nu  )  = \nu  r\, M(r;\nu  )\,,
\eqno(4.26)$$
with
$$ M(r;\nu  ) :=
 {1\over 2\pi i}\,\int_{Br}   \!\!
 \e^{\ds \sigma -r\sigma ^\nu } \,
   {d\sigma \over \sigma^{1-\nu }} \,, \q  r >0\,, \q 0<\nu  <1\,,
    \eqno(4.27)$$
and, from (4.25),
$$ t\,	\Gs(x,t;\nu  )	= F(r;\nu  )\,,
\eqno(4.28)$$
with
$$ F(r;\nu  ) :=
 {1\over 2\pi i}\,\int_{Br}   \!\!
 \e^{\ds \sigma -r\sigma ^\nu } \,
   {d\sigma} \,, \q  r >0\,, \q 0<\nu  <1\,.
    \eqno(4.29)$$
Therefore we conclude that, for $\, x > 0\,, t > 0 \,, \, r>0\,, $
$$ 2\nu  \, x\,  \Gc(x,t;\nu  )  = t\, \Gs(x,t;\nu  )
    = F(r;\nu ) = \nu  r\, M(r;\nu  )
  \,.\eqno(4.30) $$
\vsp
The above  definitions of $F(r;\nu  )$ and $M(r;\nu )$
by the {Bromwich representation} can be analytically continued
from $r>0$ to any $z \in \CC$,
by adopting suitable integral and series representations valid in all of
$\CC$.
\vsp
For this purpose, let us
 deform the Bromwich  path $Br$ into the Hankel path   $Ha$
[a contour consisting of pieces of the two rays
$\, {\rm arg}\, \sigma	= \pm \phi\,$ extending to infinity,
and of the circular arc $\, \sigma  = \epsilon	\, \e^{i\theta}\,, $
$ |\theta| \le \phi\,, $ with  $  \phi \in (\pi/2,\pi)\,$]
which is chosen to be equivalent to the original path (at least)  for
$z$ real and positive. The Hankel integral representation allows us
to obtain the series representation for each auxiliary function.
In fact,   after  expanding in series
of positive powers of $z$
the exponential function,  $\, {\rm exp} (-z \,\sigma ^\nu  )\,, $
exchanging the order between the  series and the integral
and using
the Hankel representation of  the reciprocal of
the Gamma function,
     $$ \rec{\Gamma(\zeta )} =
      \rec{2\pi i} \int_{Ha} \!\!
	   \e^{\ds \,\sigma} \,  \sigma ^{\ds -\zeta } \, d\sigma \,,
\q \zeta \in \CC\,, $$
we finally obtain the required power {series representation}.
Since the radius  of convergence of the  power series
can be proven to be infinite for $0<\nu  <1$, our auxiliary functions
turn out to be entire in $z$ and therefore the
exchange between the series and the integral  is legitimate.
\vfill\eject
\vsp
The integral and series representations
of $F(z;\nu  )$ and  $M(z;\nu  )$, valid on all of $\CC\,,$
with $0<\nu  <1$
turn out to be
$$ F(z;\nu  ) = \l\{
 \eqalign{
 & {1\over 2\pi i}\,\int_{Ha}	\!\!
 \e^{\ds \, \sigma -z\sigma ^\nu } \,
   d\sigma  \cr
& \sum_{n=1}^{\infty}{(-z)^n\over n!\, \Gamma(-\nu  n)}\cr}
       \q z\in \CC\,,\q 0<\nu  <1\,, \r.
   \eqno(4.31)$$
and
$$ M(z;\nu  ) =  \l\{
 \eqalign{
 &	  {1\over 2\pi i}\,\int_{Ha}   \!\!
 \e^{\ds \,\sigma -z\sigma ^\nu } \,
   {d\sigma\over \sigma ^{1-\nu  }}\cr
& \sum_{n=0}^{\infty}
 {(-z)^n \over n!\, \Gamma[-\nu  n + (1-\nu  )]}\cr}
 \q z\in \CC\,,\q 0<\nu  <1\,. \r.   \eqno(4.32)$$
\vsp
In Appendix we show that the two auxiliary functions turn out to be
particular examples of a {\it special entire function} known as
{\it Wright function}. We refer the reader to the Appendix for
the main properties of the auxiliary functions including
the related Laplace transform  pairs.
\vvs
\line{{\it 4.4 Plots}\hfill}\vsp
In the following figures we exhibit   the plots of the Green functions
for both the Cauchy and Signalling problems,
in the cases	$\beta  =1/2\,, \,1\,, \,3/2\,, $
($ \beta = 2 \nu$)
takining $\D=1$.
\vsp
The plots of $\Gc(x,t)$ 
versus $x$ at fixed times
($t=1,2,3\, $)
for $\beta = 1/2, 1,3/2$ are reported in Figs 4-1, 4-2, 4-3, respectively.
The plots of $\Gs(x,t)$
versus $t$ at fixed positions
($x=0.9,1,1.1\, $)
for $\beta = 1/2, 1,3/2$ are reported in Figs 4-4, 4-5, 4-6, respectively.
\vskip 1.0 truecm
 \cen{\epsfxsize=8.0truecm \epsfysize=5.0truecm \epsffile{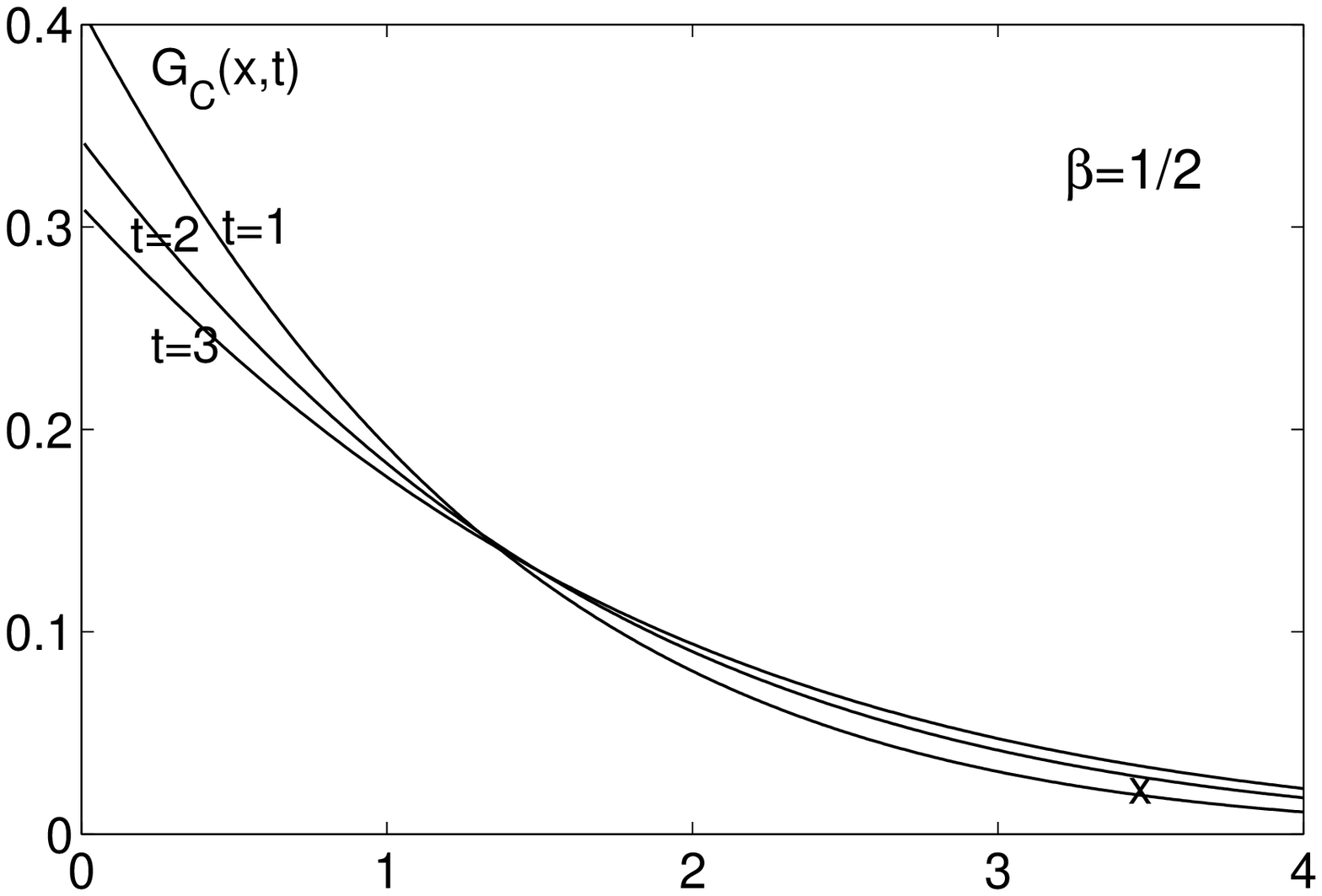}}
\centerline{{\bf Fig. 4-1}}
\vsh
\centerline{The Green function $\Gc (x,t)$ for
$\beta	= 1/2$ versus $x$ at fixed times. ({\it Cauchy problem})}
\vfill\eject
$\null$
 \cen{\epsfxsize=8.0truecm \epsfysize=5.0truecm \epsffile{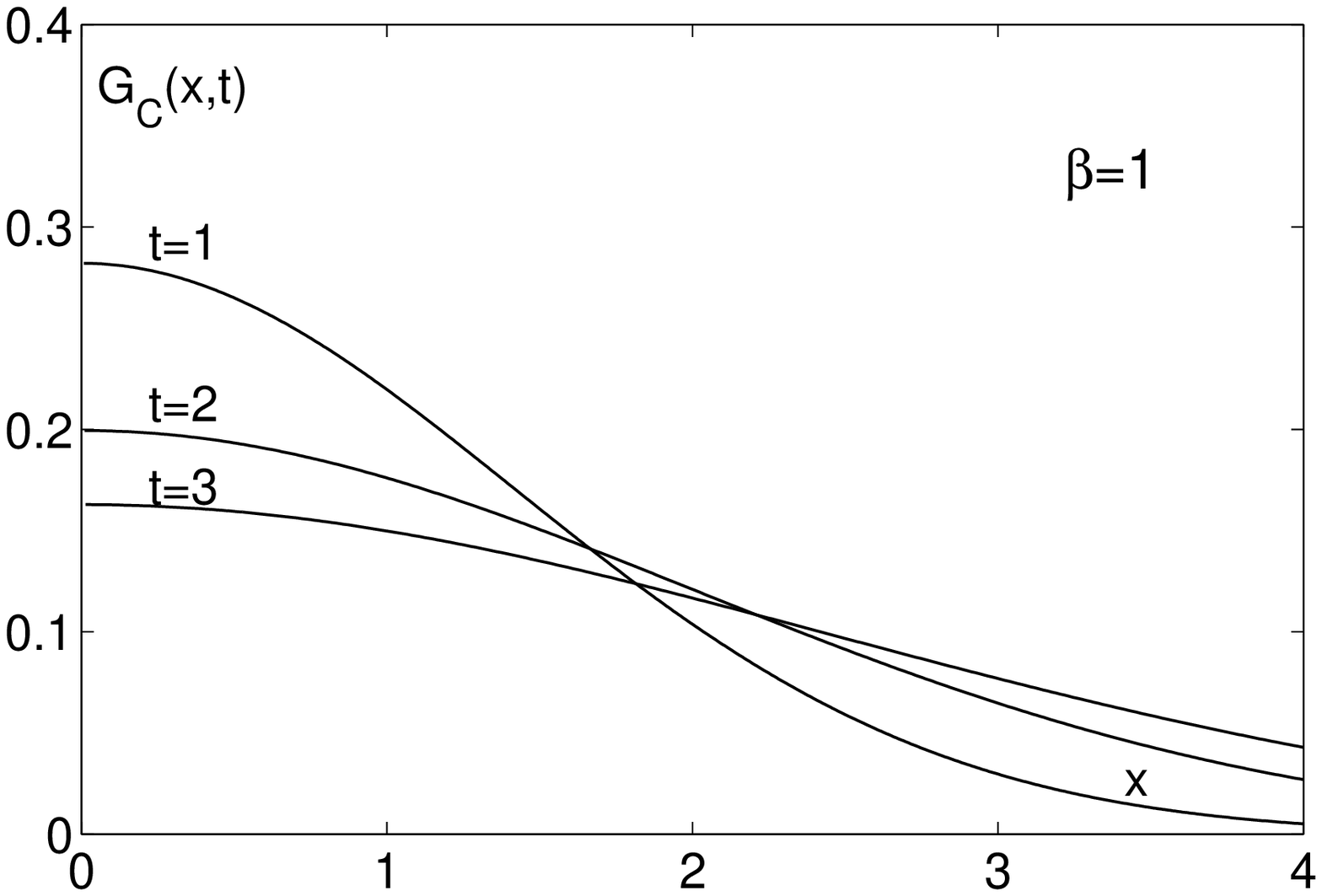}}
\centerline{{\bf Fig. 4-2.}}
\vsh
\centerline{The Green function $\Gc(x,t)$ for
$\beta = 1$ versus $x$ at fixed times. ({\it Cauchy problem})}
 \vskip 1.0 truecm
\cen{\epsfxsize=8.0truecm \epsfysize=5.0truecm \epsffile{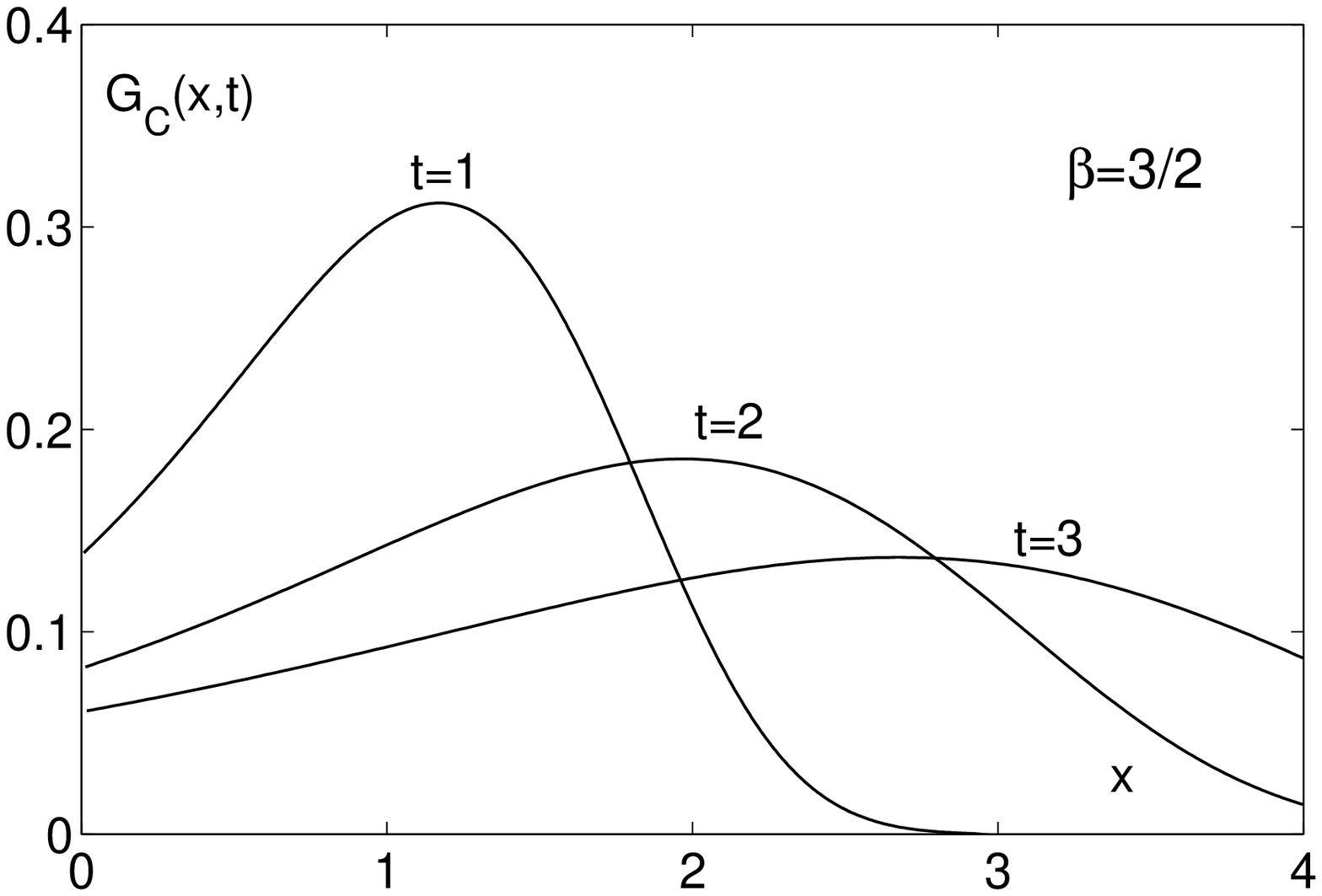}}
\centerline{{\bf Fig. 4-3.}}
\vsh
\centerline{The Green function $\Gc(x,t)$ for
$\beta	= 3/2$ versus $x$ at fixed times. ({\it Cauchy problem})}
 \vskip 1.0 truecm
\cen{\epsfxsize=8.0truecm \epsfysize=5.0truecm \epsffile{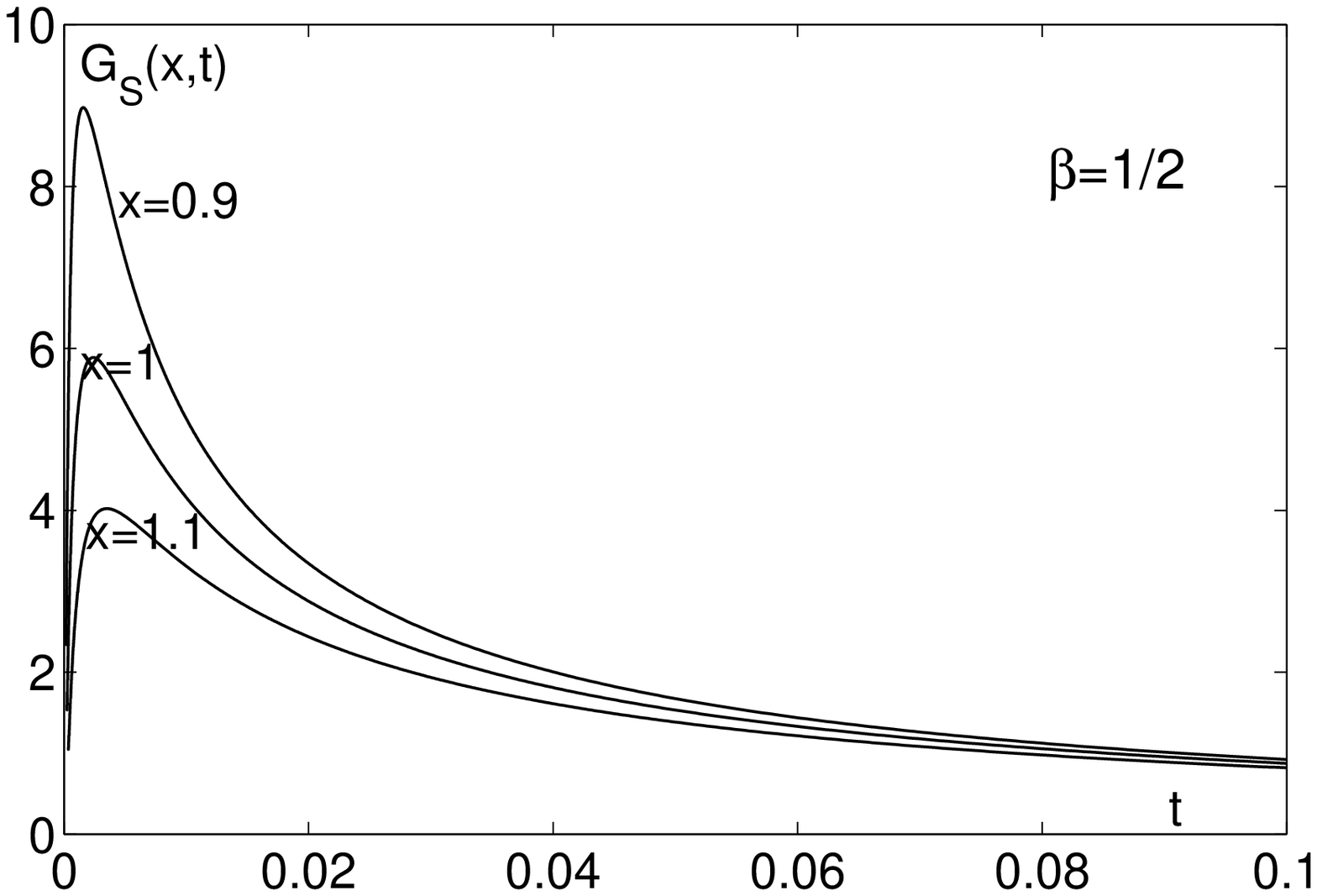}}
\centerline{{\bf Fig. 4-4}}
\vsh
\centerline{The Green function $\Gs(x,t)$ for
$\beta	= 1/2$ versuts $t$ at fixed positions. ({\it Signalling problem})}
\vfill\eject  
$\null$
 \vskip 1.0 truecm
\cen{\epsfxsize=8.0truecm \epsfysize=5.0truecm \epsffile{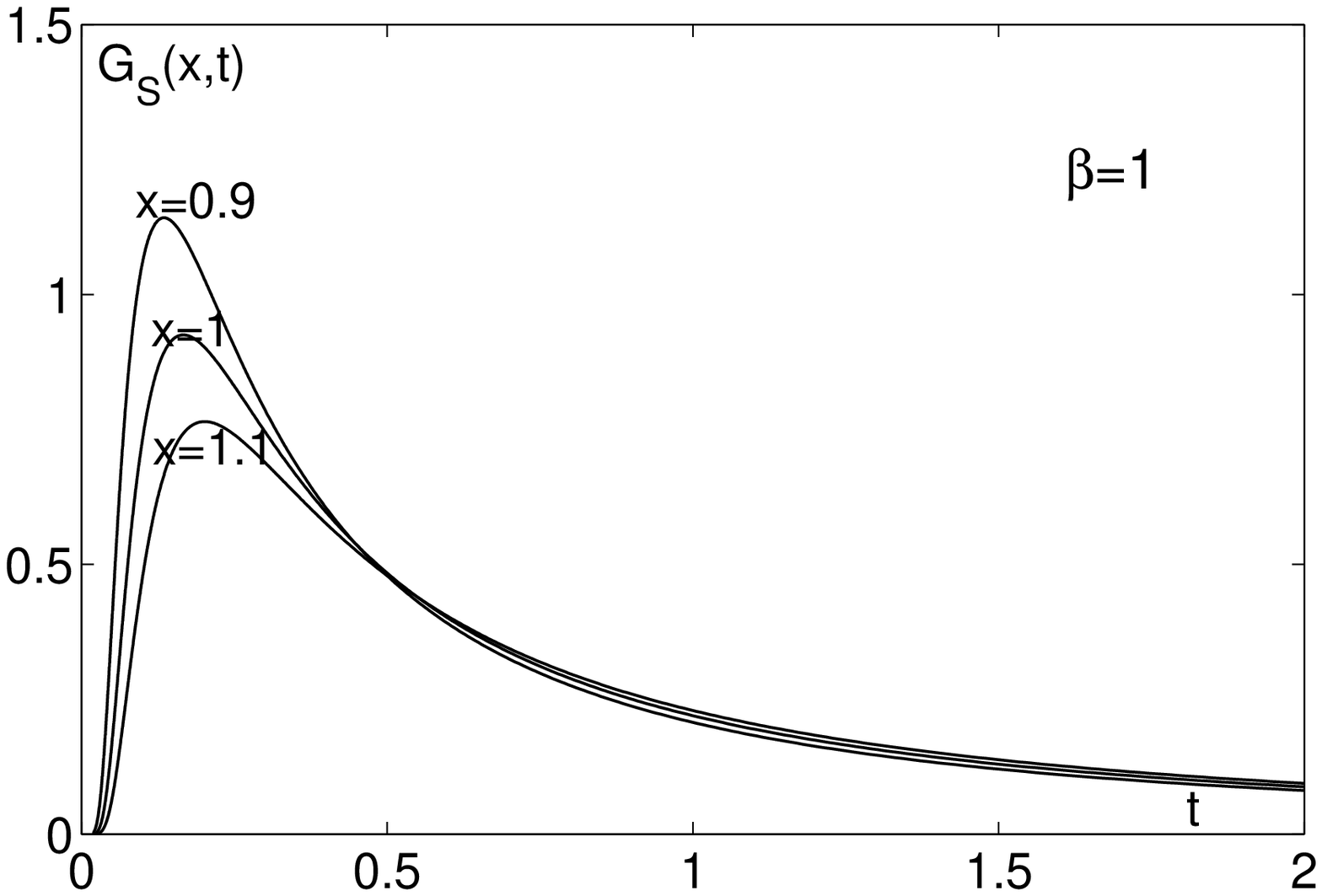}}
\centerline{{\bf Fig. 4-5}}
\vsh
\centerline{The Green function $\Gs(x,t)$ for
$\beta	= 1$ versuts $t$ at fixed positions.
({\it Signalling problem})}
\vskip 1.0 truecm
\cen{\epsfxsize=8.0truecm \epsfysize=5.0truecm \epsffile{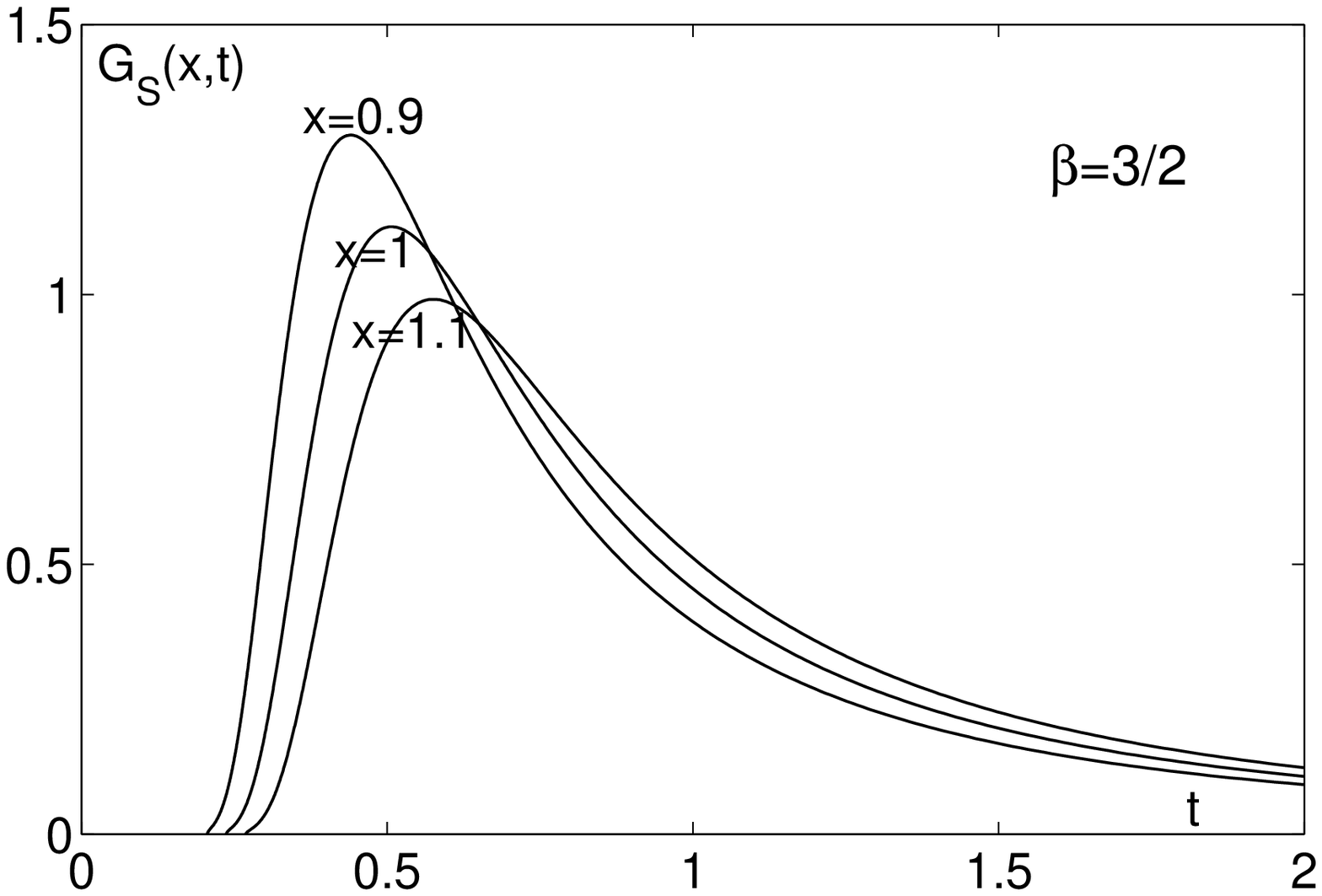}}
\centerline{{\bf Fig. 4-6}}
\vsh
\centerline{The Green function $\Gs(x,t)$ for
$\beta	= 3/2 $ versuts $t$ at fixed positions.
({\it Signalling problem})}
 \vsp
In order to gain  more insight into the phenomena governed by the
fractional diffusion wave equation (4.1), we consider a simple
Cauchy problem where the initial data are provided by a box-type
function. Precisely, taling $\D=1\,, $ we consider
$$ u(x,0^+) =  \Theta (1-|x|) \,
      \Longrightarrow\, u(x,t) = \int_{-1}^{+1}
   \Gc(x-\xi ,t;\nu   )\, d\xi	\,. \eqno(4.33)$$
\vsp
In the following 
we exhibit plots of the solution versus $x$ ($0\le x \le 3$),
 at fixed $t\,$ ($t = 0\,,\, 0.5, \, 1$), for
some fractional values of $\beta = 2 \nu \,. $
 In Fig. 4-7 we compare the cases concerning the fractional diffusion
equation ($0<\beta  \le 1$), whereas
in Fig. 4-8we compare those concerning the fractional wave
equation ($1<\beta  \le 2$).
\vfill\eject
$\null$
\cen{\epsfxsize=10truecm \epsfysize=5.0truecm \epsffile{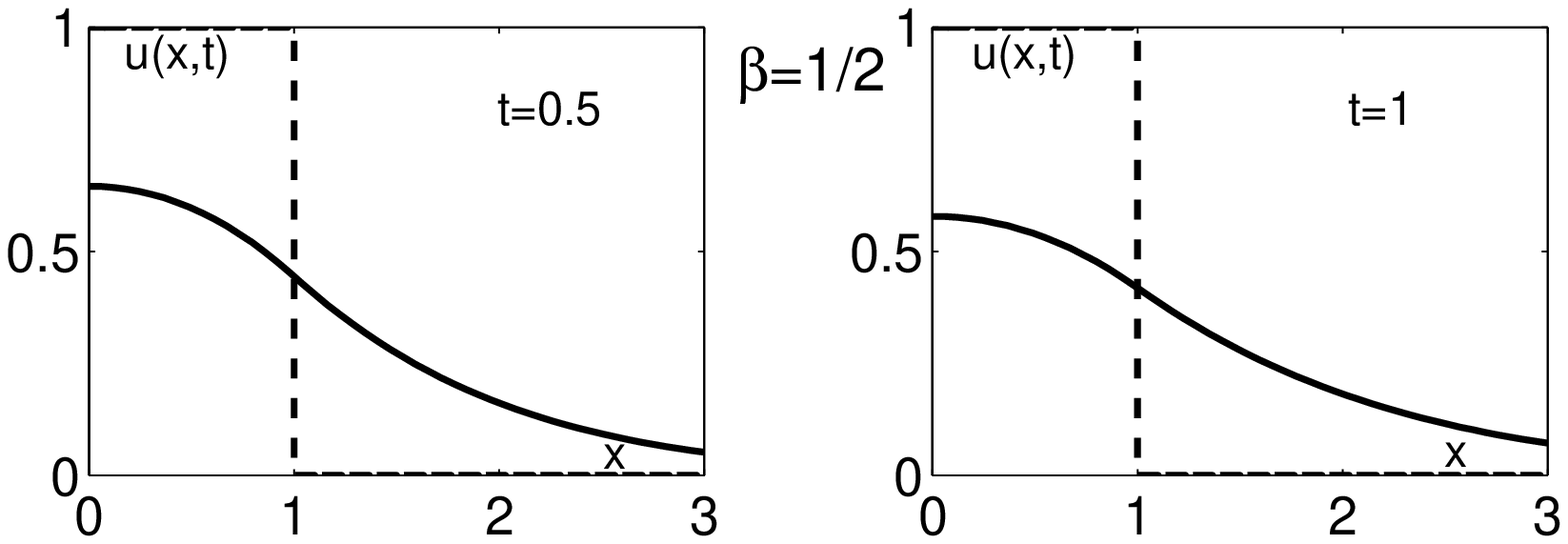}}
\cen{\epsfxsize=10truecm \epsfysize=5.0truecm \epsffile{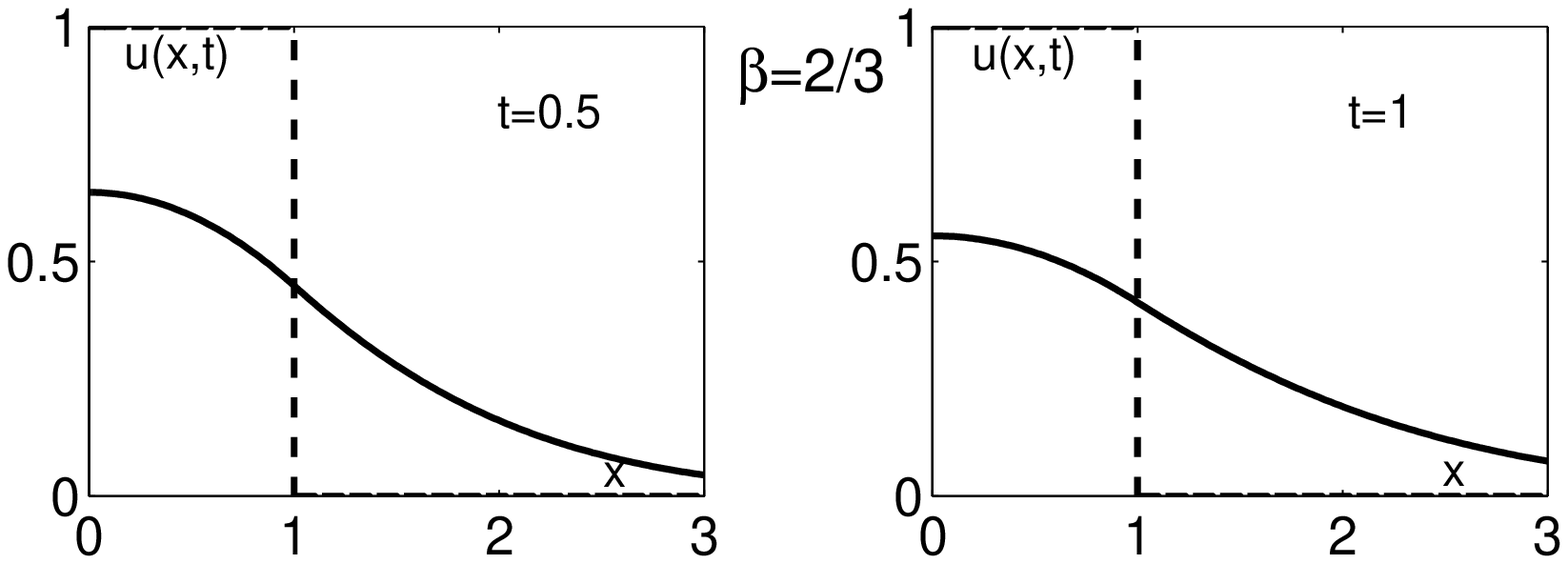}}
\cen{\epsfxsize=10truecm \epsfysize=5.0truecm \epsffile{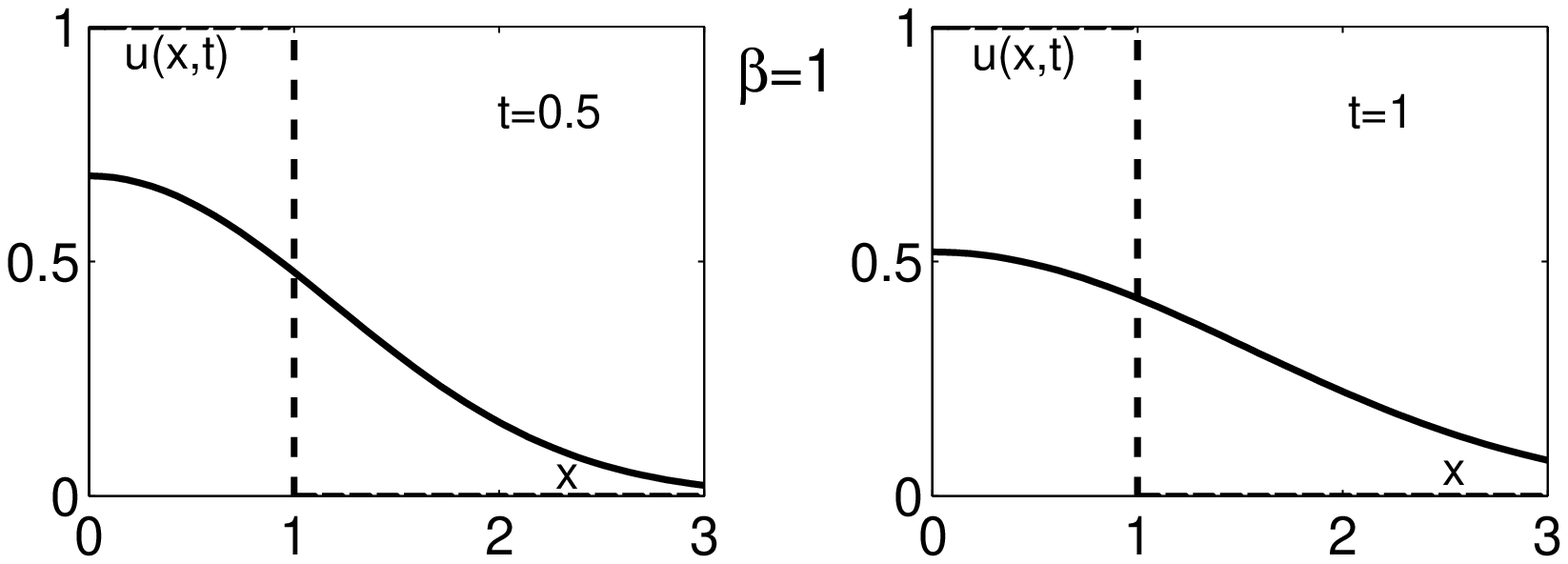}}
\centerline{{\bf Fig. 4-7}}
\vsh
\centerline{ Evolution of the initial box-signal
 (dashed line) at  $t =0.5$ (left) and
   $t = 1$ (right),}
 \centerline{versus $x\,,$ for various values of $\beta  $ :
  $\, 1/2\,,\,2/3\,,\, 1.$}
\vs\vsp
In Fig. 4-7 we thus  obtain some comparison between the fractional
diffusion and the standard diffusion.
We easily recognize for $0<\beta< 1$ a diffusive behaviour,
which  is slower with respect to the case $\beta= 1\,$
of standard diffusion:
this is consistent with a {\it slow diffusion process}.
\vfil\eject
 \null
\cen{\epsfxsize=10.0truecm \epsfysize=5.0truecm \epsffile{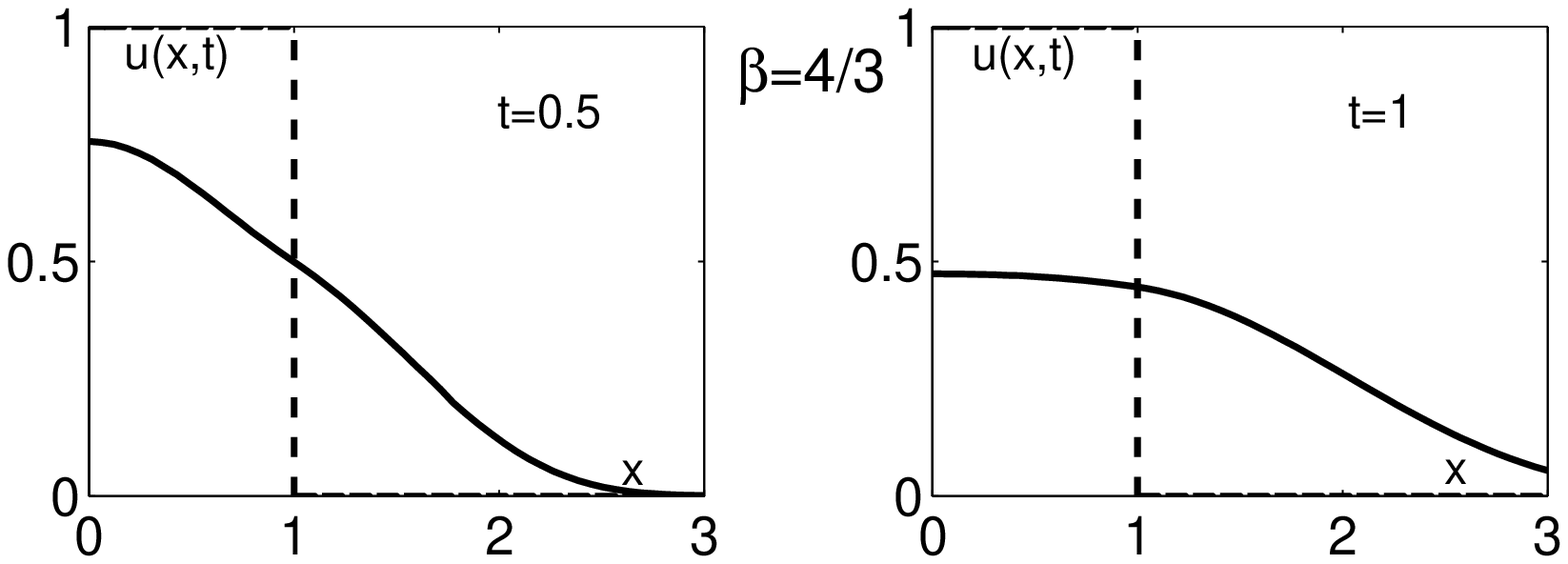}}
\cen{\epsfxsize=10.0truecm \epsfysize=5.0truecm \epsffile{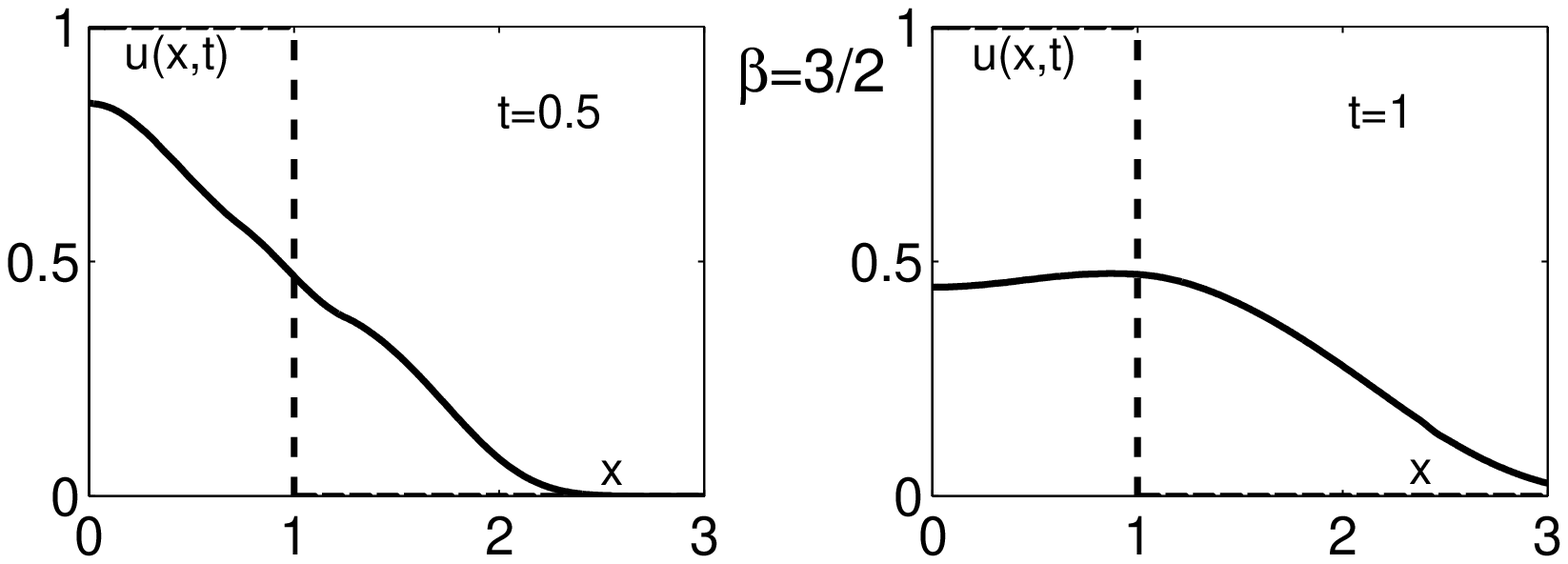}}
\cen{\epsfxsize=10.0truecm \epsfysize=5.0truecm \epsffile{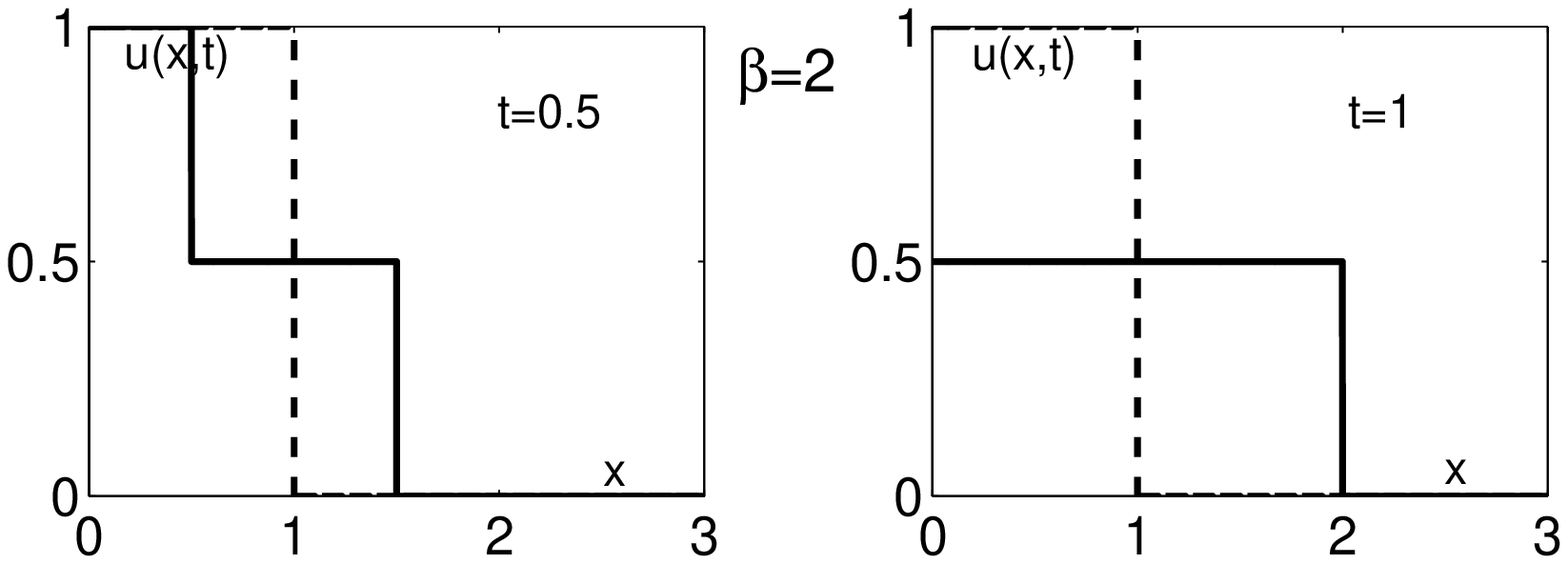}}
\centerline{{\bf Fig. 4-8}}
\vsh
\centerline{Evolution of the initial box-signal
 (dashed line) at  $t =0.5$ (left) and
   $t = 1$ (right),}
 \centerline{versus $x\,,$ for various values of $\beta  $ :
      $\,4/3\,, \,3/2\,, \, 2\,.$}
\vs\vsp
In Fig. 4-8  we recognize the {\it intermediate
process}
between standard diffusion (where discontinuities are smoothed
out) and wave propagation (where discontinuities can propagate
with finite speed).
This can be seen from the appearance of a hump,
which tends to be narrower as $\beta \to 2^-$
up to reproduce  the discontinuities of the signal for
$\beta	=2\,. $ For $1<\beta  <2$
the hump travels
with finite velocity (as in a wave process) but the
signal diffuses instantaneously (as in a diffusion process).
\vfill\eject

%% file: fmnew2A.tex
\pageno=333
\line{APPENDIX: THE WRIGHT FUNCTION  \hfill} \vsp
In this Appendix we  shall consider  the general class
of the {\it Wright}  functions
with special regard to the special functions
that we have introduced for the sake of convenience
in the treatment of
the  fractional diffusion wave equation,
the so-called {\it auxiliary} functions.
It is our purpose  to provide a review of the
main properties of these functions including their
series	and integral representations and the related
Laplace transform pairs.
We also mention their  connection
with the {\it Mittag-Leffler}  functions,
for which we refer the reader to Gorenflo and Mainardi [8].
\vvs
\line{{\it A.1 The representations for the Wright function
 $W_{\lambda ,\mu } (z)$     \hfill}}
\vsp
    The Wright function, that we denote as $W_{\lambda ,\mu } (z)\,, $
where $\lambda >-1$ and $\mu  >0\,, $
is so named from the British mathematician
E. Maitland Wright who in 1933 introduced it
in the asymptotic theory of partitions.
A list of formulas concerning this function can
be found in the handbook of the Bateman Project [125].
We note that  originally Wright considered $\lambda \ge 0$ [126]
and only later, in 1940,   he  extended to $ -1 < \lambda  < 0\,$ [127].
Relevant investigations on this functions have been carried out by
Stankovic [128-129],   by other authors
quoted by Kiryakova [130], 
and  more recently by Gorenflo, Luchko and Mainardi [142-143]
and by Wong and  Zhao [144-145].
\vsp
The Wright function  is defined
by the {\it series representation}, valid in all of $\CC\,, $
  $$ W_{\lambda ,\mu }(z ) :=
   \sum_{n=0}^{\infty}{z^n\over n!\, \Gamma(\lambda  n + \mu )}
\,, \q z\in \CC\,, \q \lambda >-1\,,\q \mu >0\,, \eqno(A.1)$$
so that it turns out to be an {\it entire} function. This
property remains valid even if $\mu$ is an arbitrary complex number.
The   {\it integral representation} reads
    $$ W_{\lambda ,\mu }(z )
 = {1\over 2\pi i}\,\int_{Ha}	\!\!
 \e^{\, \ds \sigma +z\sigma ^{-\lambda }} \,
   {d\sigma \over \sigma^{\mu}} \,,  \q z \in \CC \,,
    \q \lambda >-1\,,\q \mu >0\,,
   \eqno(A.2)$$
where $Ha$ denotes the Hankel path.
The formal equivalence between the two representations
is easily proved  using the Hankel formula for the Gamma
function
$$\rec{\Gamma(\zeta )} = \int_{Ha} \e^{\, \ds u} \, u^{-\zeta }\, du\,,
  \q \zeta \in \CC\,, $$
and performing a term-by-term integration.
In fact,
 $$ \eqalign{
W_{\lambda ,\mu }(z)
& =  {1\over 2\pi i}\,\int_{Ha}   \!\!
 \e^{\, \ds \sigma +z\sigma ^{-\lambda }} \,
   {d\sigma \over \sigma^{\mu}}
 = {1\over 2\pi i}\,\int_{Ha}	\!\! \e^{\,\ds \sigma}\,
 \l[\sum_{n=0}^{\infty}{ z^n\over n!} \,\sigma^{-\lambda  n}\r]\,
 {d\sigma  \over \sigma^{\mu }}  \cr
 &= \sum_{n=0}^{\infty} { z^n  \over n!}
 \l[ \rec{2\pi i} \int_{Ha} \!\!
 \e^{\, \ds \sigma} \, \sigma ^{-\lambda  n -\mu } \, d\sigma \r]
 = \sum_{n=0}^{\infty}
 { z^n\over n!\, \Gamma[\lambda  n + \mu ]}\,. \cr}
 $$
\vfill\eject
It is possible to prove that the Wright function is entire of order
$1/(1+\lambda)\,, $ hence
of exponential type if $\lambda \ge 0\,. $
The case $\lambda =0$ is trivial
since
$  W_{0, \mu }(z) = { \e^{\, z}/ \Gamma(\mu )}\,.$
\vsp
Wright showed in particular that in the case $\lambda =-\nu  \in (-1,0)\,$
there is the following asymptotic expansion, valid in a suitable
sector about the negative real axis
$$ W_{-\nu  ,\mu } (z) =
   Y^{\, 1/2-\mu } \, \e^{\, - Y}\,   \l(
  \sum_{m=0}^{M-1} A_m	\, Y^{-m} + O(|Y|^{-M})\r) \q
  {\rm as}\q z \to -\infty\,, \eqno(A.3)$$
with $Y = Y(z) = (1-\nu  )\, (-\nu  ^\nu  \, z)^{1/(1-\nu  )}\,, $
where the $A_m$ are certain real numbers.
\vvs
\line{{\it A.2 The  Wright functions as generalization of the Bessel
functions   \hfill}}
\vsp
For $\lambda =1$ and $\mu =\nu +1$ the Wright function
turns out to be related to the well known
Bessel functions $ J_\nu  $  and $I_\nu $  by the following identity
$$	\left({z/ 2 }\right )^\nu \,
      W_{1, \nu + 1}\left(\mp {z^2 / 4} \right)
  = \left\{ \eqalign{J_{\nu}(z) \cr
		I_{\nu}(z) \cr} \right. \,.
    \eqno(A.4)$$
In view of this property  some authors refer to the Wright function as
the  {\it Wright  generalized  Bessel function}
(misnamed also as the {\it Bessel-Maitland function})
and introduce the notation
$$J_\nu ^{(\lambda)} (z) := \left({z\over 2 }\right )^\nu \,
    \sum_{n=0}^{\infty}{(-1)^n (z/2)^{2n}\over n!\,
  \Gamma(\lambda  n + \nu +1)}\,;
   \q J_\nu ^{(1)} (z) := J_\nu  (z)\,. \eqno(A.5)$$
As a matter of fact, the Wright function
appears as the natural generalization of the  entire function
known as {\it Bessel - Clifford function},  see \eg [130],
and referred  by Tricomi [131-132]  
as  the {\it uniform Bessel function}
\footnote{*}{\baselineskip=12truept   \note 
The great Italian mathematician  denoted this function
by $  E_\nu (z)\,. $  Here we write $T(z;\nu )$
to remain in accordance with the standard notation used for
the Mittag-Leffler function [8].},
 $$ T (z;\nu) := z^{-\nu /2} \, J_\nu (2 \sqrt{z}) =
    \sum_{n=0}^{\infty}{(-1)^n z^n\over n!\,  \Gamma( n + \nu +1)}=
     W_{1, \nu +1}(-z)	   \,. \eqno(A.6)$$
\vsp
Some of the properties	which the Wright functions  share with
the most popular Bessel functions
were enumerated by Wright himself.
Hereafter, we quote some relevant relations from
the Bateman Project [125], which can easily be derived from (A.1):
$$ \lambda z \, W_{\lambda ,\lambda + \mu}(z) =
   W_{\lambda ,\mu -1}(z) + (1-\mu )\, W_{\lambda ,\mu}(z) \,,
   \eqno(A.7)$$
$$ {d \over dz}\, W_{\lambda ,\mu}(z) =
    W_{\lambda ,\lambda +\mu}(z) \,. \eqno(A.8)  $$
\vfill\eject
\line{{\it A.3 The Auxiliary Functions $F(z;\nu )$
  and $M(z;\nu	)$}\hfill}
\vsp
In our treatment of the time fractional
diffusion wave equation we have found it convenient
to introduce two {\it auxiliary functions}
 $F(z;\nu )$ and $M(z;\nu )\,, $ where $z$ is a complex variable and
$\nu  $ a real parameter $0<\nu  <1\,. $
Both functions turn out to be analytic in the whole
complex plane, \ie they are entire functions. Their
respective {\it integral  representations}  read,
see (4.31-32),
$$ F(z;\nu  ) :=
 {1\over 2\pi i}\,\int_{Ha}   \!\!
 \e^{\ds \, \sigma -z\sigma ^\nu } \,
   d\sigma \,,	\q z\in \CC\,,
   \q 0<\nu  <1   \,, \eqno(A.9)$$
$$ M(z;\nu  ) :=
 {1\over 2\pi i}\,\int_{Ha}   \!\!
 \e^{\ds \,\sigma -z\sigma ^\nu } \,
   {d\sigma\over \sigma ^{1-\nu  }} \,,   \q z\in \CC\,,
   \q 0<\nu  <1   \,. \eqno(A.10)$$
From a comparison of (A.9-10) with (A.2) we easily recognize that
these functions are  special cases of the Wright function according to
$$ F(z;\nu  ) =   W_{-\nu , 0}(-z)
  \,, \eqno(A.11)$$
and
  $$ M(z;\nu  ) =  W_{-\nu , 1-\nu }(-z)
  \,. \eqno(A.12)$$
From (A.7) and (A.11-12)  we find the relation
$$ F(z;\nu  ) = \nu  \, z \, M(z;\nu  ) \,. \eqno(A.13)  $$
This relation can be obtained directly from (A.9-10) with
an integration by parts, \ie
$$ \int_{Ha}   \!\! \e^{\ds \,\sigma -z\sigma ^\nu } \,
   {d\sigma\over \sigma ^{1-\nu  }}
  = \int_{Ha}	\!\! \e^{\ds \,\sigma} \,
 \l(-{1\over \nu  z}\, {d\over d\sigma } \e^{\ds\, -z\sigma ^\nu }\r) \,
   d\sigma = {1\over \nu  z}\,
   \int_{Ha}   \!\! \e^{\ds \, \sigma -z\sigma ^\nu } \, d\sigma\,. $$
\vsp
The {\it series representations} for our auxiliary functions
turn out to be respectively,
see (4.31-32),
 $$ F(z;\nu ) :=
   \sum_{n=1}^{\infty}{(-z)^n\over n!\, \Gamma(-\nu  n)}
  = -{1\over \pi}\,
\sum_{n=1}^{\infty}{(-z)^n\over n!}\,
 \Gamma(\nu  n +1 )\, \sin(\pi \nu   n)\,, \eqno(A.14)	 $$
and
$$ M(z;\nu  ) :=
 \sum_{n=0}^{\infty}
 {(-z)^n \over n!\, \Gamma[-\nu  n + (1-\nu  )]}  =
  \rec{\pi}\, \sum_{n=1}^{\infty}\,{(-z)^{n-1} \over (n-1)!}\,
  \Gamma(\nu  n)  \,\sin (\pi\nu  n)
  \,.	   \eqno(A.15)$$
The series at the R.H.S. have been obtained by using
the well-known reflection formula for the Gamma function
 $ \Gamma(\zeta)\,\Gamma(1-\zeta)  =\pi /\sin\,\pi \zeta\,.$
Furthermore we note  that  $F(0;\nu )= 0\,, $
$M(0,\nu  ) = 1/\Gamma(1-\nu )\,$
and that the relation (A.13) can be  derived also
from (A.14-15).
\vfill\eject
Explicit expressions of $F(z;\nu ) $ and $M(z;\nu  )$ in terms of known
functions are expected for some particular values of $\nu  $.
Mainardi \& Tomirotti [114] have shown that for  $\nu  =1/q\,,$
where $q \ge 2\,$ is a positive integer,
the auxiliary functions  can be expressed as a sum of
$(q-1)$ simpler entire functions.
In  the particular cases $q=2$ and $q=3$ we find
$$ M(z;1/2) = \rec{\sqrt{\pi}}\,
    \sum_{m=0}^\infty (-1)^m \, {\l({1\over 2}\r)}_m\,
  {z^{2m}\over (2m)!}
 = \rec{\sqrt{\pi}}\, \exp \l(-{\,z^2/ 4}\r)\,,
\eqno(A.16)$$
and
$$ \eqalign{M(z;1/3) &=
 \rec{\Gamma(2/3)}\,
   \sum_{m=0}^{\infty}{\l({1\over 3}\r)}_m\,{z^{3m}\over (3m)!}
- \, \rec{\Gamma(1/3)}\,
  \sum_{m=0}^{\infty} {\l({2\over 3}\r)}_m \,{z^{3m+1}\over (3m+1)!}
  \cr
 &= 3^{2/3} \, {\rm Ai} \l( {z/ 3^{1/3}}\r) \,,\cr}
 \eqno(A.17)$$
where $Ai$ denotes the {\it Airy function}.
\vsp
Furthermore it can be  proved [114] that
$M(z; 1/q)$  satisfies the   differential equa\-tion 
of order $q-1$
$$ {d^{q-1}\over dz^{q-1}} \, M(z;1/q) +
  {(-1)^q\over q}\, z\, M(z;1/q) =0\,, \eqno(A.18)$$
subjected to the $q-1$ initial conditions at $z=0$,
derived from (A.15),
$$ M^{(h)}(0;1/q) =
   {(-1)^h\over {\pi}}\,
  \Gamma[(h + 1)/q]\,\sin [\pi\, (h+1)/q]
  \,,  \q h = 0,\,1,\, \ldots \, q-2\,.
   \eqno(A.19)$$
We note that, for $q\ge 4\,,$ Eq. (A.18) is  akin to the
{\it hyper-Airy}  differential equation  of order $ q-1\,, $
see \eg [133].
Consequently, in view of the above considerations, the auxiliary function
$M(z;\nu  )$ can be referred to as the {\it generalized hyper-Airy
function}.
\vsp
Let us now consider the problem of the asymptotic evaluation
of the function $M(z;\nu  )$ as $|z| \to \infty$ in the complex plane.
Referring to a preliminary report of ours [134]
for the detailed asymptotic analysis in the whole complex plane,
which includes the phenomenon of Stokes lines,
 here we limit ourselves to provide
the asymptotic representation as $z=r$ is real and positive
by using the ordinary saddle-point method.
 Choosing as a variable $r/\nu	$ rather than $r$ the computation is
 easier and yields, see [114],
 $$ M(r/\nu  ;\nu  ) \sim
   a(\nu  )\, r^{\ds{(\nu  -1/2)/(1-\nu )}}  \;
   \exp \, \l[- b(\nu )\,r^{\ds {1/(1-\nu )}}\r]\,,
 \q  r\to +\infty\,, \eqno(A.20)$$
 where $$ a(\nu ) = {1\over  \sqrt{2\pi\,(1-\nu )}}    >0 \,,  \q
  b(\nu ) = {1-\nu  \over \nu  }    >0 \,. \eqno(A.21)$$
The above evaluation is consistent with the first term in
Wright's asymptotic expansion (A.3) after having used (A.12).
\vfill\eject
The exponential decay for $r \to +\infty$ ensures that $M(r;\nu  )$
is absolutely integrable in $\RR^+\,. $
We can easily prove the normalization property in $\RR^+$
  $$\int_0^{+\infty} \!\!\! M(r;\nu  )\, dr = 1\,, \eqno(A.22)$$
and more generally we can compute all the moments in $\RR^+$
  $$ \int_0^{+\infty} \!\!\! r^{\, n}\, M(r;\nu  )\, dr
   = {\Gamma(n+1)\over \Gamma(\nu  n+1)}\,,\q n\in \NN\,.
\eqno(A.23)$$
The results (A.22-23) can be formally derived by using the Laplace
transform of $M(r;\nu  )\,, $ as shown later.
Analogously we can compute  all the moments in $\RR^+$
for $F(r;\nu  )\,. $
\vvs
\line{{\it A.4 The Laplace transform pairs related to the Wright
   function}  \hfill}
\vsp
Let us consider the Laplace transform of the Wright function
using the following notation
$$ W_{\lambda ,\mu } (\pm r)	\,\div\,
{\cal{L}} \,\l[ W_{\lambda ,\mu } (\pm r)\r] :=
  \int_0^\infty \!\! \e^{\ds\, -s\,r} \, W_{\lambda ,\mu } (\pm r)\, dr
   \,, $$
where $r$ denotes a non negative real variable, \ie $0\le r<+\infty\,, $
and $s$ is the Laplace complex parameter.
\vsp
When  $\lambda > 0$  the series representation of Wright function can
be transformed term-by-term. In fact, for a known theorem
of the theory of the Laplace transforms, see \eg Doetsch [109],
the  Laplace transform of an entire function
of exponential type can be obtained  by transforming term-by-term the
Taylor expansion of the
original function around the origin.
In this case the resulting Laplace transform turns out
to be	analytic and vanishing at infinity.
As a consequence we obtain  the Laplace transform pair
$$   W_{\lambda ,\mu } (\pm r) \,\div \,
    {1\over s}\, E_{\lambda ,\mu }\l(\pm {1\over s}\r) \,,
  \q \lambda > 0\,, \q |s| > \rho> 0\,, \eqno(A.24)$$
where $ E_{\lambda ,\mu }$ denotes  the generalized Mittag-Leffler function
in two parameters, and $\rho$ is an arbitrary positive number.
The proof is  straightforward noting that
$$ \sum_{n=0}^\infty  {(\pm r)^n\over n!\, \Gamma (\lambda n+\mu )}
\,\div \, {1\over s}\,
\sum_{n=0}^\infty  {(\pm 1/s)^n\over  \Gamma (\lambda n+\mu )}	\,,
$$
and recalling the series representation
of the generalized Mittag-Leffler function,
$$ E_{\alpha, \nu  } (z ) := \sum_{n=0}^\infty {z^n\over
 \Gamma(\alpha n+ \nu	)}\,,\quad \alpha >0\,,\;\, \nu  \in \CC\,,
   \quad z \in\CC\,. $$
\vfill\eject
For $\lambda \to 0^+\,$ (A.24) we obtain the Laplace transform pair
$$ W_{0^+,\mu } (\pm r) =  { \e^{\ds\, \pm r}\over \Gamma(\mu )}\,
 \div \, {1\over \Gamma(\mu)}\, {1\over s\mp 1}
  = {1 \over s}\, E_{0 ,\mu }\l(\pm {1\over s}\r) \,,
    \q |s| > 1\,, \eqno(A.25)$$
where, to remain in agreement with (A.24), we have formally put
$$E_{0,\mu } (z) := \sum_{n=0}^\infty {z^n \over
 \Gamma(\mu  )} := {1\over \Gamma(\mu)}\, E_0(z) :=
  {1\over \Gamma(\mu)}\,   {1\over 1-z}\,,
   \quad |z |<1 \,. $$
We recognize that in this special case the
Laplace transform exhibits a simple pole at
$s = \pm 1$ while for $\lambda >0$ it exhibits an essential
singularity at $s = 0\,. $
\vsp
For $ -1<\lambda <0$ the Wright function  turns out to be
an entire function of order greater than 1, so that
care is  required in establishing the existence
of its	Laplace transform, which necessarily must tend to
zero as $s \to \infty$ in its half-plane of convergence.
For the sake of convenience we limit ourselves to derive the
Laplace transform  for the special case of $M(r;\nu  )\,; $
the exponential decay	as $r \to \infty\, $  of the {\it original}
function  provided by (A.20) ensures the existence of
the {\it image} function.
From the integral representation (A.10) we obtain
$$ \eqalign{
M(r;\nu  ) &\,\div\,
 {1\over 2\pi i}\,\int_0^\infty \e^{\ds\, -s\, r} \,
  \l[ \int_{Ha}   \!\! \e^{\ds \sigma -r\sigma ^\nu } \,
   {d\sigma \over \sigma^{1-\nu }}\r]\, dr \cr
& = {1\over 2\pi i}\,\int_{Ha}	 \!\! \e^{\ds \sigma} \,
    \sigma^{\nu -1}\,\l[
 \int_0^\infty \e^{\ds\, - r(s+\sigma ^\nu  )} \, dr \r]\, d\sigma
= {1\over 2\pi i}\,\int_{Ha}   \!\!
 { \e^{\ds \sigma} \, \sigma^{\nu -1}	\over \sigma ^\nu  +s}\, d\sigma
 \,. \cr}$$
Then, by recalling the integral representation of the Mittag-Leffler
function,
  $$ { E_\alpha (z) =
  \rec{2\pi i}\,
 \int_{Ha} {\zeta ^{\alpha -1}
 \, \e^{\,\zeta} \over \zeta ^\alpha -z}\,
	d\zeta \,,  \quad  \alpha  >0\,,
 \quad z\in \CC
 }\,, $$
we obtain the Laplace transform pair
$$  M(r;\nu  ) \,\div\,  E_\nu	(-s)\,, \q 0<\nu <1\,.	\eqno(A.26)$$
In this case, transforming term-by-term
 the Taylor series of
$M(r;\nu )\, $	yields a series
of negative powers of $s\,, $ which represents the asymptotic
expansion
of  $E_\nu  (-s)$ as $s\to \infty\,$ in a sector around the positive
real axis. We note that (A.26) contains the well-known Laplace
transform pair, see \eg [109],
$$  M(r;1/2 ) := \rec{\sqrt{\pi}}\, \exp \l(-{\,r^2/ 4}\r)
\,\div\,  E_{1/2} (-s) := \exp \l(s^2\r)\, {\rm erfc}\l( s\r)\,,
   \q s \in \CC\,.  $$
We also note that (A.26) allows us to derive (A.22-23) by accounting
for the property
  $$\int_0^{+\infty} \!\!\! r^{\, n}\, M(r;\nu	)\, dr
     = \lim_{s\to 0} \, (-1)^n \, {d^n\over ds^n}\, E_\nu  (-s)
\,. $$
\vfill\eject
Analogously, because of (A.3), we can prove that  in the case $\lambda
=-\nu  \in (-1,0)$ we get
$$     W_{-\nu	,\mu }(-r) \,\div\,
   E_{\nu   , \mu +\nu	 }(-s)\,, \q 0<\nu   <1\,. \eqno(A.27)$$
In the limit as $\lambda \to 0^-$
we formally obtain the Laplace transform pair
$$ W_{0^-, \mu }(-r) := {\e^{\ds\, -r} \over \Gamma(\mu )}
   \,\div\, {1\over \Gamma(\mu )}\, {1\over s+1} := E_{0 ,\mu }(-s)\,.
\eqno(A.28)$$
Therefore, as $\lambda \to  0^\pm \,, $ we note a sort of continuity
in the formal results (A.25)  and (A.28)    since
 $${1\over (s+1)} = \l\{
  \eqalign{
(1/s)\, E_0(-1/s)\,,  \q |s|>1\,; \cr
  E_0(-s)\,,  \q |s|<1\,. \cr }\r. \eqno(A.29)$$
\vsp
A quite relevant  Laplace transform pair related to the
{\it auxiliary} functions of argument $\,r^{-\nu }\,$ is
$$ {1\over r}\, F\l( 1/{r^\nu  };\nu  \r) =
   {\nu  \over r^{\nu  +1}}\,  M\l( 1/{r^\nu  };\nu  \r)\,\div\,
    \exp \l(\ds -s^\nu \r)\,, \q  0<\nu  <1\,. \eqno(A.30)$$
  We recall that a rigorous proof of (A.30) was formerly given by
Pollard  [135], based on a formal result by Humbert [136].
The Laplace transform pair was also obtained  by
Mikusi\'nski [137] and, albeit unaware of  the previous results
by  Buchen \& Mainardi [138] in a formal way.
\vsp
After noting that the  pair (A.30) can be easily deduced  (with $r=t$) from
(4.16) and (4.30),  hereafter we like to provide  two independent
proofs carrying out
the inversion of $\exp (-s^\nu )\,, $ either by  the complex integral
formula or  by	the formal series method.
We obtain
$$ \eqalign{
{\cal{L}}^{-1} \,\l[  \exp \l(\ds -s^\nu \r)\r] &=
   {1\over 2\pi i}\,\int_{Ha}	\!\!
 \e^{\ds \, sr - s^\nu } \,   ds   =
   {1\over 2\pi i\, r }\,\int_{Ha}   \!\!
 \e^{\ds \, \sigma -(\sigma/r) ^\nu } \,   d\sigma\cr
 &=  {1\over r}\, F\l( 1/{r^\nu  };\nu	\r)
  =  {\nu  \over r^{\nu  +1}}\,  M\l( 1/{r^\nu	};\nu  \r)
 \,, \cr}$$
and
$$ \eqalign{
{\cal{L}}^{-1} \, \l[ \exp \l(\ds -s^\nu \r)\r] &=
   \sum_{n=0}^\infty {(-1)^n  \over n!}\,
    {\cal{L}}^{-1} \, \l[ s^{\nu  n}\r]
   =  \sum_{n=1}^\infty {(-1)^n  \over n!}\,
    { r^{-\nu  n-1}\over \Gamma(-\nu  n)}\cr
   &= {1\over r}\, F\l( 1/{r^\nu  };\nu  \r) =
   {\nu  \over r^{\nu  +1}}\,  M\l( 1/{r^\nu  };\nu  \r) \,.
  \cr}$$
\vsp
Last, but not the least, we would like to mention the relevance
of our auxiliary functions in  probability theory. In fact,
as shown by Engler [124], they turn out to be related
with the probability density functions of
the so-called  stable distributions. For this interesting topic
we refer the reader to our recent works [139-141].
\vfill\eject

%% file: fmnew2R.tex
\pageno=340
\line{REFERENCES \hfill} \vsn
\vs
\rf{1}	Gross, B.:
 {\it Mathematical Structure of the Theories of Viscoelasticity},
     Hermann, Paris, 1953.
\vsp
\rf{2} Bland, D.R.: {\it The Theory of Linear Viscoelasticity},
   Pergamon, Oxford 1960.
 \vsp
\rf{3} Caputo, M. and  F. Mainardi:
  Linear models of dissipation in  anelastic solids,
  {\it Riv. Nuovo Cimento} (Ser. II), {\bf 1} (1971), 161-198.
\vsp
\rf{4} Christensen, R.M.: {\it Theory of Viscoelasticity},
  Academic Press, New York 1982.
\vsp
\rf{5} Pipkin, A.C.: {\it Lectures on Viscoelastic Theory},
  Springer Verlag, New York 1986.
\vsp
\rf{6} Berg, C. and G. Forst: {\it Potential Theory on Locally Compact
   Abelian Groups}, Springer Verlag, Berlin 1975, \S 9., pp. 61-72.
\vsp
\rf{7}	Zener, C.M.:
  {\it Elasticity and Anelasticity of Metals},
  Chicago University Press, Chicago 1948.
\vsp
\rf{8} Gorenflo, R. and F. Mainardi:
 Fractional calculus: integral and differential
 equations of fractional order, in this book,
 {\it Fractals and Fractional Calculus in Continuum Mechanics}
 (Ed. A. Carpinteri and F. Mainardi), Springer Verlag, Wien 1997,
 223-276. 
[Reprinted in NEWS 010101,  {\tt http://www.fracalmo.org}].
\vsp
\rf{9}	Scott-Blair, G.W.:
{\it Survey of General and Applied Rheology},
 Pitman, London 1949.
\vsp
\rf{10} Caputo, M. and	F. Mainardi:
   A new dissipation model based on memory mechanism,
  {\it Pure and Appl. Geophys.}, {\bf 91} (1971), 134-147.
\vsp
\rf{11} Gemant, A.: {On fractional differentials},
 {\it Phil. Mag.} [Ser. 7], {\bf 25} (1938), 540-549.
 \vsp
\rf{12} Gemant, A.: {\it Frictional Phenomena},
Chemical Publ. Co, Brooklyn N.Y. 1950.
\vsp
\rf{13} Scott-Blair, G.W. and J.E. Caffy:
 An application  of the theory of quasi-properties to the treatment
 of anomalous stress-strain relations,
 {\it Phil. Mag.} [Ser. 7], {\bf 40} (1949), 80-94.
 \vsp
\rf{14}  Rabotnov, Yu.N.:
Equilibrium of an elastic medium with after effect,
{\it Prikl. Matem. i Mekh.}, {\bf 12} (1948), 81-91. [in Russian]
\vsp
\rf{15}  Rabotnov, Yu.N.:
{\it Elements of Hereditary Solid Mechanics},
 MIR, Moscow 1980.
\vsp
\rf{16} Meshkov, S.I. and Yu.A.  Rossikhin:
Sound wave propagation in a viscoelastic medium whose hereditary
properties are determined by weakly singular kernels,
in {\it Waves in Inelastic Media} (Ed. Yu.N. Rabotnov), Kishinev 1970,
162-172. [in Russian]
\vsp
\rf{17} Lokshin, A.A. and Yu.V. Suvorova:
{\it Mathematical Theory of Wave Propagation in Media with Memory},
 Moscow University Press, Moscow 1982. [in Russian]
\vsp
\rf{18} Caputo, M.:
   Linear models of dissipation whose Q is almost frequency
 independent,	{\it Annali di Geo\-fisica},
 {\bf 19} (1966), 383-393.
\vsp
\vfill\eject
\rf{19}  Caputo, M.:
  Linear models of dissipation whose Q is almost frequency
  independent,	 Part II.,
{\it Geophys. J. R. Astr. Soc.}, {\bf 13}  (1967), 529-539.
\vsp
\rf{20}  Caputo, M.: {\it Elasticit\`a e Dissipazione},
     Zanichelli, Bologna 1969.
\vsp
\rf{21} Caputo, M.:
  Vibrations of an infinite viscoelastic layer with a dissipative memory,
 {\it J. Acoust. Soc. Am.}, {\bf 56} (1974), 897-904.
\vsp
\rf{22} Caputo, M.:
 A model for the fatigue in elastic materials with frequency
 independent Q,
{\it J. Acoust. Soc. Am.}, {\bf 66} (1979), 176-179.
\vsp
\rf{23}  Caputo, M:
  Generalized rheology and geophysical consequences,
{\it Tectonophysics}, {\bf 116} (1985), 163-172.
\vsp
\rf{24}  Caputo, M.:
     Modern rheology and electric induction: multivalued
  index of refraction, splitting of eigenvalues and fatigues,
  {\it An\-nali di Geo\-fi\-sica}, {\bf 39} (1996),  941-966.
\vsp
\rf{25}  Mainardi, F. and  E. Bonetti:
The appli\-cation of real-order derivatives in linear
visco\-elasticity,
{\it  Rheol. Acta}, {\bf 26} Suppl. (1988),  64-67.
\vsp
\rf{26}  Mainardi,  F.:
 Fractional relaxation in anelastic solids,
{\it J. Alloys and Compounds}, {\bf 211/212} (1994), 534-538.
\vsp
\rf{27}  Smith, W. and	H. de Vries:
 Rheological models containing	fractional derivatives,
  {\it	Rheol. Acta}, {\bf 9}  (1970), 525-534.
\vsp
\rf{28} Scarpi, G.B.:
 Sui modelli reologici intermedi per liquidi viscoelastici,
 {\it Atti Accademia Scienze Torino, Classe Sci. fis. mat. nat.},
 {\bf 107} (1973), 239-243.
\vsp
\rf{29} Stiassnie, M.:
  On the application of fractional calculus for the formulation of
  viscoelastic models, {\it Appl. Math. Modelling}, {\bf 3}
    (1979), 300-302.
\vsp
\rf{30}  Bagley, R.L. and  P.J. Torvik:
  A generalized derivative model for an elastomer damper,
  {\it Shock Vib. Bull.}, {\bf 49}  (1979), 135-143.
\vsp
\rf{31} Bagley, R.L. and  P.J. Torvik:
 A theoretical basis for the application  of fractional calculus,
  {\it J. Rheology}, {\bf 27}  (1983), 201-210.
\vsp
\rf{32} Torvik, P.J. and  R.L. Bagley:
On the appearance of the fractional derivatives in the behavior of
   real materials,
{\it J. Appl. Mech.}, {\bf 51} (1984), 294-298.
\vsp
\rf{33} Bagley R.L. and  P.J. Torvik:
 On the fractional calculus model of viscoelastic behavior,
  {\it J. Rheology}, {\bf 30} (1986),  133-155.
\vsp
\rf{34} Rogers, L.:
   Operator and fractional derivatives for viscoelastic constitutive
  equations,
  {\it J. Rheology}, {\bf 27} (1983), 351-372.
\vsp
\rf{35}  Koeller,  R.C.:
  Applications of fractional calculus to the theory of
 viscoelasticity,
{\it J. Appl. Mech.}, {\bf 51} (1984),	299-307.
\vsp
\rf{36}  Koeller, R.C.:
 Polynomial operators, Stieltjes convolution and fractional calculus
 in hereditary mechanics,
 {\it Acta Mech.}, {\bf 58} (1986), 251-264.
\vsp
\vfill\eject
\rf{37} Koh, C.G. and J.M. Kelly:
   Application of fractional derivatives to seismic analysis of
   base-isolated models,
  {\it Earthquake Engineering and Structural Dynamics},
  {\bf 19} (1990), 229-241.
\vsp
\rf{38} Friedrich,  C.:
  Mechanical stress relaxation in polymers: fractional integral
  model versus	fractional differential model,
  {\it J.  Non-Newtonian Fluid Mechanics},
 {\bf 46} (1993), 307-314.
\vsp
\rf{39}  Nonnenmacher, T.F. and  W.G. Gl\"ockle:
 A fractional model for mechanical stress relaxation,
    {\it Phil. Mag. Lett.}, {\bf 64} (1991), 89-93.
\vsp
\rf{40}  Gl\"ockle, W.G. and  T.F. Nonnenmacher:
  Fractional relaxation and the time-temperature superposition
  principle,  {\it Reologica Acta}, {\bf 33} (1994), 337-343.
\vsp
\rf{41} Makris, N. and M.C. Constantinou:
 Models of viscoelasticity with complex-order derivatives,
 {\it J. Eng. Mech.}, {\bf 119} (1993), 1453-1464.
\vsp
\rf{42} Heymans, N. and J.-C. Bauwens:
   Fractal rheological models and fractional differential equations
 for viscoelastic  behavior,
  {\it Rheologica Acta}, {\bf 33} (1994), 219-219.
\vsp
\rf{43}  Schiessel, H.,  Metzler, R.,  Blumen, A. and
   T.F. Nonnenmacher:
  Generalized  viscoelastic models: their fractional equations
  with solutions,
 {\it J. Physics A: Math. Gen.}, {\bf 28} (1995), 6567-6584.
\vsp
\rf{44} Gaul, L., Klein, P. and S. Kempfle:
   Damping description involving fractional operators,
  {\it Mechanical Systems and Signal Processing}, {\bf 5} (1989), 81-88.
\vsp
\rf{45} Beyer, H. and S. Kempfle:
   Definition of physically consistent damping laws with
  fractional derivatives,
  {\it ZAMM}, {\bf 75} (1995), 623-635.
\vsp
\rf{46} Fenander, {\AA}.:
   Modal synthesis when modelling damping by use of fractional
  derivatives,
  {\it AIAA Journal}, {\bf 34} (1996), 1051-1058.
\vsp
\rf{47} Pritz, T.:
   Analysis of four-parameter fractional derivative model
   of real solid materials,
  {\it J. Sounds Vibr.}, {\bf 195} (1996), 103-115.
\vsp
\rf{48} Rossikhin, Yu.A. and M.V. Shitikova:
 Applications of fractional calculus to dynamic problems of linear
 and non linear hereditary mechanics of solids,
 {\it Appl. Mech. Rev.}, {\bf 50} (1997), 16-67.
\vsp
\rf{49} Rossikhin  Yu.A. and  Shitikova, M.V.:
 Application of fractional operators to the  analysis of damped
vibrations of viscoelastic single-mass systems,
  {\it Journal of Sound and Vibration},  {\bf 199} No 4 (1997), 567-586.
\vsp
\rf{50} Lion, A.:
 On the thermodynamics of fractional damping elements within
 the framework of rheological models,
{\it Continuum Mechanics and Thermodynamics}, {\bf 9} (1997), 83-96.
\vsp
\vfill\eject
\rf{51} Boussinesq, J.:
 Sur la r\'esistance qu'oppose un liquid ind\'efini en repos,
 san pesanteur, au mou\-vement vari\'e d'une sph\`ere solide qu'il
 mouille sur toute sa surface, quand les vi\-tesses restent bien
 continues et assez faibles pour que leurs carr\'es et
 produits soient n\'egligea\-bles,
 {\it C.R. Acad. Paris}, {\bf 100} (1885), 935-937.
\vsp
\rf{52}  Basset, A. B.:
  {\it A Treatise on Hydrodynamics}, Vol.2,
  Deighton Bell, Cambridge 1888, Chap. 22.
\vsp
\rf{53}  Stokes, G.G.:
  On the effect of the internal friction of fluids on the motion
   of pendulums,
{\it Cambridge Phil. Trans.} (Ser. VIII), {\bf 9} (1851), ???-???,
reprinted in
{\it Mathematical and Physical Papers}, Vol. III, pp. 1-141,
Cambridge Univ. Press, 1922.
 \vsp
\rf{54}  Picciati, G.:
 Sul moto di una sfera in un liquido viscoso,
 {\it Rend. R. Acc. Naz. Lincei} (ser. 5), {\bf 16} (1907), 943-951.
 [1-st sem.]
\vsp
\rf{55}  Boggio, T.:
 Integra\-zione dell'equa\-zione fun\-ziona\-le che reg\-ge la
  ca\-du\-ta di una  sfera in un li\-quido vi\-scoso,
 {\it Rend. R. Acc. Naz. Lincei} (ser. 5),
 {\bf 16} (1907), 613-620, 730-737. [2-nd sem.]
\vsp
\rf{56} Basset, A.B.:
 On the descent of a sphere in a viscous liquid,
 {\it Quart. J. Math}, {\bf 41} (1910), 369-381.
\vsp
\rf{57}   Hughes, R.R. and E.R. Gilliand: The mechanics of drops,
 {\it Chem. Engng. Progress}, {\bf 48} (1952), 497-504.
\vsp
\rf{58}  Odar, F. and W.S. Hamilton:
 Forces on a sphere accelerating in a viscous fluid,
  {\it J. Fluid Mech.}, {\bf 18} (1964), 302-314.
\vsp
\rf{59}  Odar, F.:
 Verification of the proposed equation for calculation of forces on a
 sphere accelerating in a viscous fluid,
 {\it J. Fluid Mech.}, {\bf 25} (1966), 591-592.
\vsp
\rf{60}  Maxey, M.R. and J.J. Riley:
  Equation of motion for a small rigid sphere in a nonuniform flow,
  {\it Phys. Fluids}, {\bf 26} (1983), 883-889.
\vsp
\rf{61}  McKee, S. and	A. Stokes:
 Product integration methods for the nonlinear Basset equation,
 {\it SIAM J. Numer. Anal.}, {\bf 20} (1983), 143-160.
\vsp
\rf{62}  Reeks, M.V. and S. McKee:
 The dispersive effects of Basset history forces
  on particle motion in a turbulent flow,
 {\it Phys. Fluids}, {\bf 27} (1984), 1573-1582.
\vsp
\rf{63}  Lovalenti, P.M. and J.F. Brady:
 The hydrodynamic force on a rigid particle undergoing arbitrary
  time-dependent motion at small Reynolds number,
 {\it J. Fluid Mech.}, {\bf 256} (1993), 561-605.
\vsp
\rf{64}   Mei, R.:
 History forces on a sphere due to a step change in the free-stream
  velocity,
  {\it Int. J. Multiphase Flow}, {\bf 19} (1993), 509-525.
\vsp
\rf{65}  Lawrence, C.J. and  R. Mei:
 Long-time behaviour of the drag on a body in impulsive motion,
 {\it J. Fluid. Mech.}, {\bf 283} (1995), 307-327.
\vsp
\rf{66}  Lovalenti, P.M. and J.F. Brady:
 The temporal behaviour of the hydrodynamic force on a body in response
to an abrupt change in velocity at small but finite Reynolds number,
 {\it J. Fluid Mech.}, {\bf 293} (1995), 35-46.
\vsp
\rf{67}  Mainardi, F.,	Pironi, P. and F. Tampieri:
  On a generalization of the Basset problem via  fractional calculus,
   in {\it Proceedings	CANCAM 95}
  (Eds.  Tabarrok, B and S. Dost),    Vol. II (1995),  836-837.
   [15-th Canadian Congress of Applied Mechanics,
   Victoria, British Columbia, Canada, 28 May - 2 June 1995].
\vsp
\rf{68} Mainardi, F.,  Pironi, P. and F. Tampieri:
  A numerical approach to the generalized Basset problem for a sphere
  accelerating in a viscous fluid, in
  {\it Proceedings  CFD 95}
   (Eds.  Thibault, P. A. and D.M. Bergeron),
  Vol. II (1995),  105-112.
   [3-rd Annual Conference of the Computational Fluid Dynamics Society
   of Canada, Banff, Alberta, Canada, 25-27 June 1995].
\vsp
\rf{69} Tatom, F.B.: The Basset term as a semiderivative,
  {\it Appl. Sci. Res.}, {\bf 45} (1988), 283-285.
\vsp
\rf{70}   Wax, N. (Ed.):
   {\it Selected Papers on Noise and Stochastic Processes},
   Dover, New-York 1954.
\vsp
\rf{71}  Fox, R.F. and G.E. Uhlenbeck:
     Contributions to non-equilibrium thermodynamics.
     I. Theory of hydro dynamical fluctuations,
     {\it Phys. Fluids}, {\bf 13} (1970), 1893-1902.
\vsp
\rf{72}  Fox, R.F.: Gaussian Stochastic Processes in Physics,
  {\it Physics Reports}, {\bf 48} (1978), 179-283.
\vsp
\rf{73}  Kubo, R., Toda, M., and  N. Hashitsume:
   {\it Statistical Physics II, Nonequilibrium Statistical Mechanics},
   Springer Verlag, Berlin 1991.
\vsp
\rf{74}  Alder, B.J. and  T.E. Wainwright:
 Decay of velocity autocorrelation function,
   {\it Phys. Rev. A}, {\bf 1} (1970), 18-21.
\vsp
\rf{75} Ernst, M.H.,  Hauge, E.H. and J.M.J. Leenwen:
   Asymptotic time behavior of correlation functions,
  {\it Phys. Rev. Lett.}, {\bf 25} (1970),  1254-1256
\vsp
\rf{76}  Dorfman, J.R. and E.G. Cohen:
  Velocity correlation functions in two and three dimensions,
  {\it Phys. Rev. Lett.}, {\bf 25} (1970),   1257-1260
\vsp
  \rf{77} Zwanzig, R. and  M. Bixon:
   Hydrodynamic theory of the velocity correlation function,
   {\it Phys. Rev. A}, {\bf 2} (1970),	2005-2012.
\vsp
\rf{78}  Kawasaki, K.:
  Long time behavior of the velocity autocorrelation function
  {\it Phys. Lett.}, {\bf 32A} (1971),	379-380.
\vsp
\rf{79}   Widom, A.:
  Velocity fluctuations of a hard-core Brownian particle,
   {\it Phys. Rev. A}, {\bf 3} (1971), 1394-1396.
\vsp
\rf{80}  Case, K.M.:
   Velocity fluctuations of a body in a fluid,
   {\it Phys. Fluids}, {\bf 14} (1971),  2091-2095.
\vsp
\rf{81}  Mazo,	R.M.:
   Theory of Brownian motion IV; a hydrodynamic model for the friction
   factor,
   {\it J. Chem. Phys.}, {\bf 54} (1971), 3712-3713.
\vsp
\rf{82} Nelkin, M.:
   Inertial effects in motion driven by hydrodynamic fluctuations,
   {\it  Phys. Fluids}, {\bf 15} (1972),  1685-1690.
\vsp
\rf{83} Chow, Y.S. and	J.J. Hermans:
   Effect of inertia on the Brownian motion of rigid particles
   in a viscous fluid,
   {\it J. Chem. Phys.}, {\bf 56} (1972),  3150-3154.
\vsp
\rf{84}  Hynes, J.T.:
   On hydrodynamic models for Brownian motion,
   {\it J. Chem. Phys.}, {\bf 57} (1972), 5612-5613.
\vsp
\rf{85}  Pomeau, Y.:   Low-frequency behavior of transport coefficients
 in fluids,   {\it Phys. Rev. A}, {\bf 5} (1972),   2569-2587.
\vsp
\rf{86}  Keizer, J.:
   Comment on effect of inertia on  Brownian motion,
   {\it J. Chem. Phys.}, {\bf 58} (1973), 824-825.
\vsp
\rf{87}  Davis, H.T. and G. Subramian:
   Velocity fluctuation of a Brownian particle: Widom's model
   {\it J. Chem. Phys.}, {\bf 58} (1973), 5167-5168.
\vsp
\rf{88}  Hauge, E.H. and  A. Martin-L\"of:
   Fluctuating hydrodynamics and Brownian motion,
   {\it J. Stat. Phys.}, {\bf 7} (1973),  259-281.
\vsp
\rf{89} Dufty, J.W.:
     Gaussian model for fluctuation of a Brownian particle,
     {\it Phys. Fluids}, {\bf 17} (1974), 328-333.
\vsp
\rf{90}  Szu H.H.,  Szu, S.C. and J.J. Hermans:
     Fluctuation-dissipation theorems on the basis of hydrodynamic
     propagators,
     {\it Phys. Fluids}, {\bf 17} (1974), 903-907.
\vsp
\rf{91} Bedeaux, D. and P. Mazur:
  Brownian motion and fluctuating hydrodynamics,
  {\it Physica}, {\bf 76} (1974) 247-258.
\vsp
\rf{92}  Hinch,  E.J.:
   Applications of the Langevin equation to fluid suspension,
   {\it J. Fluid. Mech.}, {\bf 72} (1975),  499-511.
\vsp
\rf{93} Y. Pomeau  and P. R\'esibois:
   Time dependent correlation functions and mode-mode coupling theories,
   {\it Physics Reports}, {\bf 19} (1975), 63-139.
\vsp
\rf{94}  Warner,  M.:
   The long-time fluctuations of a Brownian sphere,
   {\it J. Phys. A: Math. Gen.}, {\bf 12} (1979),  1511-1519.
\vsp
\rf{95}  Paul, G.L. and P.N. Pusey:
   Observation of a long-time tail in Brownian motion,
   {\it J. Phys. A: Math. Gen.}, {\bf 14} (1981),  3301-3327.
\vsp
\rf{96}  Reichl,   L.E.:
   Translation Brownian motion in a fluid with internal  degrees of
   freedom,
   {\it Phys. Rev.}, {\bf 24} (1981),  1609-1616.
\vsp
\rf{97} Clercx, H.J.H. and P.P.J.M. Schram:
   Brownian particles in shear flow and harmonic potentials:
   a study of long-time tails,
   {\it  Phys. Rev. A}, {\bf 46} (1992),  1942-1950.
\vsp
\rf{98} Muralidhar, R., Ramkrishna, D., Nakanishi, H.,
   and D.J. Jacobs:
  Anomalous diffusion: a dynamic perspective,
   {\it Physica A}, {\bf 167} (1990),  539-559.
\vsp
\rf{99}  Bouchaud, J.-P. and  A. Georges:
   Anomalous diffusion in disordered media: sta\-tistical mechanisms,
models	  and physical applications,
   {\it Physics Reports}, {\bf 195} (1990),  127-293.
\vsp
\rf{100}  Wang,  K.C.:
   Long-time correlation effects and biased anomalous diffusion,
   {\it Phys. Rev. A}, {\bf 45} (1992),  833-837.
\vsp
\rf{101}  Giona, M.  and  H.E. Roman:
  Fractional diffusion equation for transport phenomena in random
  media,  {\it Physica A}, {\bf 185} (1992),  82-97.
\vsp
\rf{102}  Metzler, R.,	Gl\"ockle, W.G. and T.F. Nonnenmacher:
   Fractional model equation for anomalous diffusion,
   {\it Physica A}, {\bf 211} (1994),  13-24.
\vsp
\rf{103}  Kubo, R.:
   The fluctuation-dissipation theorem,
   {\it Rep. on Progress in Physics}, {\bf 29} (1966),	255-284.
\vsp
\rf{104}  Felderhof, B.U.:
 On the derivation of the fluctuation-dissipation theorem,
  {\it J. Phys. A: Math. Gen.}, {\bf 11} (1978),  921-927.
\vsp
\rf{105} Mainardi, F. and P. Pironi:
 The fractional Langevin equation: the Brownian motion revisited,
 {\it Extracta Mathematicae}, {\bf 11} (1996),	140-154.
\vsp
\rf{106}  Mainardi, F. and F. Tampieri:  
Diffusion regimes in Brownian induced by the Basset history force,
{\it Technical Report No 1, ISAO-TR1/99} ISAO-CNR, Bologna (Italy),
March 1999, pp. 25.
[Reprinted as FRACALMO PRE-PRINT 0102, see {\tt http://www.fracalmo.org}]
\vsp
\rf{107}  Gel'fand, I.M.  and  G.E. Shilov:
  {\it Generalized Functions}, Vol. 1,
  Academic Press, New York 1964.
\vsp
\rf{108} Zemanian, A.H.:
 {\it Distribution Theory and Transform Analysis}, McGraw-Hill,
 New York 1965.
\vsp
\rf{109}   Doetsch, G.:
  {\it Introduction to the Theory and Application of the Laplace
  Transformation}, Springer Verlag, Berlin 1974.
\vsp
\rf{110}  Nigmatullin, R.R.:
  The realization of the generalized transfer equation in a medium with
  fractal geometry,
   {\it  Phys. Stat. Sol. B}, {\bf 133} (1986), 425-430.
  [English transl. from Russian]
\vsp
\rf{111} Mainardi,  F.:
  Fractional diffusive waves  in viscoelastic solids
  in  {\it IUTAM Symposium - Nonlinear Waves in Solids}
  (Ed. J. L. Wegner  and F. R. Norwood), ASME/AMR, Fairfield NJ 1995,
   93-97.  [Abstract in {\it Appl. Mech. Rev.}, {\bf 46} (1993), 549]
\vsp
\rf{112}  Mainardi,  F.:
 On the initial value problem for the fractional
  diffusion-wave equation, in
  {\it Waves and Stability in Continuous Media}
  (Ed. S. Rionero and T. Ruggeri),
  World Scientific, Singapore 1994, 246-251.
\vsp
\rf{113} Mainardi, F.: The time fractional diffusion-wave equation,
  {\it Radiofisika}, {\bf 38} (1995), 20-36.
  [English Translation: Radiophysics and Quantum Electronics]
\vsp
\rf{114} Mainardi, F. and M. Tomirotti: On a special function
  arising  in  the time fractional diffusion-wave equation,
  in {\it Transform Methods and Special Functions, Sofia 1994}
    (Ed. P. Rusev, I. Dimovski and V. Kiryakova),
  Science Culture Technology, Singapore 1995, 171-183.
\vsp
\rf{115} Mainardi, F.: Fractional relaxation-oscillation and fractional
  diffusion-wave phenomena,
  {\it Chaos, Solitons and Fractals}, {\bf 7} (1996), 1461-1477.
\vsp
\rf{116} Mainardi, F. :
  The fundamental solutions for the fractional diffusion-wave
   equation,
   {\it Applied Mathematics Letters}, {\bf 9}, No 6 (1996), 23-28.
\vsp
 \rf{117}  Wyss, W.:  Fractional diffusion  equation,
  {\it J. Math. Phys.}, {\bf 27} (1986), 2782-2785.
\vsp
\rf{118}  Schneider, W.R. and  W. Wyss:
    Fractional diffusion and wave equations,
   {\it J. Math. Phys.}, {\bf 30} (1989), 134-144.
\vsp
\rf{119}  Schneider, W.R:
    Fractional diffusion,
   in: {\it Dynamics and Stochastic Processes, Theory and Applications}
   (Eds. R. Lima, L. Streit and D. Vilela Mendes),
  Lecture Notes in Physics \# 355,
  Springer Verlag, Heidelberg 1990,  276-286.
\vsp
\rf{120} Fujita, Y.:
  Integrodifferential equation which interpolates the heat equation
 and the wave equation,
  {\it Osaka J. Math.}, {\bf 27} (1990), 309-321, 797-804.
\vsp
\rf{121} Kochubei, A.N.:
  A Cauchy problem for evolution equations of fractional order,
  {\it J. Diff. Eqns}, {\bf 25} (1989), 967-974.
  [English transl. from Russian]
\vsp
\rf{122} Kochubei, A.N.:
  Fractional order diffusion,
  {\it J. Diff. Eqns}, {\bf 26} (1990), 485-492.
  [English transl. from Russian]
\vsp
\rf{123} El-Sayed, A.M.A.:
 Fractional-order diffusion-wave equation,
 {\it Int. J. Theor. Phys.}, {\bf 35} (1996), 311-322.
\vsp
\rf{124} Engler, H.:
 Similarity solutions for a class of hyperbolic integrodifferential
 equations,
  {\it Differential Integral Eqns}, {\bf 10} (1997),  815-840.
\vsp
\rf{125}  Erd\'elyi, A., Magnus, W., Oberhettinger and F.G. Tricomi:
  {\it Higher Transcendental Functions},
  Bateman Project,  Vol. 3,  McGraw-Hill, New York 1955, Ch. 18.
\vsp
\rf{126}  Wright, E.M.:
 On the coefficients of power series having exponential singularities,
 {\it J. London Math. Soc.}  {\bf 8} (1933), 71-79.
\vsp
\rf{127}  Wright, E.M.:
  The generalized Bessel function of order greater than one,
  {\it Quart. J. Math.} (Oxford ser.) {\bf 11} (1940), 36-48.
\vsp
\rf{128}  Stankovi{\'c}, B.:
 On the function of E.M. Wright,
 {\it Publ. Institute Math. Beograd} (Nouv. serie)
 {\bf 10}, No 24 (1970), 113-124.
\vsp
\rf{129}  Gaji{\'c}, Lj.  and  B. Stankovi{\'c}, \
  Some properties of Wright's function,
  {\it Publ. Institute Math. Beograd} (Nouv. serie)
  {\bf 20}, No 34 (1976), 91-98.
\vsp
\rf{130}  Kiryakova, V.:
    {\it Generalized Fractional Calculus and Applications},
     Pitman Research Notes in Mathematics \# 301,
     Longman, Harlow 1994.
\vfill\eject
\rf{131} Tricomi, F.G.:
 {\it Fonctions Hyperg\'eometriques Confluentes},
  M\'em. Sci. Math. \# 140, Gauthier-Villars, Paris 1960.
\vsp
\rf{132} Gatteschi, L.: {\it Funzioni Speciali},
 UTET, Torino 1973, pp. 196-197.
\vsp
\rf{133}  Bender, C.M.	and  S.A. Orszag:
 {\it Advanced Mathematical Methods for Scientists and Engineers},
  McGraw-Hill, Singapore  1987, Ch 3.
\vsp
\rf{134}  Mainardi, F.	 and  M. Tomirotti:
  The asymptotic representation of the generalized hyper-Airy function
   in the complex plane,
  {\it PRE-PRINT}, Dept. of Physics, University of Bologna, 1996.
  \vsp
\rf{135} Pollard, H.:
 The representation of $\,{\rm exp}\,( -x^\lambda)\,$  as a Laplace
 integral,  {\it Bull. Amer. Math. Soc.}, {\bf 52} (1946), 908-910.
\vsp
\rf{136} Humbert, P.:
 Nouvelles correspondances symboliques,
{\it Bull. Sci. Math\'em.} (Paris, II ser.), {\bf 69} (1945), 121-129.
\vsp
\rf{137}  Mikusi{\'n}ski, J.:
    On the function whose Laplace transform
    is $\,{\rm exp}\, (-s^\alpha \lambda )\,,$
    {\it Studia Math.}, {\bf 18} (1959), 191-198.
\vsp
\rf{138}  Buchen, P.W. and F. Mainardi:
  Asymptotic expansions for transient  viscoelastic waves,
  {\it Journal de M{\'e}canique}, {\bf 14} (1975), 597-608.
\vsp
\rf{139}  Mainardi, F. and M. Tomirotti:
  Seismic pulse propagation with constant $Q$ and stable probability
  distributions,
  {\it Annali di Geofisica}, {\bf 40} (1997),  1311-1328.
\vsp
\rf{140}  Mainardi, F.,  Paradisi, P and R. Gorenflo:
  Probability distributions generated by  fractional
  diffusion equations,
   in  {\it Econophysics: an Emerging Science}
  (Eds.  Kertesz, J. and I. Kondor),
  Kluwer, Dordrecht ????, pp. 39, to appear.
[Pre-print available at {\tt http://www.fracalmo.org}].
 \vsp
\rf{141}    Gorenflo, R. and F. Mainardi:
  Fractional calculus and stable probability distributions,
 {\it Archives of Mechanics}, {\bf 50} (1998), 377-388.
\vsp
\rf{142}
{Gorenflo, R.,	Luchko, Yu. and   F. Mainardi}:
Analytical properties and applications of the Wright function.
{\it Fractional Calculus and Applied Analysis}, {\bf 2} (1999), 383-414.
\vsp
\rf{143}
{Gorenflo, R.,	Luchko, Yu. and   F. Mainardi}:
Wright functions as scale-invariant solutions of the diffusion-wave
 equation,
{\it J. Computational and Appl. Mathematics\/},
     {\bf 118}	(2000), 175-191.
\vsp
\rf{144}
 Wong, R. and Y.-Q. Zhao:
Smoothing of Stokes' discontinuity for the generalized Bessel function,
{\it Proc. R. Soc. London A}, {\bf 455} (1999), 1381--1400.
\vsp
\rf{145}
Wong, R. and Y.-Q. Zhao:
Smoothing of Stokes' discontinuity for the generalized Bessel function II,
{\it Proc. R. Soc. London A}, {\bf 455} (1999), 3065--3084.
